\documentclass{aa}

\usepackage[varg]{txfonts}
\usepackage{graphicx}
\usepackage{multirow}

\usepackage{url,twoopt,natbib}
\usepackage{hyperref}                
\usepackage{pdfcomment,acronym}      
\hypersetup{
  colorlinks=true,   
  urlcolor=blue,     
  linkcolor=red,     
}

\makeatletter
\newcommand{\bibnote}[2]{\@namedef{#1note}{#2}}
\newcommand{\biblink}[2]{\@namedef{#1link}{#2}}
\makeatother

\makeatletter
 \newcommandtwoopt{\citeads}[3][][]{%
   \nonstopmode
   \href{http://adsabs.harvard.edu/abs/#3}%
        {\def\hyper@linkstart##1##2{}%
         \let\hyper@linkend\@empty\citealp[#1][#2]{#3}}
   \biblink{#3}{\href{http://adsabs.harvard.edu/abs/#3}{ADS}}%
   \errorstopmode}            
 \newcommandtwoopt{\citepads}[3][][]{%
   \nonstopmode
   \href{http://adsabs.harvard.edu/abs/#3}%
        {\def\hyper@linkstart##1##2{}%
         \let\hyper@linkend\@empty\citep[#1][#2]{#3}}
   \biblink{#3}{\href{http://adsabs.harvard.edu/abs/#3}{ADS}}%
   \errorstopmode}            
 \newcommandtwoopt{\citetads}[3][][]{%
   \nonstopmode
   \href{http://adsabs.harvard.edu/abs/#3}%
        {\def\hyper@linkstart##1##2{}%
         \let\hyper@linkend\@empty\citet[#1][#2]{#3}}
   \biblink{#3}{\href{http://adsabs.harvard.edu/abs/#3}{ADS}}%
   \errorstopmode}            
 \newcommandtwoopt{\citeyearads}[3][][]{%
   \nonstopmode
   \href{http://adsabs.harvard.edu/abs/#3}%
        {\def\hyper@linkstart##1##2{}%
         \let\hyper@linkend\@empty\citeyear[#1][#2]{#3}}
   \biblink{#3}{\href{http://adsabs.harvard.edu/abs/#3}{ADS}}%
   \errorstopmode}            
\makeatother

\begin{document}

\title{Investigating the rotational evolution of very low-mass stars
and brown dwarfs in young clusters using Monte Carlo simulations}

\author{M. J. Vasconcelos\inst{\ref{UESC},\ref{IPAG}} \and 
J. Bouvier\inst{\ref{IPAG}}}

\institute{LATO - DCET, Universidade Estadual de Santa Cruz, UESC, \\
Rodovia Jorge Amado, km 16, Ilh\'eus/BA, 45662-900, Brazil
\email{mjvasc@uesc.br}\label{UESC}
\and
Univ. Grenoble Alpes, CNRS, IPAG, F-38000 Grenoble, France\label{IPAG}}

\date{Received / Accepted}

\abstract{Very low-mass (VLM) stars and brown dwarfs (BDs) present
a different rotational behaviour from their solar mass counterparts.}
{We investigate the rotational evolution of young VLM stars and BDs
using Monte Carlo simulations under the hypothesis of disk locking
and stellar angular momentum conservation.}  {We built a set of
objects with masses ranging from 0.01 M$_\odot$ to 0.4 M$_\odot$
and considered models with single- and double-peaked initial period
distributions with and without disk locking. An object is considered
to be diskless when its mass accretion rate is below a given
threshold.} {Models with initial  single-peaked period distributions
reproduce  the observations well given that BDs rotate faster than
VLM stars.  We observe a correlation between rotational period and
mass when we relax the disk locking hypothesis, but with a shallower
slope compared to some observational results.  The angular momentum
evolution of diskless stars is flatter than it is for stars with a
disk which occurs because the moment of inertia of objects less
massive than 0.2 M$_\odot$ remains pratically constant for a time
scale that increases with decreasing stellar mass.}{Comparing our
results with the available observational data we see that  disk
locking is not as important in the low-mass regime and that the
rotational behaviour of VLM stars and BDs is different from what
is seen in their solar mass counterparts.}


\titlerunning{Rotational evolution of VLM stars and BDs}
\authorrunning{Vasconcelos \& Bouvier}

\maketitle

\section{Introduction}

\bibnote{2015A&A...578A..89V}{(Paper~I)}
\bibnote{2010A&A...515A..13R}{(RL10)}
\bibnote{2014MNRAS.444.1157D}{(D14)}
\bibnote{2004A&A...419..249S}{(S04)}
\bibnote{2005A&A...429.1007S}{(S05)}
\bibnote{2015ApJ...809L..29S}{(S15)}
\bibnote{2010ApJS..191..389C}{(C10)}
\bibnote{2008MNRAS.384..675I}{(I08)}

\def\PaperI{\href{http://adsabs.harvard.edu/abs/2015A&A...578A..89V}{Paper~I}}
\def\RLX{\href{http://adsabs.harvard.edu/abs/2010A&A...515A..13R}{RL10}}
\def\DXIV{\href{http://adsabs.harvard.edu/abs/2014MNRAS.444.1157D}{D14}}
\def\SIV{\href{http://adsabs.harvard.edu/abs/2004A&A...419..249S}{S04}}
\def\SXV{\href{http://adsabs.harvard.edu/abs/2015ApJ...809L..29S}{S15}}
\def\CX{\href{http://adsabs.harvard.edu/abs/2010ApJS..191..389C}{C10}}
\def\IVIII{\href{http://adsabs.harvard.edu/abs/2008MNRAS.384..675I}{I08}}

Young stars are subjected to several phenomena that affect their
overall evolution. Accretion, ejection, and angular momentum evolution
are connected in a complex way (see e.g. \citeads{2014prpl.conf..433B}).
For solar-type stars, the disk seems to play an important role in
preventing the star from spinning up to its break-up speed, but it
has a short duration. Even so, the disk influence on the rotational
properties of young stellar clusters is remarkable.
\citetads{2004AJ....127.1029R} and \citetads{2005ApJ...633..967H}
investigated with observational data and simple models the influence
of accretion disks on the evolution of rotational rates in young
clusters. In Vasconcelos \& Bouvier (\citeyearads{2015A&A...578A..89V},
hereafter Paper~I), Monte Carlo simulations were used to show that
indeed the rotational properties of clusters younger than 20 Myr
can be explained by the co-existence of stars with disks prevented
from spinning up and diskless stars that are being rotationally
accelerated while conserving angular momentum.  The double-peaked
period distributions seen in the Orion nebula cluster (ONC), in NGC
2264 \citepads{2007ApJ...671..605C}, and in h Per
\citepads{2013A&A...560A..13M} show both populations present in the
peaks, but there is a predominance of stars with disks in the
long-period peak. This was seen not only in the period distributions,
but also when we examine disk fractions and mass accretion rates
as a function of the period.  In \PaperI\ we also explain the
existence of slowly rotating diskless stars as stars that have
long-lasting disks and that had no time to spin up.  Another finding
of \PaperI\ is that the specific angular momentum evolution is not
very different for disk and diskless stars in a young cluster. The
reason for this is that the disk lifetime is not the same for all
stars and thus there is a sequential release from the disk locking
condition for a number of stars with a disk.  Individual diskless
stars will start to spin up and conserving angular momentum, but
the average specific angular momentum of the diskless sample as a
whole will decrease.

However, is this rotational behaviour the same for very low-mass
stars or brown dwarfs? Several works show that these low-mass objects
rotate faster than their solar mass counterparts (Scholz et al.
2015). Also, there is a debate in the literature about the disk
fraction at this mass range, with some studies pointing to a higher
disk fraction (Luhman et al. 2008; Luhman \& Mamajek 2012 and Downes
et al. 2015 - see also Table \ref{tabdiskfracBD}). There is also
evidence that disk locking is not as efficient (Lamm et al. 2005).

In this work, we investigate the main variables that can influence
the spin rate evolution of a cluster of very low-mass stars and
brown dwarfs using Monte Carlo simulations, and compare the results
to observations available in the literature. We consider the evolution
of disk and diskless stars and brown dwarfs from 1 Myr to 12.1 Myr.
In section \ref{method}, we explain the main assumptions of the
simulations. In section \ref{results}, we present and discuss the
different models considered in this work. In section \ref{conclusions},
we draw our conclusions.

\section{Monte Carlo simulations \label{method}}

We follow the same set-up as described in detail in \PaperI.  However,
we consider another mass range, from 0.01 M$_\odot$ to 0.4 M$_\odot$,
encompassing the BD (0.01 M$_\odot \leq$ M$_\ast \leq 0.07$ M$_\odot$)
and the very low-mass (VLM) (0.07 M$_\odot <$ M$_\ast \leq 0.4$
M$_\odot$) regimes, separated into bins of 0.01 M$_\odot$ from 0.01
M$_\odot$ to 0.1 M$_\odot$, and into bins of 0.1 M$_\odot$ for
higher masses. We simulate more than 200,000 stars and for each of
them we assign a mass $M_\ast$, a mass accretion rate
$\dot{M}_\mathrm{acc}$, and a rotational period $P$ and we evolve
the system from 1 Myr to 12.1 Myr.  The number of stars per mass
bin is calculated from the canonical initial mass function (IMF)
by \citetads{2013pss5.book..115K}.  The mass accretion rate values
are randomly chosen from 13 different log-normal distributions, one
per each mass bin, all with $\sigma = 0.8$ dex and with mean values
given by

\begin{align}
\langle \dot{M}_\mathrm{acc} (t_0, M_\ast) \rangle = 1 \times 10^{-8}
\, \left(\frac{M_\ast}{M_\odot}\right)^{1.8} \, ,
\end{align}
where $t_0 = 1.0$ Myr is the initial age of the simulations.  The
dispersion of the mass accretion rate was chosen in order to reproduce
the observed spread of more than 2 orders of magnitude in
$\dot{M}_\mathrm{acc}$ seen, for example, in NGC 2264 by
\citetads{2014A&A...570A..82V} and in the ONC by
\citetads{2012ApJ...755..154M}. The mass dependency of the mean
mass accretion rate follows from recent results obtained for Lupus
and $\rho$ - Ophiucus (\citeads{2014A&A...561A...2A},
\citeads{2015A&A...579A..66M}) for stars less massive than 0.5
M$_\odot$. In \PaperI\ we have used $\dot{M}_\mathrm{acc} \propto
M_\ast^{1.4}$ from the results of \citetads{2014A&A...570A..82V}
for NGC 2264.  However, as stated there, our results -- namely the
rotational distributions, the disk fraction as a function of time,
and even the behaviour of the mass accretion rate as a function of
the rotational period -- do not depend on the choice of this exponent.
In Table \ref{tabMac} we show the different values of $\langle
\dot{M}_\mathrm{acc} (t_0, M_\ast) \rangle$ and also the maximum
mass accretion rate value, which is between 4$\sigma$ - 5 $\sigma$,
at each mass bin.

\begin{table}[t]
\caption{Mass accretion rate values \label{tabMac}}
\centering
{\small
\begin{tabular}{cccc}
\hline\hline
Mass & $\langle \dot{M}_\mathrm{acc} (t_0, M_\ast) \rangle$ &
$\dot{M}_\mathrm{acc, max} (t_0, M_\ast)$ & $\dot{M}_\mathrm{acc,
th} (M_\ast)$ \\
(M$_\odot$) & \multicolumn{3}{c}{$(\times$ M$_\odot$ yr$^{-1})$} \\
\hline
0.01 & 2.5 (-12) & 1.9 (-9) & 7.7 (-13) \\
0.02 & 8.7 (-12) & 9.7 (-9) & 2.7 (-12) \\
0.03 & 1.8 (-11) & 9.1 (-9) & 5.6 (-12) \\
0.04 & 3.0 (-11) & 3.2 (-8) & 9.3 (-12) \\
0.05 & 4.5 (-11) & 3.1 (-8) & 1.4 (-11) \\
0.06 & 6.3 (-11) & 4.9 (-8) & 1.9 (-11) \\
0.07 & 8.3 (-11) & 1.2 (-7) & 2.6 (-11) \\
0.08 & 1.1 (-10) & 1.2 (-7) & 3.2 (-11) \\
0.09 & 1.3 (-10) & 9.3 (-8) & 4.0 (-11) \\
0.1  & 1.6 (-10) & 4.0 (-7) & 4.9 (-11) \\
0.2  & 5.5 (-10) & 2.9 (-6) & 1.7 (-10) \\
0.3  & 1.1 (-9)  & 9.0 (-7) & 3.5 (-10) \\
0.4  & 1.9 (-9)  & 1.9 (-5) & 5.9 (-10) \\
\hline
\end{tabular}
\tablefoot{The numbers inside the parentheses give the power of ten of
the mass accretion rate values.}}
\end{table}

Since we want to disentangle the contributions of disk and diskless
stars on the rotational evolution of young very low-mass stars and
brown dwarfs, a key ingredient of our work is the disk lifetime of
individual stars. This parameter is related to the stellar mass
accretion rate value, a quantity that changes with time. As in
\PaperI, we set a mass accretion rate threshold $\dot{M}_\mathrm{acc,th}$
given by

\begin{align} \label{Maccth}
\dot{M}_\mathrm{acc,th} (M_\ast) = 10^{-8} \,
\left(\frac{M_\ast}{M_\odot}\right)^{1.8} \, \left(\frac{t_\mathrm{th}}{t_0}
\right)^{-1.5} \, ,
\end{align}
where $t_\mathrm{th}$ is the threshold time scale. The time dependency
of the mass accretion rate comes from the self-similar accretion
theory by \citetads{1998ApJ...495..385H}. If a star has
$\dot{M}_\mathrm{acc} \geq \dot{M}_\mathrm{acc, th}$ it will have
a disk. Otherwise, it will be diskless. In Table \ref{tabMac} we
show the threshold values obtained assuming $t_\mathrm{th} = 2.2$
Myr.

The evolution equations for the stars assumed to be disk locked are
as follows,

\begin{align} 
P (t) & =  P (t_0) \, , \label{Pdisk}\\
\dot{M}_\mathrm{acc} (t, M_\ast)  & = \dot{M}_\mathrm{acc} (t_0, M_\ast)
\left(\frac{t}{t_0} \right)^{-1.5}  \, , \label{Macdisk}\\
j (t) & =  \frac{I (t) \, \omega (t_0)}{M_\ast} \, , \label{jdisk}
\end{align}
for $t < t_\mathrm{disk}$, where $t_\mathrm{disk}$ is the disk
lifetime, $I (t) = M_\ast k (t) R(t)^2$ is the moment of inertia,
$k$ is the gyration radius, $R$ is the radius, and $\omega = 2\pi
/ P$ is the angular velocity of the star. As in \PaperI, the stellar
parameters are obtained from the stellar evolutionary models of
\citetads{1998A&A...337..403B}.  Equation (\ref{Pdisk}) expresses
the constancy of the rotational period for a star with a disk, a
behaviour known as the ``disk locking hypothesis'' proposed initially
in the context of T Tauri stars by \citetads{1991ApJ...370L..39K}
and required by stellar angular momentum evolution models (e.g.
\citeads{2013A&A...556A..36G}, \citeyearads{2015A&A...577A..98G}).
According to  this hypothesis, even though the star is accreting
it does not spin up owing to its magnetic interaction with its
surrounding accretion disk; however,  the exact mechanism behind
this interaction is not well known (see e.g.
\citeads{2014prpl.conf..433B}).  Equation (\ref{Macdisk}) gives the
temporal evolution of the mass accretion rate derived from fits of
different young clusters (see e.g.  \citeads{1998ApJ...495..385H},
\citeads{2014A&A...570A..82V}, and \citeads{2014A&A...572A..62A}).
With this time-dependency, we were able to reproduce well the
observed disk fraction as a function of time (see Fig. 4 of \PaperI).
However, there are controversies about the right value of the time
exponent  (e.g.  \citeads{2012ApJ...755..154M} and
\citeads{2011A&A...525A..47R}). Equation (\ref{jdisk}) gives the
specific angular momentum for a star with a disk. Since the stellar
angular velocity is constant, $j$ decreases with time as a result
of  the contraction of the stellar radius.  We are also assuming
solid body rotation up to 12.1 Myr.

For diskless stars (with $\dot{M}_\mathrm{acc} < \dot{M}_\mathrm{acc, th}$),
the equations are 

\begin{align}
P (t) & =  P(t - \Delta t) \frac{I(t)}{I(t - \Delta t)} \, ,
\label{Pdiskl} \\
\dot{M}_\mathrm{acc} (t, M_\ast)  & = \dot{M}_\mathrm{acc, th}
(M_\ast) \, , \label{Macdiskl}\\
j (t) & =  j (t_\mathrm{disk}) \, , \label{jdiskl}
\end{align}
valid for $t \geq t_\mathrm{disk}$. We impose conservation of angular
momentum (equation \ref{jdiskl}) since there is no mechanism able
to remove it until the star reaches the zero age main sequence
\citepads{2009MNRAS.400.1548S} at a time beyond the maximum time
considered in our simulations (12.1 Myr). The star is no
longer locked to its disk and it is free to spin up (equation
\ref{Pdiskl}).  The stellar rotational period will decrease because
of the radius contraction. The mass accretion rate is set equal to
the mass accretion rate threshold for the corresponding mass bin
(equation \ref{Macdiskl}). This is an arbitrary choice, but it mimics
the lower mass accretion rate limit of the observations. In the
initial mass accretion rate distributions there are stars that have
$\dot{M}_\mathrm{acc} < \dot{M}_\mathrm{acc, th}$, and for these
stars we also  impose equation (\ref{Macdiskl}).  This causes a bump
in the initial mass accretion rate distributions (see Fig. 3 in
\PaperI).


In the next section, we  analyse the disk fraction as a function
of time obtained with the set-up presented above.

\section{Results} \label{results}

\subsection{Disk fraction} \label{secDF}

The disk fraction as a function of time is an important diagnostic
for disk dispersal mechanisms, disk lifetimes, and hence planetary
formation. Disk fractions have been measured in different young
clusters. The disk detection is based on accretion indicators, such
as H$\alpha$ emission, and/or dust indicators, such as near- and
mid- IR excess, with a recent predominance of mid-IR IRAC colours.
According to \citetads{2009AIPC.1158....3M}, it can be described
by an exponential law with a characteristic time of 2.5 Myr.
\citetads{2014ApJ...793L..34P} raise several concerns\footnote{Basically,
\citetads{2014ApJ...793L..34P} argue that the samples contain mostly
massive and extended clusters and that the observations are based
on just a small fraction of the original cluster population, the
remaining being lost in the course of the cluster evolution. The
observed stars were located originally in the central cluster regions
being then more prone to both internal and external disk dissipation
processes than the population outside this region.} about the
relevance and the reliability of the data used to build the disk
fraction and consider that disk lifetimes can be much longer than
estimated from the observational samples.  Also, the disk fraction
as a function of time depends on the correct determination of the
age of the star or of the cluster to which it belongs, which can
be quite uncertain (\citeads{2013MNRAS.434..806B};
\citeads{2015ApJ...807....3K}; Herczeg \& Hillenbrand
\citeyearads{2015ApJ...808...23H}).

With these uncertainties in mind, we compare disk fractions obtained
from our simulations with available observational values shown in
Table \ref{tabdiskfracBD}. We have assembled disk fractions for
nine young stellar clusters, associations, and groups with ages in
the range from 1 Myr to approximately 10 Myr. We focus on M stars
mostly later than M3 and we show disk fractions for BDs, VLM stars,
and all the stars (BD + VLM) together.  We note that the total disk
fraction for the ONC obtained by \citetads{2010A&A...515A..13R} is
much smaller than the values for the ONC itself (found by
\citeads{2014MNRAS.444.1157D}), Taurus, and Cha I, all regions
with ages around 2 Myr. This is probably due to the use of near-IR
excesses as a disk indicator which could lead to a low level of
disk detections for VLM stars and BDs (see e.g. the discussion in
Cody \& Hillenbrand \citeyearads{2010ApJS..191..389C}). The values
for 25 Ori and TW Hya, on the other hand, can suffer from low number
statistics, and for Upper Scorpius the problem is the great uncertainty
in its age.

\begin{figure}[!htb]
\centerline{\includegraphics[width=8.5cm]{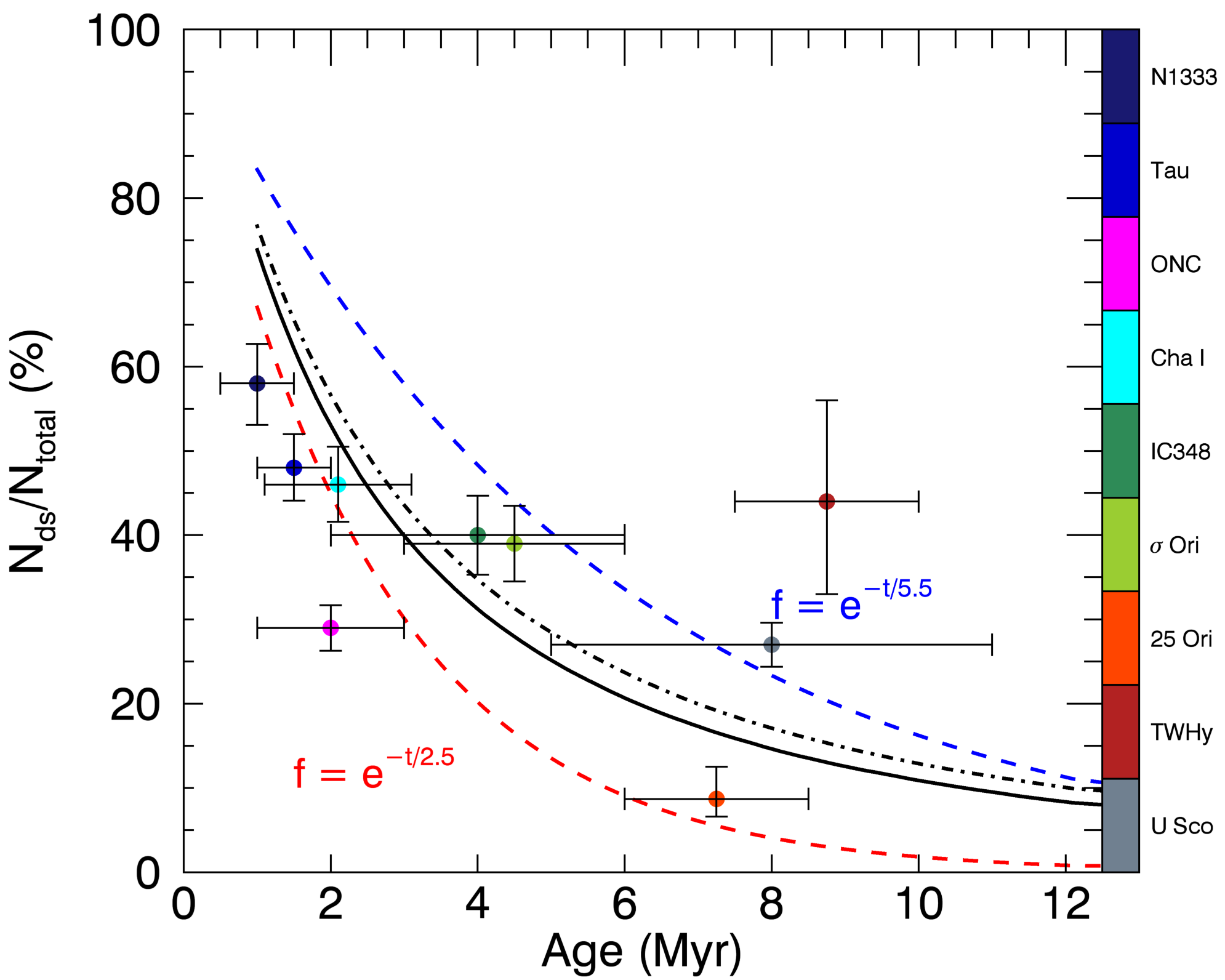}}
\bigskip
\caption{Disk fraction as a function of time obtained from our
simulations taking $t_\mathrm{th} = 2.2$ Myr (solid black line) for
VLM stars and BDs, and assuming different mass accretion rate
threshold time scales for VLM stars $t_\mathrm{th, VLM} = 2.2$ Myr
and for BDs $t_\mathrm{th, BD} = 3.2$ Myr (black dash-dotted line).
The coloured dashed lines are exponential decay laws expected from
disk e-folding times 2.5 Myr (red line) and 5.5 Myr (blue line).
The coloured dots are data for nine young nearby clusters, associations,
and groups taken from Table \ref{tabdiskfracBD}.} \label{diskfrac}
\end{figure}

The relevant parameters to constrain the disk fraction at a given
age in our simulations are the time exponent of the mass accretion
rate (Equation \ref{Macdisk}), the value of the mass accretion rate
threshold (Equation \ref{Maccth} and Table \ref{tabMac}), and the
dispersion of the mass accretion rate distribution (0.8 dex in our
simulations). In Fig. \ref{diskfrac} we plot two numerical disk
fractions as a function of time superimposed on the total disk
fractions from the regions listed in Table \ref{tabdiskfracBD} and
on two exponential laws with short (2.5 Myr) and long (5.5 Myr)
characteristic times. We note that they lie between the two theoretical
curves although they cannot be fitted by a simple exponential law.
One of them was obtained taking $t_\mathrm{th} = 2.2$ Myr. Its
initial disk fraction is  74\% and a disk fraction of 50\% is reached
at 2.2 Myr.  The curve is very similar to that  obtained in \PaperI\
since the relevant parameters are the same (the time exponent of
the mass accretion rate and its dispersion) or are scaled in the
same manner (mass accretion rate values and mass accretion rate
threshold).

\begin{figure}[!htb]
\centerline{\includegraphics[width=8.5cm]{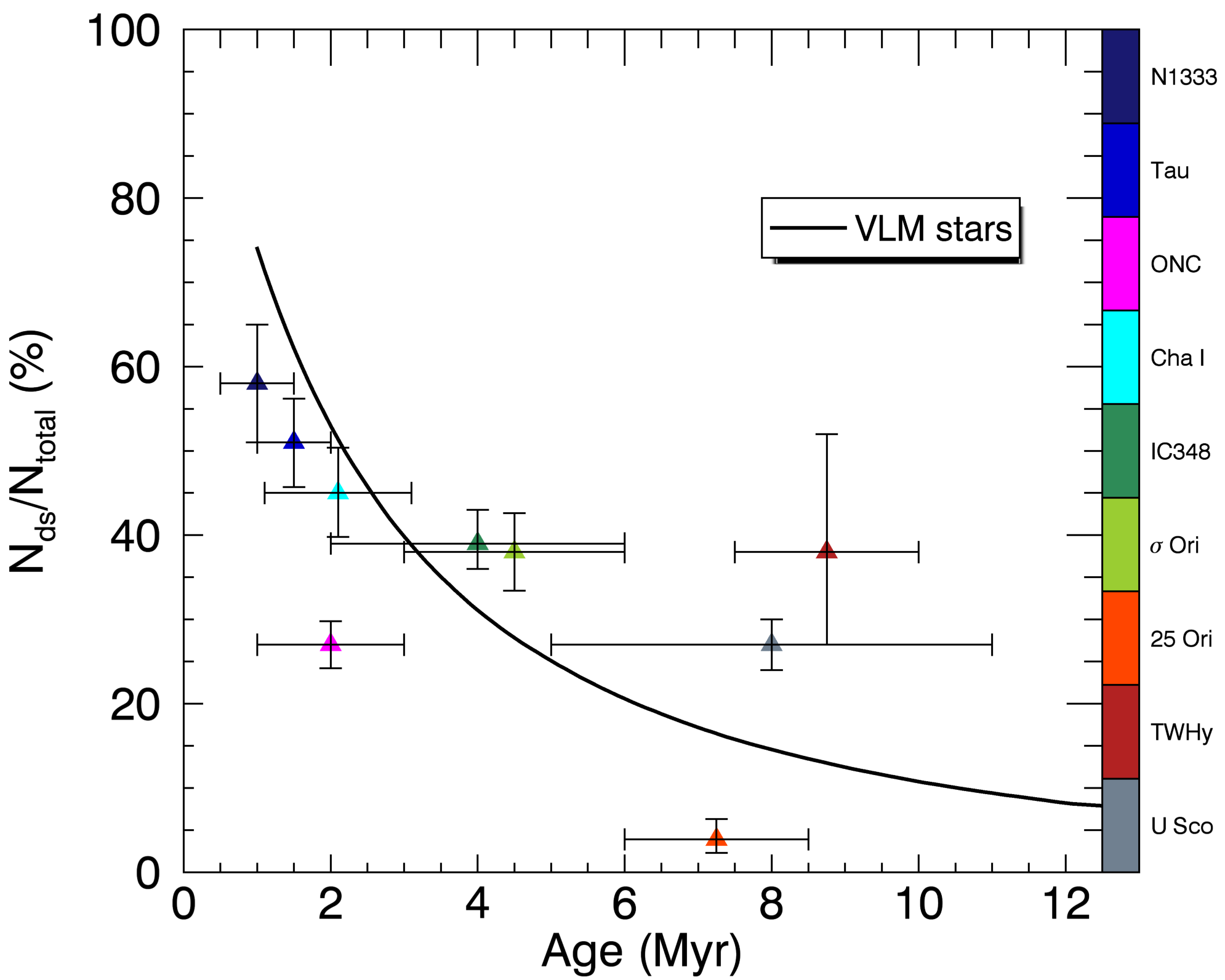}}
\centerline{\includegraphics[width=8.5cm]{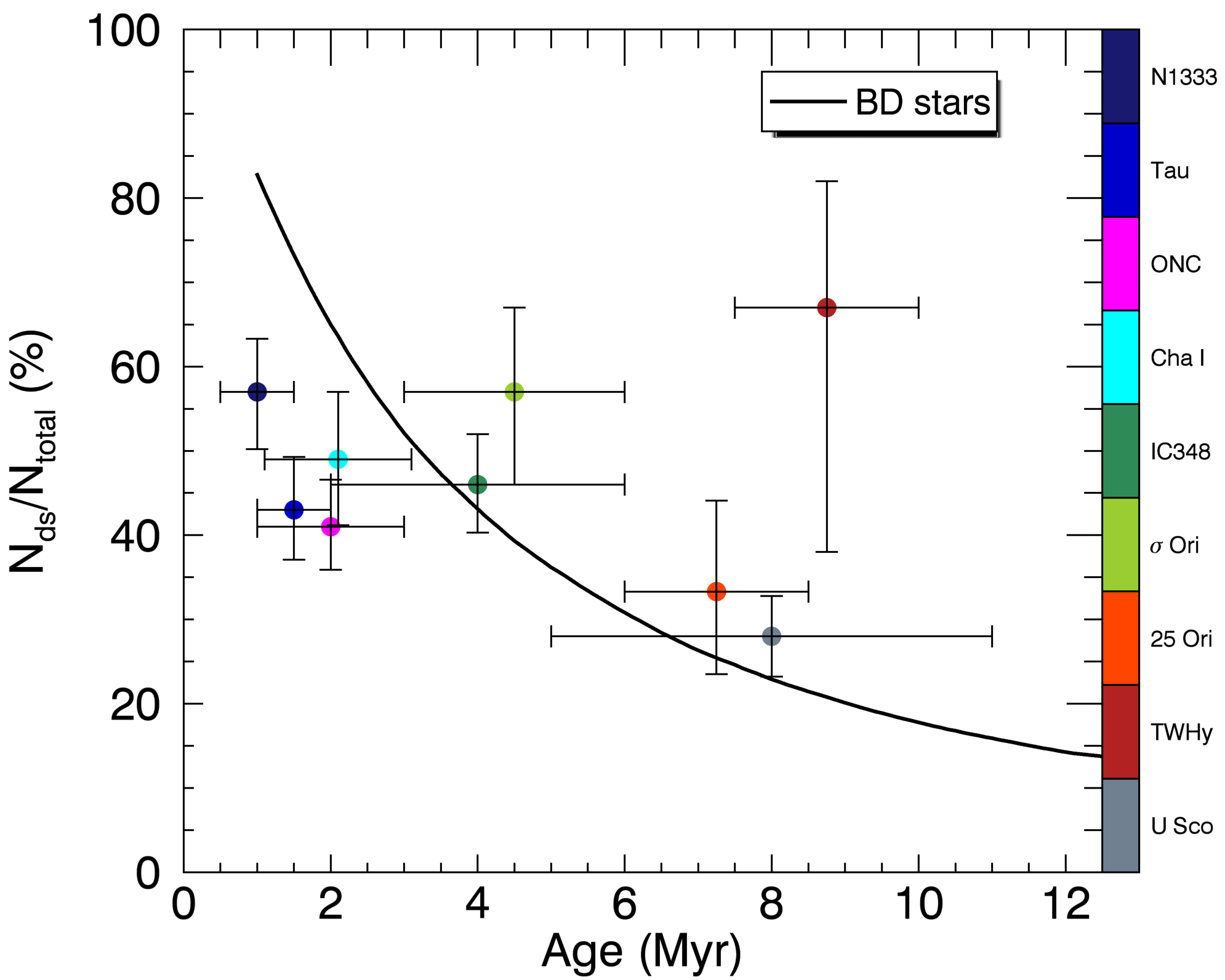}}
\bigskip
\caption{Disk fraction as a function of time for VLM stars (top
panel) and BDs (bottom panel). Symbols are for disk fraction of BDs
(circles) and VLM stars (triangles) in nine young clusters.}
\label{diskfrac3}
\end{figure}

When we observe the disk fractions for BDs and VLM stars shown in
Table \ref{tabdiskfracBD} we note that for the ONC, $\sigma$ Ori,
25 Ori, and TW Hya the values differ significantly. In the literature
there are some claims that low-mass stars actually have lower disk
lifetimes when compared to BDs (e.g. Luhman et al.
\citeyearads{2008ApJ...675.1375L}, Luhman \& Mamajek
\citeyearads{2012ApJ...758...31L}, and Downes et al.
\citeyearads{2015MNRAS.450.3490D}) and some claims against this
idea (e.g. Luhman et al. \citeyearads{2005ApJ...631L..69L}, Damjanov
et al. \citeyearads{2007ApJ...670.1337D}, and Scholz \& Jayawardhana
\citeyearads{2008ApJ...672L..49S}).

\begin{table*}[t]
\caption{Observational disk fraction for BDs and low-mass stars
\label{tabdiskfracBD}}
\setlength{\tabcolsep}{2.5pt}
\renewcommand{\arraystretch}{2}
\centering
{\small
\begin{tabular}{lccccp{3cm}c}
\hline\hline
Region & Age (Myr) & BD (\%) & VLM (\%) & Total (\%) & Disk criteria
& References \\
\hline
NGC 1333\tablefootmark{a} & 1 - 2 & $31/54 = 57^{+6.3}_{-6.8}$ &
$28/48 = 58^{+7}_{-7}$ & $59/102 = 58^{+4.7}_{-4.9}$ & SED spectral
slope & 1 \\
ONC & 1 - 3 & 33/81 = 41$^{+5.6}_{-5.1}$ & $121/445 = 27^{+2.8}_{-2.8}$
& $154/526 = 29^{+2.7}_{-2.7}$ & near-IR excess & 2 \\
ONC & 1 - 3 &            -               & 62/125 = 50$ \pm 4.4$ &
62/125 = 50$ \pm 4.4$ & Spitzer colour-mag diagrams & 3 \\
Taurus\tablefootmark{b} & 1 - 2 & 27/63 = 43$^{+6.3}_{-5.9}$ & 44/86
= 51$^{+5.2}_{-5.3}$ & 71/149 = 48$^{+4.0}_{-3.9}$ & IR colour-colour
diagrams and SED spectral slope & 4 \\
Cha I\tablefootmark{b} & 1 - 3 & $18/37 = 49^{+8.0}_{-7.8}$ &
$38/85 = 45^{+5.4}_{-5.2}$ & $56/122 = 46^{+4.5}_{-4.4}$ & SED
spectral slope & 5, 6 \\
IC 348\tablefootmark{a} & 2 - 6 & $32/70 = 46^{+6.0}_{-5.7}$ &
$71/184 = 39^{+4}_{-3}$ & $103/254 = 40^{+4.7}_{-4.7}$ & SED spectral
slope & 1  \\
$\sigma$ Ori\tablefootmark{c} & 3 - 6 & $13/22 = 57^{+10}_{-11}$ &
$95/252 = 38^{+4.6}_{-4.6}$ & $108/274 = 39^{+4.5}_{-4.5}$ & Spitzer
colour-mag diagrams & 7 \\
25 Ori\tablefootmark{d} & 6 - 8.5 & $5/15 = 33.3^{+10.8}_{-9.8}$ &
$3/77 = 3.9^{+2.4}_{-1.6}$ & $8/92 = 8.7^{+3.8}_{-2.1}$ & SED
spectral slope and H$\alpha$ EW & 8 \\
TW Hya\tablefootmark{e} & 7.5 - 10 & 2/3 = 67$^{+15}_{-29}$ & 5/13
= 38$^{+14}_{-11}$ & $7/16 = 44^{+12}_{-11}$ & SED excess emission
& 9 \\
Upper Sco\tablefootmark{f} & 5 - 11 & 45/152 = 28$^{+4.8}_{-4.8}$
& 97/364 = 27$^{+3.0}_{-3.0}$ & $139/516 = 27^{+2.6}_{-2.6}$ & IR
colour-colour diagrams & 10 \\
\hline
\end{tabular}}
\tablebib{(1) \citetads{2016arXiv160508907L}; (2)
\citetads{2010A&A...515A..13R}; (3) \citetads{2014MNRAS.444.1157D};
(4) \citetads{2010ApJS..186..111L}; (5) \citetads{2008ApJ...675.1375L};
(6) \citetads{2008ApJ...684..654L}; (7) \citetads{2008ApJ...688..362L};
(8) \citetads{2015MNRAS.450.3490D}; (9) \citetads{2016AJ....152....3K};
(10) \citetads{2012ApJ...758...31L}}
\tablefoot{\tablefoottext{a}{VLM stars with spectral types $\geq$
M3.75 and $\leq$ M5.75. BDs with spectral types $\geq$ M6.}
\tablefoottext{b}{VLM stars between M3.5 and earlier than M6. BD
spectral types $\geq$ M6.} \tablefoottext{c}{Calculated from the
combination of disk fractions of VLM stars with M$_\mathrm{J}$
between 4 - 6 and 6 - 8 and of BDs with M$_\mathrm{J}$ between 8 -
10 and 10 - 12. See \citetads{2008ApJ...688..362L} for more details.}
\tablefoottext{d}{The VLM regime contains stars from M0.5 to M5.5
with four stars with spectral types earlier than M3 out of
77.}\tablefoottext{e}{BDs with spectral types $\geq$ M6.  VLM stars
with spectral types between M3.5 and M5.} \tablefoottext{f}{The BD
($\geq$ M6) and VLM (M3.5 - M6) disk fractions were calculated based
on the authors' disk classification from the paper's Table 1 and
encompass full, transitional, and evolved disks.}}\tablefoot{ All
the uncertainties were calculated according to
\citetads{2003ApJ...586..512B}.}
\end{table*}

We take this into account in one of our models (model M3, see next
sections), setting $t_\mathrm{th, VLM} = 2.2$ Myr for VLM stars
(the same value considered above) and $t_\mathrm{th, BD} = 3.2$ Myr
for BDs.  The resultant disk fraction is the other black curve shown
in Figure \ref{diskfrac}. The initial disk fraction in this case
is slightly higher,  77\%, and a disk fraction of 50\% is reached
at $\sim$ 2.4 Myr. In Figure \ref{diskfrac3} we plot separately the
disk fractions of VLM stars and BDs. As expected, the BD disk
fraction curve falls off more slowly than the VLM disk fraction
curve. In the analysis of model M3, we  verify whether this difference
in disk lifetimes is able to better describe the rotational periods
observed for the lowest mass population of young clusters.

\subsection{Models}

We  analyse four models (Table \ref{paramod}). These models differ
by their period distributions, the disk lifetimes of very low mass
stars and brown dwarfs, and the disk locking hypothesis.  Our final
goal is to reproduce the rotational properties of the least massive
population of young clusters which rotate faster in general than
their solar-type counterparts.  We want to investigate how the
initial period conditions and the presence of a locking disk influence
these properties. We  compare our results with the period distributions
and median period values of the ONC (\citeads{2010A&A...515A..13R};
\citeads{2014MNRAS.444.1157D}), the $\sigma$ Ori cluster
(\citeads{2004A&A...419..249S}; \citeads{2010ApJS..191..389C}), NGC
2362 \citepads{2008MNRAS.384..675I}, and the Upper Sco association
(\citeads{2015ApJ...809L..29S}). We also compare the disk fraction
as a function of the period to verify whether VLM stars and BDs
with disks rotate more slowly than diskless objects, as is the case
for solar-type stars. We  also investigate whether the period-mass
relation holds and how the angular momentum evolves with time at
this mass range.  In general, the rotational properties of very
low-mass objects are different from those of their more massive
counterparts. They rotate faster than solar-type stars and their
period distributions are not bimodal (e.g. Scholz et al.
\citeyearads{2015ApJ...809L..29S} and Downes et al.
\citeyearads{2014MNRAS.444.1157D}), although \citetads{2010A&A...515A..13R}
find two distinct period peaks for the VLM stars of the ONC.
Therefore, it is not clear whether the disk locking is as efficient
at these mass ranges as it is for more massive stars
\citepads{2005A&A...430.1005L}.

\begin{table*}[t]
\caption{Model properties \label{paramod}}
\centering
{\small
\begin{tabular}{c|ccccc|ccccc|c}
\hline\hline
Model & \multicolumn{5}{c|}{BD} & \multicolumn{5}{c|}{VLM stars} & \\
\hline
      & $\bar{P}_\mathrm{d}$ & $\sigma_\mathrm{d}$ & $\bar{P}_\mathrm{dl}$
      & $\sigma_\mathrm{dl}$ &
$t_\mathrm{th}$ & $\bar{P}_\mathrm{d}$ & $\sigma_\mathrm{d}$ &
$\bar{P}_\mathrm{dl}$ & $\sigma_\mathrm{dl}$ & $t_\mathrm{th}$ &
D-L \\
 & (days) & (dex) & (days) & (dex) & (Myr) & (days) & (dex) & (days)
 & (dex) & (Myr) & (Y/N) \\
\hline
M1 & 1.0 & 3.0 & 1.0 & 3.0 & 2.2 & 2.0 & 4.0 & 2.0 & 4.0 & 2.2 & Y
\\ 
M2 & 1.0 & 3.0 & 1.0 & 3.0 & 2.2 & 3.0 & 4.0 & 2.0 & 2.0 & 2.2 & Y
\\ 
M3 & 1.0 & 3.0 & 1.0 & 3.0 & 2.2 & 2.0 & 4.0 & 2.0 & 4.0 & 4.2 & Y
\\ 
M4 & 1.0 & 3.0 & 1.0 & 3.0 & 2.2 & 2.0 & 4.0 & 2.0 & 4.0 & 2.2 & N
\\ 
\hline
\end{tabular}
\tablefoot{$\bar{P}_\mathrm{d}$, $\bar{P}_\mathrm{dl}$ and
$\sigma_\mathrm{d}$, $\sigma_\mathrm{dl}$ are the mean values and
dispersions of the rotational period distributions of disk and
diskless stars, respectively; $t_\mathrm{th}$ is the threshold mass
accretion rate time scale; and the D-L parameter specifies whether
the model is taking into account the disk locking hypothesis (Y)
or not (N).}}
\end{table*}

\subsubsection{Model M1: reference model} \label{M1}

We start with the simplest possible model, M1, which has analogous
properties to those of the  M1 model from     \PaperI\ (cf. Table \ref{paramod}).  Both
disk and diskless stars share the same initial period distribution,
the disk lifetime is the same for VLM stars and BDs, and all the disk
stars are locked, meaning that their periods are constant.  However,
we took different period distributions for VLM stars and BDs since
it seems that they rotate faster \citepads{2014prpl.conf..433B}.

\begin{figure*}[!htb]
\centerline{\includegraphics[width=0.35\textwidth]{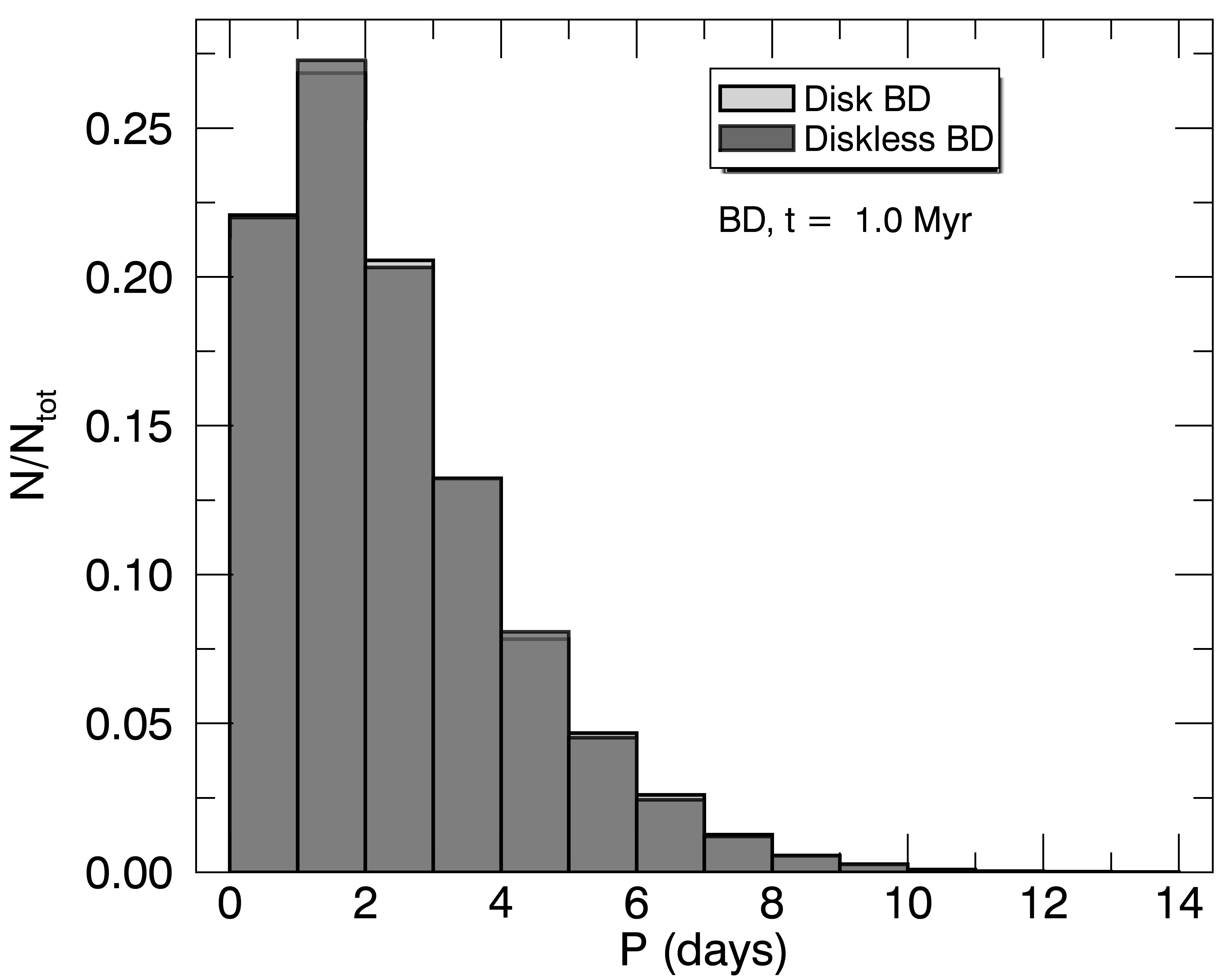}
\includegraphics[width=0.35\textwidth]{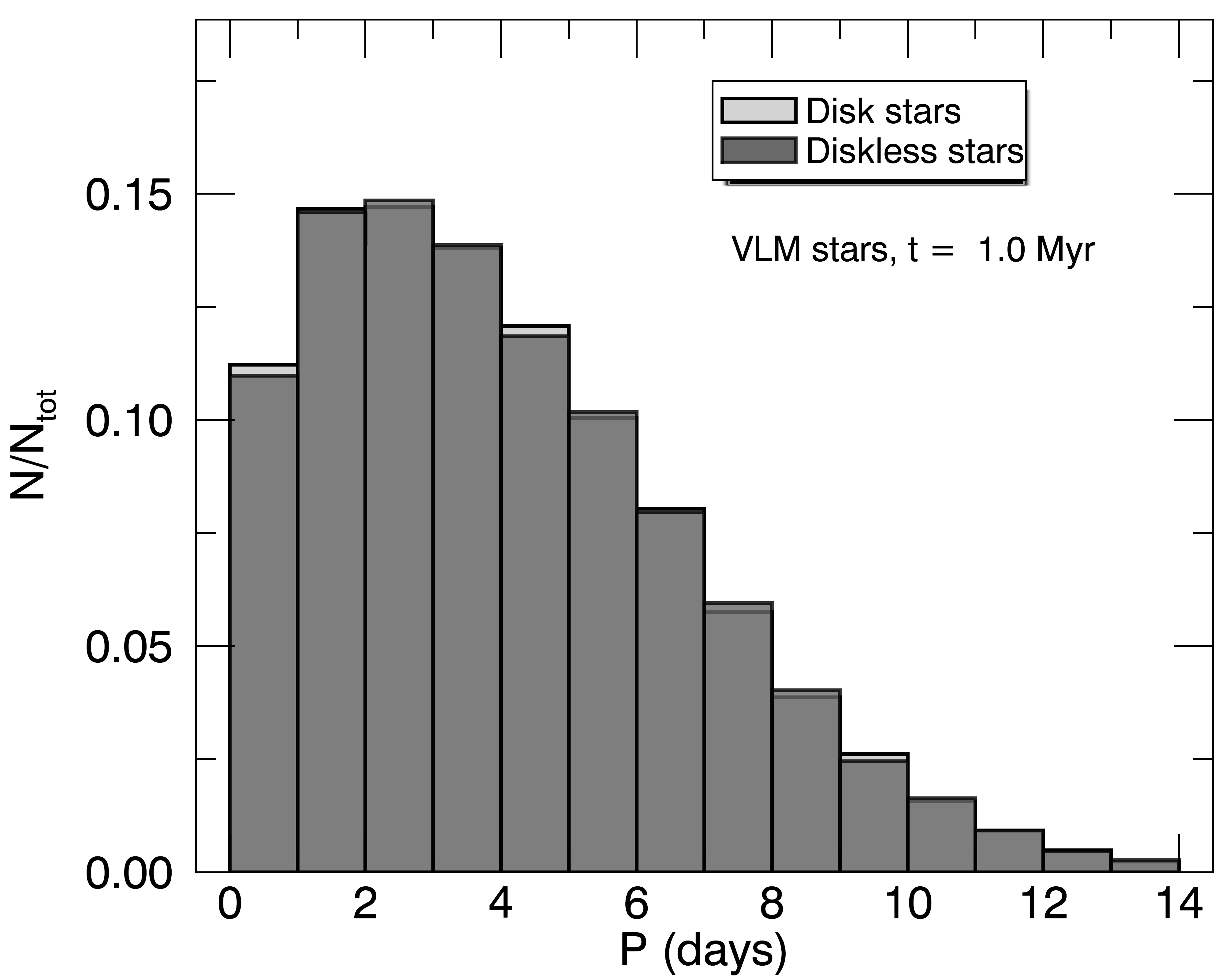}}
\centerline{\includegraphics[width=0.35\textwidth]{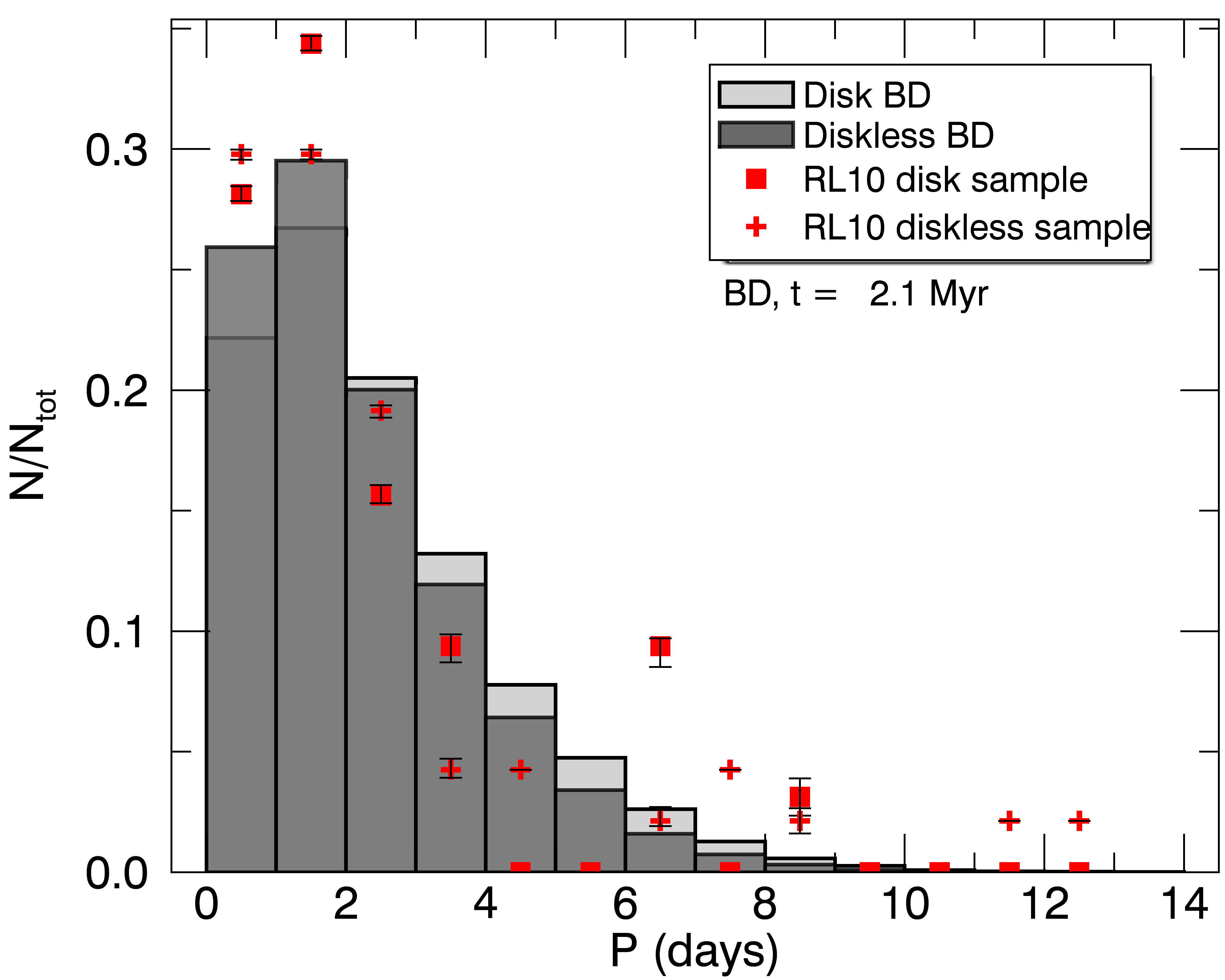}
\includegraphics[width=0.35\textwidth]{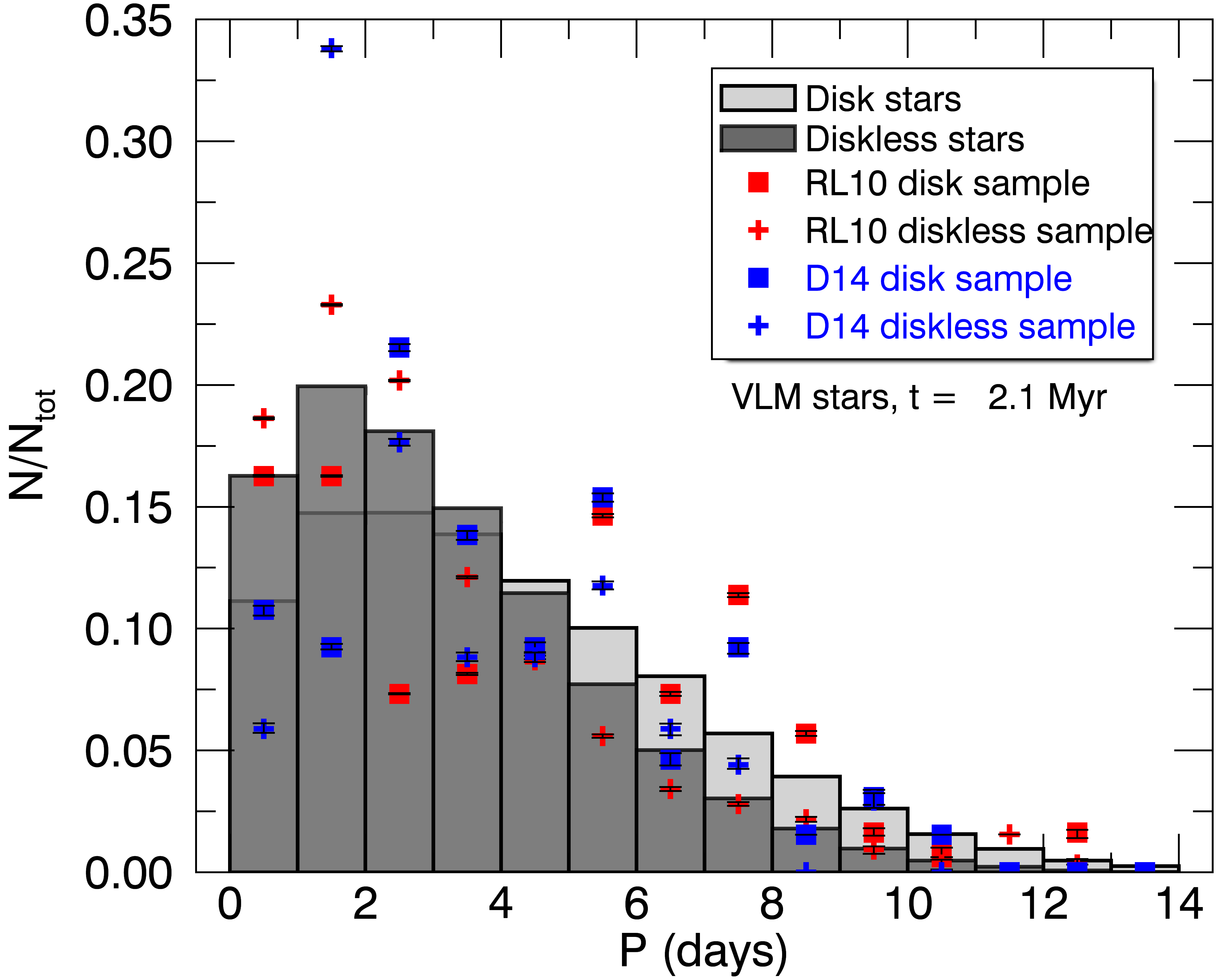}}
\centerline{\includegraphics[width=0.35\textwidth]{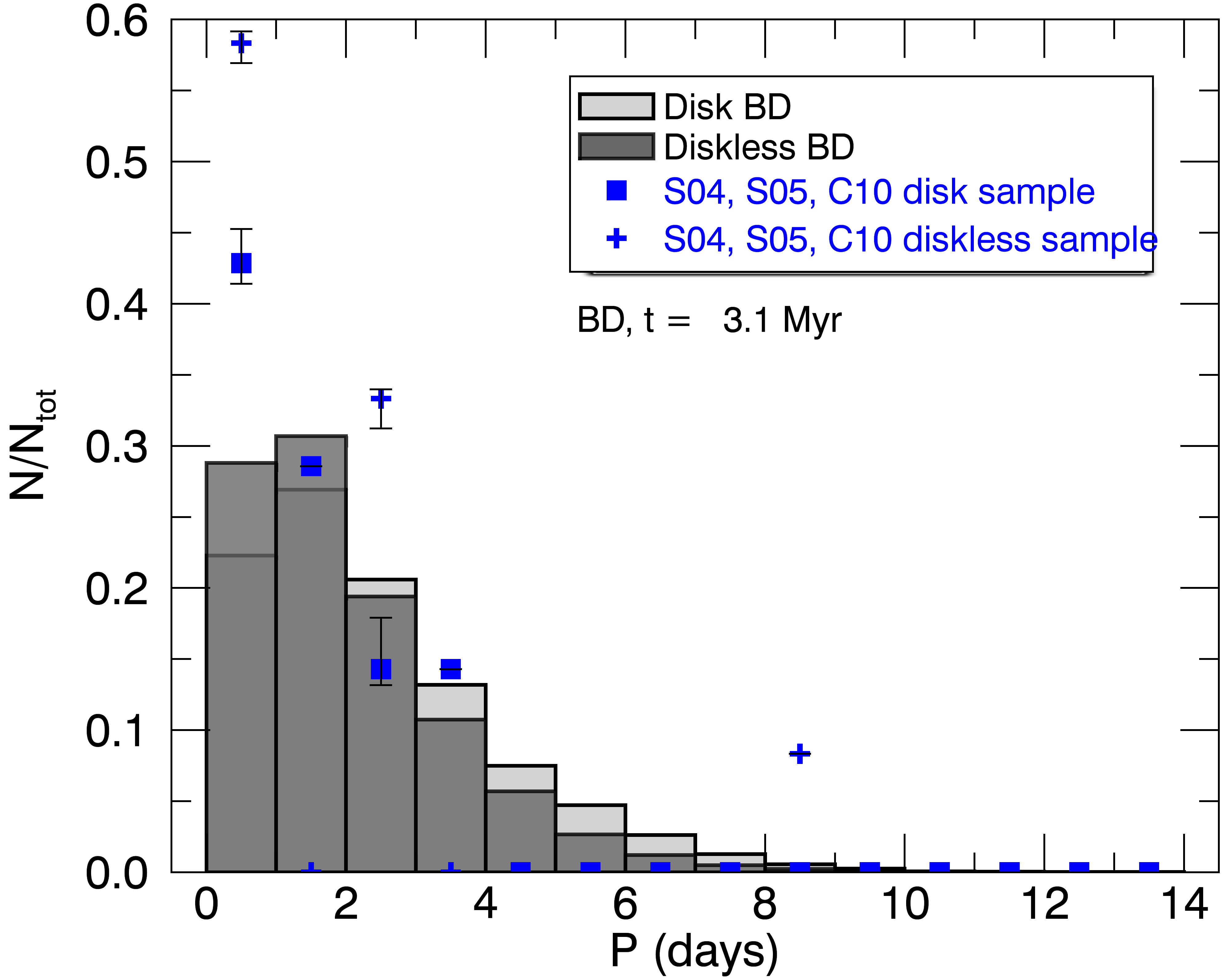}
\includegraphics[width=0.35\textwidth]{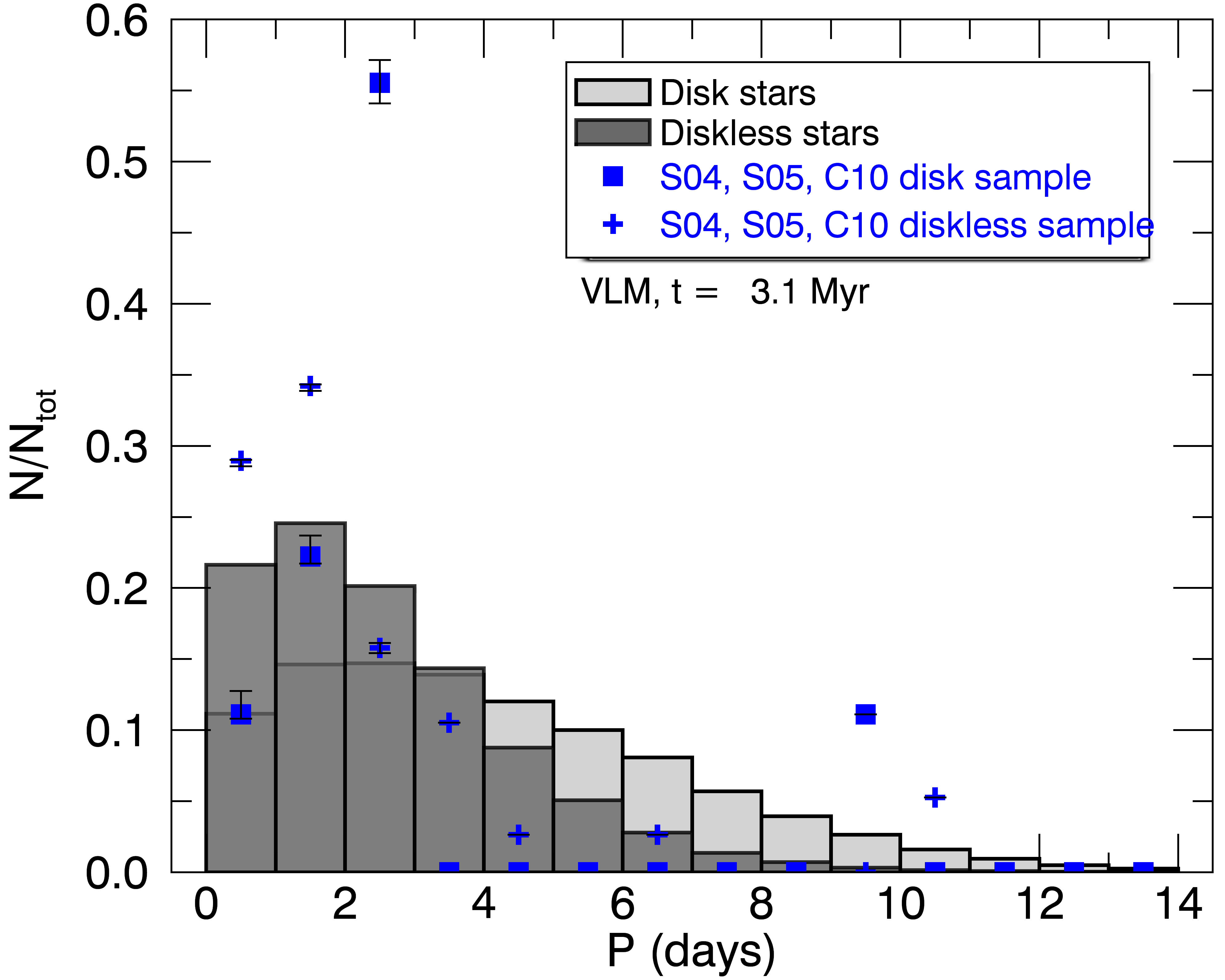}}
\centerline{\includegraphics[width=0.35\textwidth]{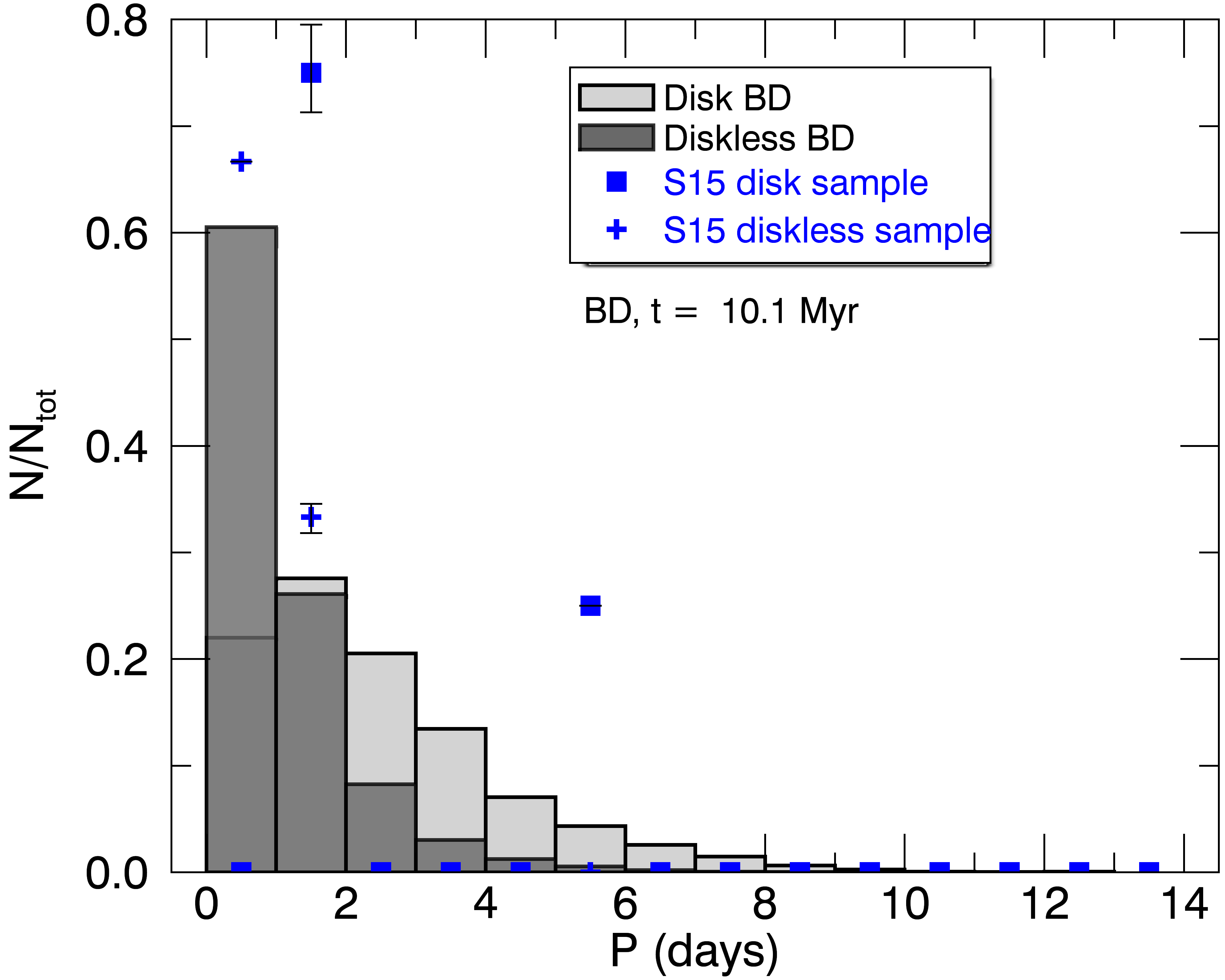}
\includegraphics[width=0.35\textwidth]{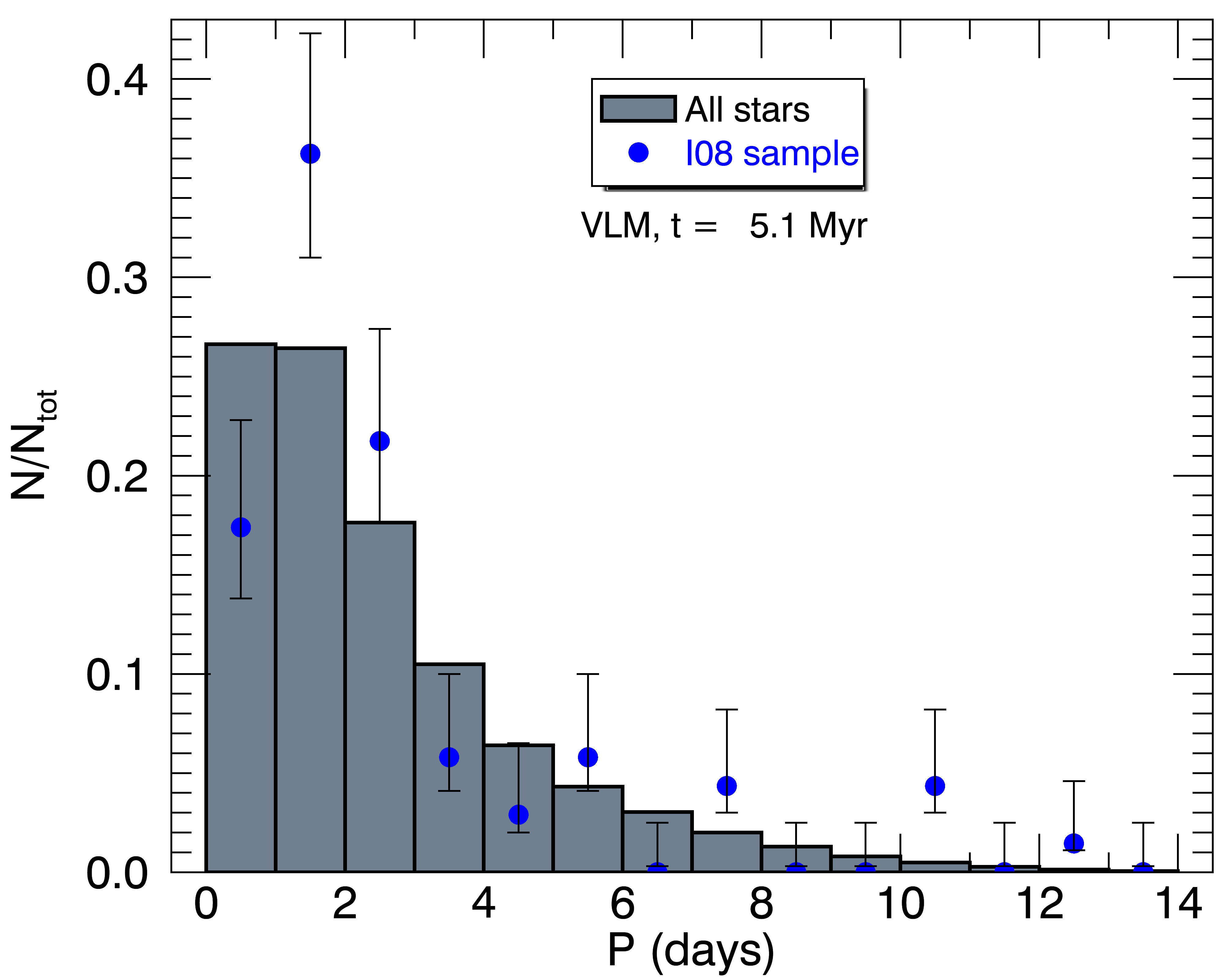}}
\caption{Period distributions obtained from model M1 for disk (light
grey), diskless (grey), and disk and disless stars (bluish grey)
for BDs (left panels) and VLM stars (right panels). The distributions
are shown at 1.0 Myr (1st row), 2.1 Myr (2nd row), 3.1 Myr (3rd
row), and at 5.1 Myr and 10.1 Myr (4th row). The binsize is   1.0
day. Superimposed on the histograms are observational data --\
squares for disk objects, crosses for diskless objects, and filled
circles for objects not classified as disk or diskless -- at 2.1
Myr  from \RLX\ (red symbols) and  \DXIV\ (blue symbols) for the
ONC; at 3.1 Myr using the combined sample of \SIV\ and \CX\ (blue
symbols) for $\sigma$ Ori; at 5.1 Myr from \IVIII\ (blue circles)
for NGC 2362; and at 10.1 Myr from \SXV\ (blue symbols) for Upper
Sco. Error bars for all observational data were calculated following
\citetads{2003ApJ...586..512B}.  \label{distPM1}}
\end{figure*}

In Fig. \ref{distPM1} we show the period distributions for BDs and
VLM stars at 1.0 Myr, 2.1 Myr, 3.1 Myr and at 5.1 Myr and 10.1 Myr.
We superimpose observational period distributions obtained from
several works in the literature.  We use data from Rodríguez-Ledesma
et al. (\citeyearads{2010A&A...515A..13R} - \RLX) and Davies et al.
(\citeyearads{2014MNRAS.444.1157D} - \DXIV) for the ONC; for the
$\sigma$ Ori cluster, we take data from the combined sample of
Scholz \& Eislöffel (\citeyearads{2004A&A...419..249S} - \SIV) and
Cody \& Hillenbrand (\citeyearads{2010ApJS..191..389C} - \CX); for
NGC 2362 we use the sample from Irwin et al.
(\citeyearads{2008MNRAS.384..675I} - \IVIII); and for the Upper Sco
association we use data from Scholz et al.
(\citeyearads{2015ApJ...809L..29S} - \SXV).  The data are summarized
in Table \ref{data}.

\begin{table}[t]
\caption{Observational data for VLM stars and BDs used for
period comparison \label{data}}
\setlength{\tabcolsep}{8.0pt}
\centering
{\small
\begin{tabular}{lcccccc}
\hline\hline
Region & Age   & \multicolumn{2}{c}{Numbers} & Mass range   & Ref. \\
       & (Myr) &       VLM & BD              & (M$_\odot$)  &      \\
\hline
ONC & 1 - 3 & 445 & 79 & 0.02 - 1.5\tablefootmark{a} & 1 \\
ONC & 1 - 3 & 135 & -  & 0.1 - 3.0\tablefootmark{a} & 2 \\
$\sigma$ Ori & 3 - 6 & 10 & 9 & 0.02 - 0.7\tablefootmark{a} & 3 \\
$\sigma$ Ori & 3  - 6 & 37 & 10 & 0.02 - 0.5\tablefootmark{a} & 4 \\
NGC 2362 & 5 & 70 & - & 0.1 - 1.2\tablefootmark{a} & 5 \\
Upper Sco & 5 - 11 & - & 16 & 0.02 - 0.09 & 7 \\
\hline
\end{tabular}
\tablebib{(1) Rodríguez - Ledesma et al. (2010, \RLX); (2) Davies
et al. (2014, \DXIV); (3) Scholz \& Eislöffel (2004, \SIV); (4)
Cody \& Hillenbrand (2010, \CX); (5) Irwin et al. (2008, \IVIII);
(7) Scholz et al. (2015, \SXV).} \tablefoot{\tablefoottext{a}{We
restrict the mass range up to 0.4 M$_\odot$.}}}
\end{table}

At 1.0 Myr, the  fraction of disk objects relative to diskless
objects is the same for both mass regimes,   $\sim 74\%$, as
established in the model (see also Fig. \ref{diskfrac}). The overall
properties of the rotational period distributions (see Table
\ref{paramod}) are the same for the two classes of objects (disk
and diskless), but are different for BDs and VLM stars; the bulk of
BDs rotate faster than VLM stars. The period medians for disk and
diskless objects have the same value,  2.0 days for BDs and  3.7
days for VLM stars. As the system evolves stars with disks maintain
their rotational rate, but diskless stars are expected to spin up.
Then the diskless distribution should move to the left toward low
periods, but the median of the period for disk objects should not
change.  The number of diskless stars is supposed to increase and
the rate of disk to diskless stars should decrease.

Our results (Fig. \ref{distPM1}) confirm all these features for
both mass regimes. The period distribution of BDs at 2.1 Myr is
compared to ONC data from \RLX. Their median values are  1.6 days
for disk objects and 1.7 days for diskless objects against 2.0 days
and 1.8 days in our simulations (cf. Table \ref{PmKol}). We ran a
$\chi^2$ test between our data  and the \RLX\ data, and the results
show that the histogram values for stars with disks match, but they
do not agree for diskless stars.  At the VLM mass regime, they
obtain very different values for the period medians of disk (4.2
days) and diskless stars (2.4 days), which disagrees with  our
results and with the data from \DXIV. Our simulations give period
medians of 3.7 days for disk stars and 2.7 days for diskless stars,
while \DXIV\ obtain 3.7 days and 2.6 days for disk and diskless
stars, respectively.  The $\chi^2$ test between the model and the
\RLX\ VLM data show that the probability of the points matching is
smaller than a 0.01 confidence level. We ran K-S tests that give
probabilities of 59\% and 18\% that our disk and diskless samples
and those from \DXIV\ are derived from the same population. It
should be noted that although the period distributions are not
bimodal and the peak values for disk and diskless objects are the
same, the medians present different values but not as separated as
the results from \RLX\ indicate.

At 3.1 Myr, the medians for diskless stars have moved to even smaller
values (1.6 days for BDs and 2.2 days for VLM stars). When we compare
our results with data for the combined catalogue of the $\sigma$
Ori cluster we see a poorer agreement. The K-S tests give probabilities
of  58\% for disk  and 10\% for diskless BD objects and 2\% for
disk and 8\% for diskless VLM stars  that the samples share the
same population. They also obtain lower medians than we do (Table
\ref{PmKol}).

At 5.1 Myr, data from \IVIII\ for VLM stars are not classified in
terms of disk and diskless stars. The median value of their sample
is  1.9 days, the same value we obtain for the median of the combined
population of disk and diskless VLM stars. However, the probability
that our data and \IVIII\ for NGC 2362 share the same population
is only 16\%.

Finally, at 10.1 Myr, the median values obtained from our results,
2.0 days for disk and 0.8 day for diskless BDs, are respectively
around 20\% smaller and 13\% greater than the median values obtained
by \SXV\ for the Upper Sco association of 1.8 days for stars with
disks and 0.96 day for diskless.  The K-S tests indicate  probabilities
of 64\% (for stars with disks) and 91\% (for diskless stars) that
the two samples have their origin in the same population.

\begin{figure*}
\centerline{\includegraphics[width=0.35\textwidth]{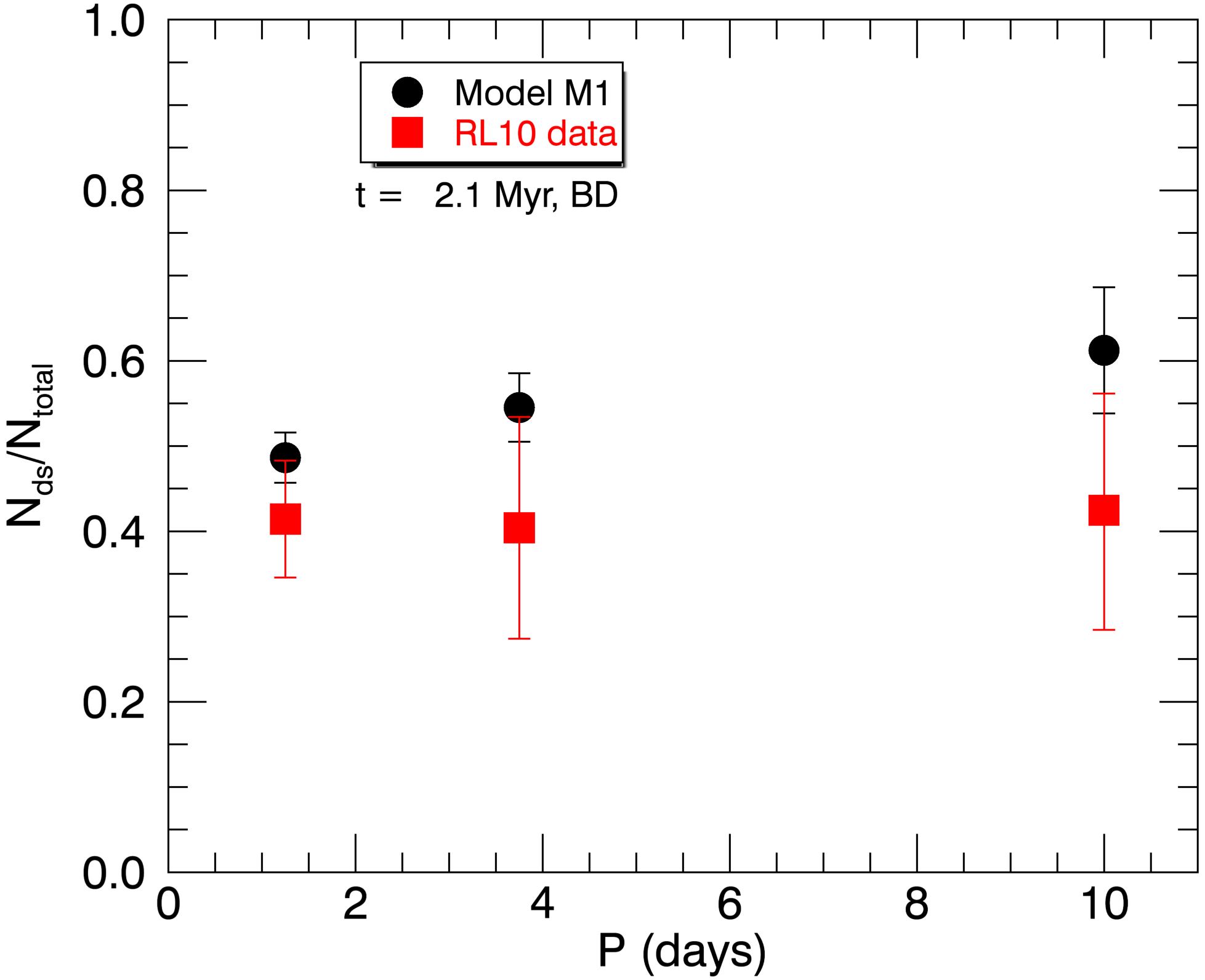}
\includegraphics[width=0.35\textwidth]{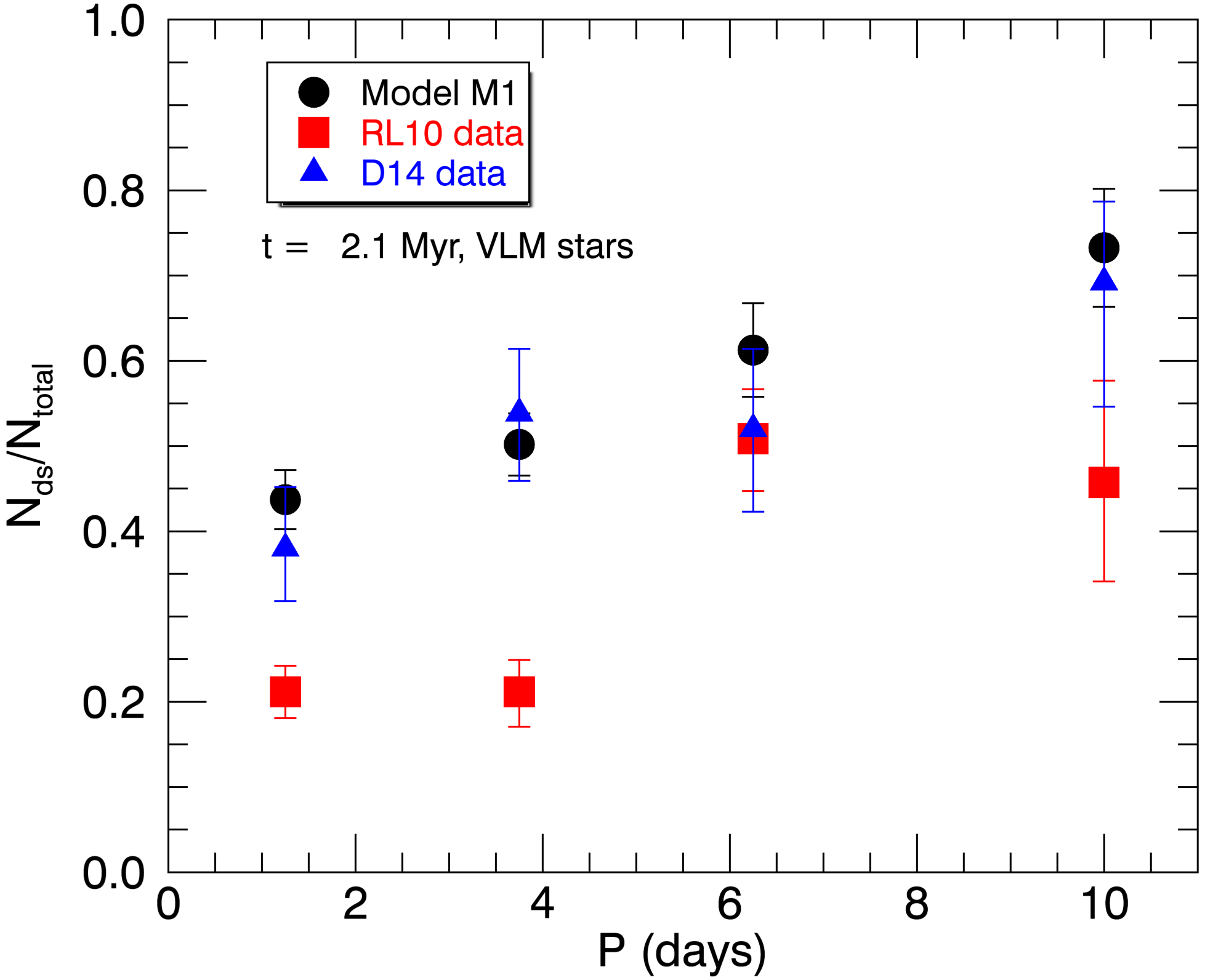}}
\centerline{\includegraphics[width=0.35\textwidth]{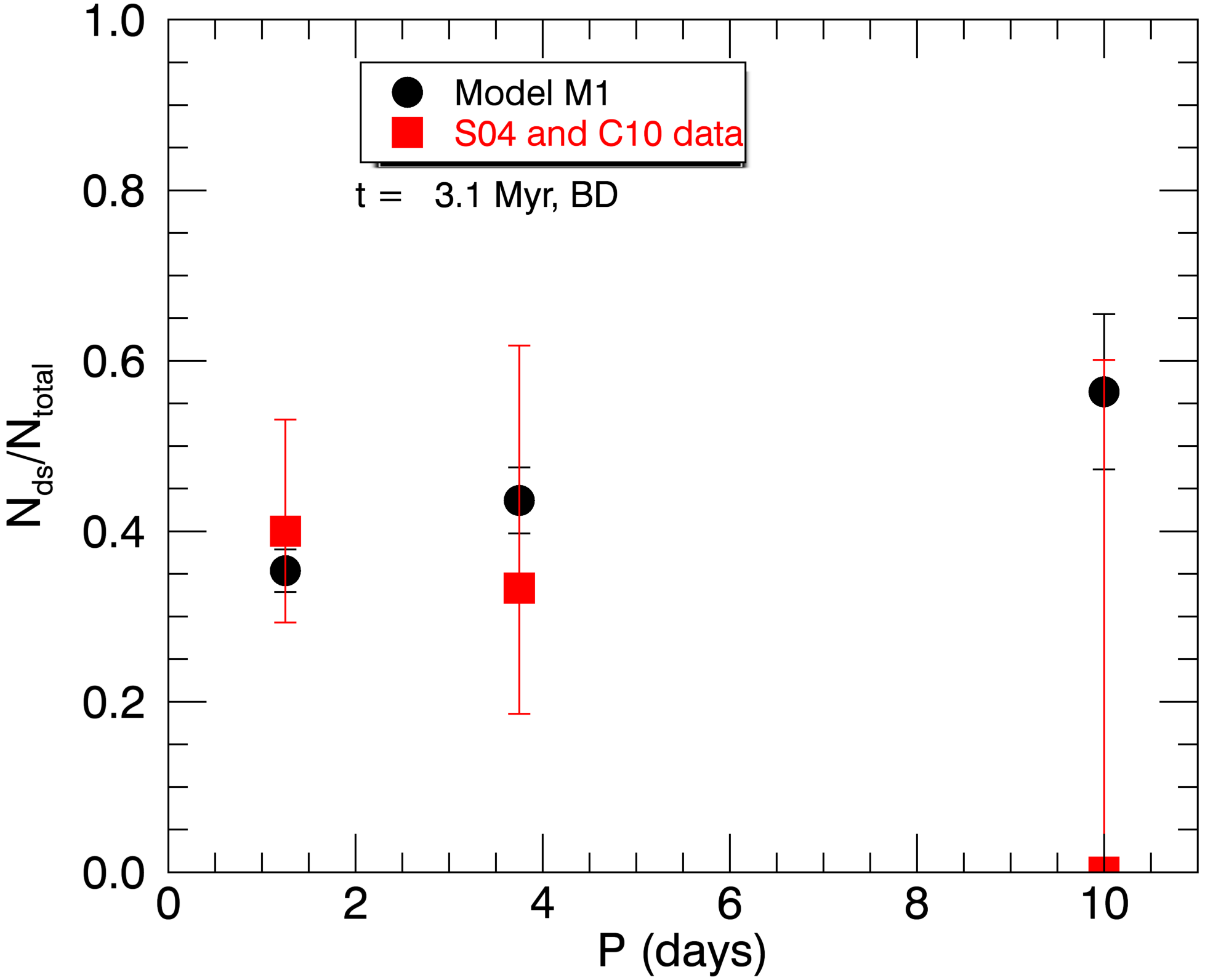}
\includegraphics[width=0.35\textwidth]{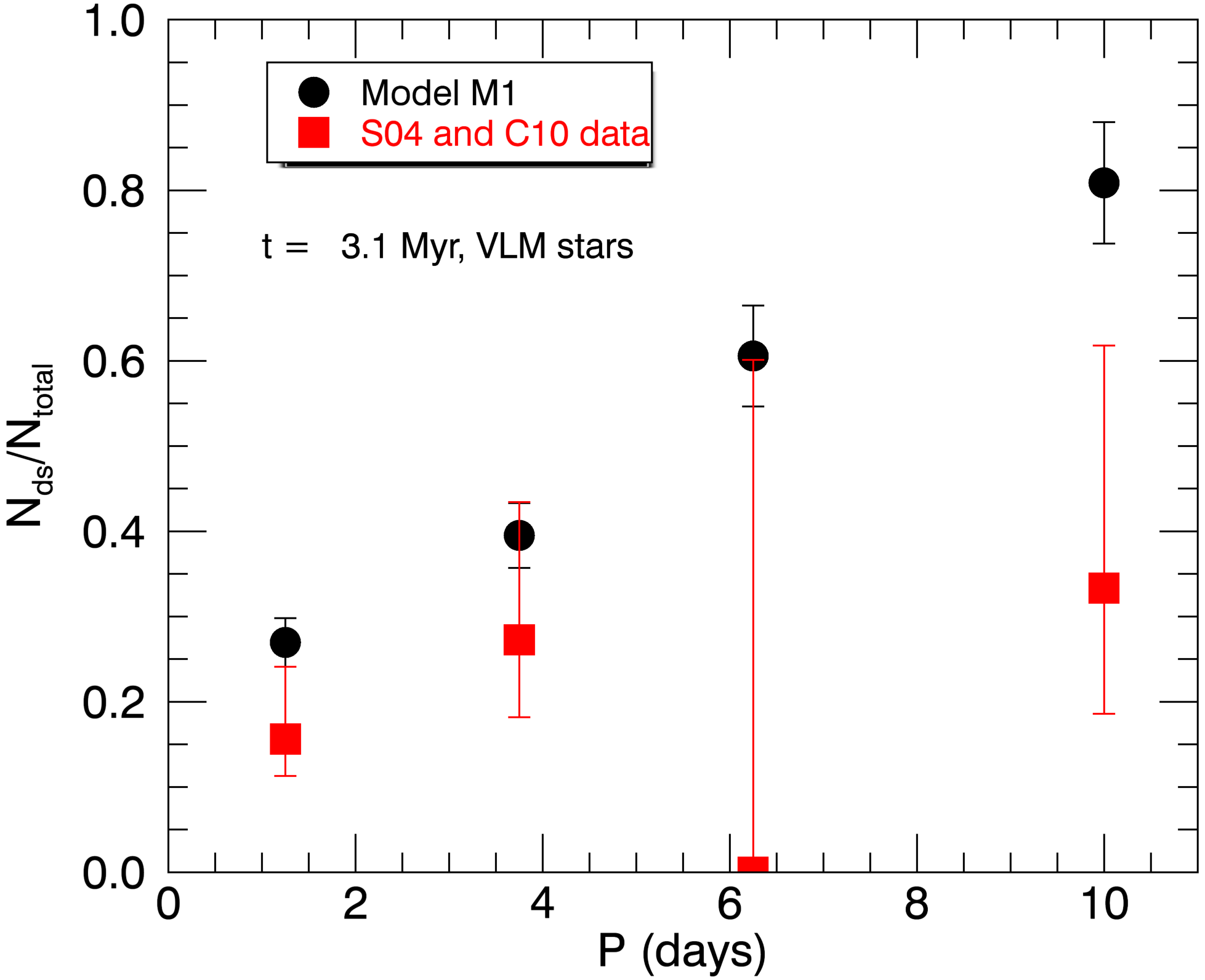}}
\caption{Disk fraction as a function of period (black circles) for
BDs (left panels) and VLM stars (right panels) at 2.1 Myr (top
panels) and at 3.1 Myr (bottom panels). The period bins are $P \leq
2.5 $ d, 2.5 d $< P \leq 5.0$ d, and $P > 5.0$ d for BDs and $P
\leq 2.5 $ d, 2.5 d $< P \leq 5.0$ d, 5.0 d $< P \leq 7.5$ d, and
$P > 7.5$ d for VLM stars. Our simulations at 2.1 Myr are compared
to observations of the ONC by \RLX\ (red squares) and \DXIV\ (blue
triangles) and, at 3.1 Myr, they are compared to the combined
$\sigma$ Ori sample of \SIV\ and \CX\ (red squares) The model's
error bars are equal to $1\sigma$ above and below the fraction value
and they were calculated through resampling. The \RLX\ error bars
were obtained from \citetads{2010A&A...515A..13R}.  For the
observational sample at 3.1 Myr, the error bars were calculated
following the Bayesian approach of \citetads{2003ApJ...586..512B}.
\label{diskfPM1}}
\end{figure*}

\begin{table}[t]
\caption{Comparison between the medians and the distributions of
the rotational periods of some observational samples and of the models
at different ages \label{PmKol}}
\centering
{\small
\begin{tabular}{l|cc|cc}
\hline \hline
              & \multicolumn{2}{c|}{BD} & \multicolumn{2}{c}{VLM} \\ 
Region/Model  & P$_\mathrm{D}$ (days) & P$_\mathrm{DL}$ (days) &
P$_\mathrm{D}$ (days) & P$_\mathrm{DL}$ (days) \\
\hline
ONC (RL10)    & 1.6 & 1.7 & 4.2 & 2.4 \\
ONC (D14)     &  \multicolumn{2}{c|}{-} & 3.7 & 2.6 \\
M1 (2.1 Myr)  & 2.0 & 1.8 & 3.7 & 2.7 \\
M2 (2.1 Myr)  & 2.0 & 1.8 & 4.3 & 2.1 \\
M4 (2.1 Myr)  & 1.7 & 1.7 & 2.8 & 2.6 \\
\hline
$\sigma$ Ori  & 1.6 & 0.88 & 2.4 & 1.7 \\
M1 (3.1 Myr)  & 2.0 & 1.6  & 3.7 & 2.2 \\
M2 (3.1 Myr)  & 2.0 & 1.6 & 4.3 & 2.1 \\
M4 (3.1 Myr)  & 1.5 & 1.5 & 2.2 & 2.0 \\
\hline
NGC 2362      &  \multicolumn{2}{c|}{-}  & \multicolumn{2}{c}{1.9} \\
M1 (5.1 Myr)  & 2.0 & 1.3 & \multicolumn{2}{c}{1.9} \\
M2 (5.1 Myr)  & 2.0 & 1.3 & \multicolumn{2}{c}{1.9} \\
M4 (5.1 Myr)  & 1.1 & 1.1 & \multicolumn{2}{c}{1.4} \\
\hline
Upper Sco     & 1.8 & 0.96 & \multicolumn{2}{c}{-} \\
M1 (10.1 Myr) & 2.0 & 0.8 & 3.6 & 1.0 \\
M2 (10.1 Myr) & 2.0 & 0.8 & 4.4 & 1.0 \\
M4 (10.1 Myr) & 0.6 & 0.6 & 0.9 & 0.8 \\
\hline
\end{tabular}
\tablefoot{P$_\mathrm{D}$ is the median of the rotational
period in days for disk BDs and VLM stars while P$_\mathrm{DL}$
is the same quantity but for diskless BDs and VLM stars.}}
\end{table}

In Fig. \ref{diskfPM1} we analyse the dependency of disk fractions
on the rotational period. We show our results at 2.1 Myr and at 3.1
Myr and compare them to the observational data of \RLX\ and \DXIV\
for the ONC at 2.1 Myr and to the combined sample of \SIV\ and \CX\
for $\sigma$ Ori at 3.1 Myr. The period bins are the same as those
proposed by \RLX. The disk fractions of model M1 increases smoothly
towards longer periods implying that  stars with disks indeed rotate
more slowly than diskless ones on average.

From $\chi^2$ tests, we find that at 2.1 Myr our results agree with
the ONC sample from \RLX\ for BDs and from \DXIV\ for VLM stars,
but not with the results of \RLX\ at this mass range, within a 0.01
confidence level most probably owing to their much lower disk
fraction values at shorter periods. This can be due to their use
of near-IR excess as an indication of the presence of disks which,
as we have mentioned before, misses many of the detections.

When we compare our results at 3.1 Myr with the $\sigma$ Ori data
the agreement is not as good, although $\chi^2$ tests do not reject
the hypothesis that the two data sets match within a 0.01 confidence
level. In spite of this, the maximum value of the period observed
in $\sigma$ Ori is  3.1 d, while in our simulations there
are stars with longer periods. For VLM stars, we also observe the
same trend of increasing disk fractions with period, again except
for the 3$^\mathrm{rd}$ data point which, however, exhibits a very
large error bar\footnote{Fractions and error bars were calculated
following Burgasser et al. (2003).}.

\begin{figure}
\centerline{\includegraphics[width=0.4\textwidth]{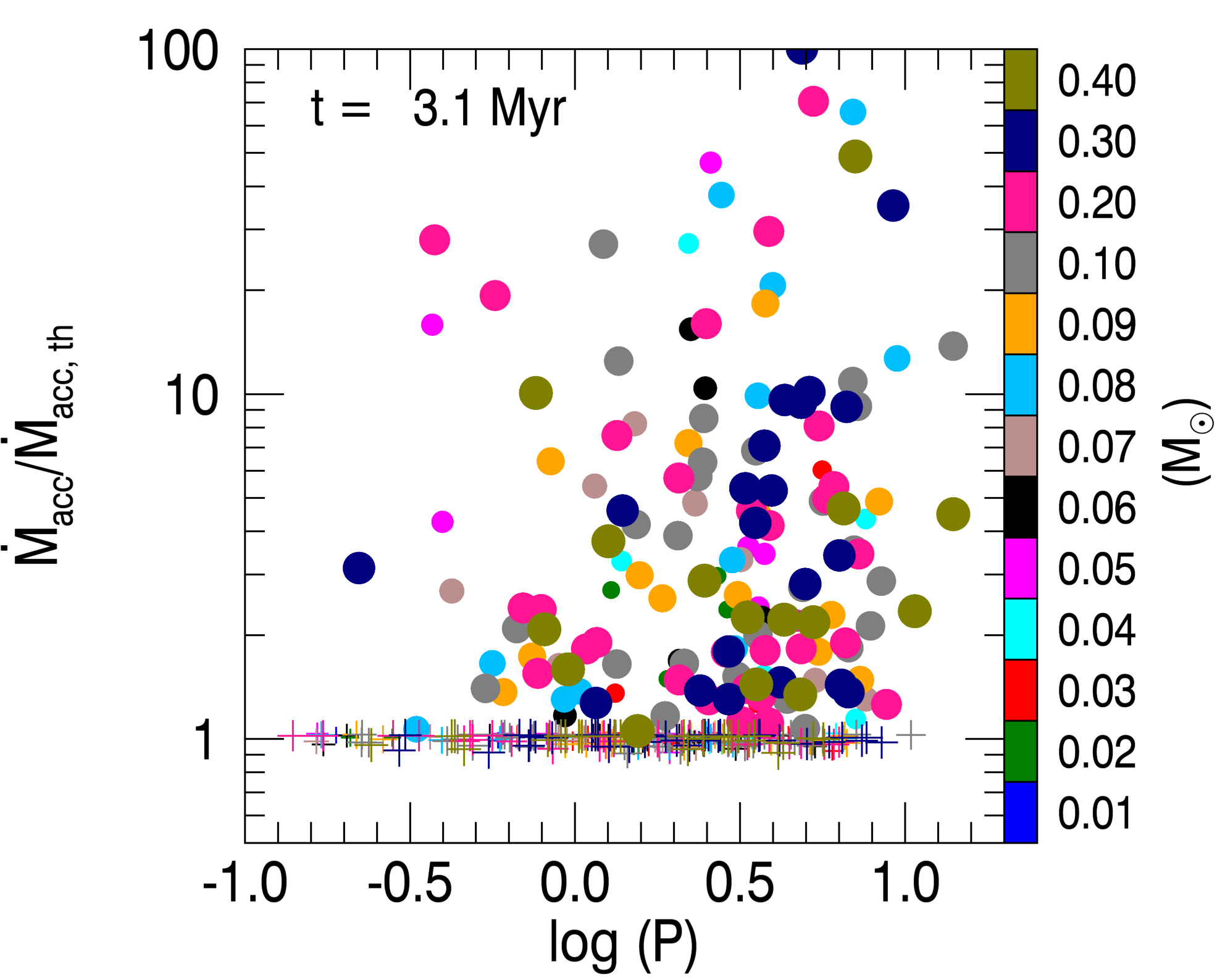}}
\caption{Rotational periods {versus} mass accretion rate
normalized to the mass accretion threshold at 3.1 Myr. Circles are
for disk objects, while crosses are for diskless objects.  Each colour
corresponds to a different object mass, shown on the colourbar at
the right of the plot. \label{MaccP}}
\end{figure}

Furthermore, no clear bimodality appears when we analyse the
rotational period as a function of the mass accretion rate. In Fig.
\ref{MaccP} we plot these quantities at an age of 3.1 Myr to make
the comparison with the data of the $\sigma$ Ori cluster from \CX\
easier.  The mass accretion rate values are normalized to the mass
accretion rate threshold, which has a different value depending on
the mass of the object (see Equation \ref{Maccth}). Because of this
and because $\dot{M}_\mathrm{acc} (M_\ast, t) = \dot{M}_\mathrm{acc,
th} (M_\ast)$ for all diskless stars, these will be clustered at
$\dot{M}_\mathrm{acc}/\dot{M}_\mathrm{acc, th} = 1$; however, for
stars with disks this ratio will be higher than 1.0. We see no
correlation of $\dot{M}_\mathrm{acc}$ with the period even for stars
with disks, the same conclusion reached by \CX\ based on their
observations. We observe no paucity of stars with disks at short
periods, as can be seen in Fig. 9 of \PaperI\ where there are no
stars with disks below $P \sim 1.0$ d at the same age.

\begin{figure*}
\begin{center}
\includegraphics[width=0.4\textwidth]{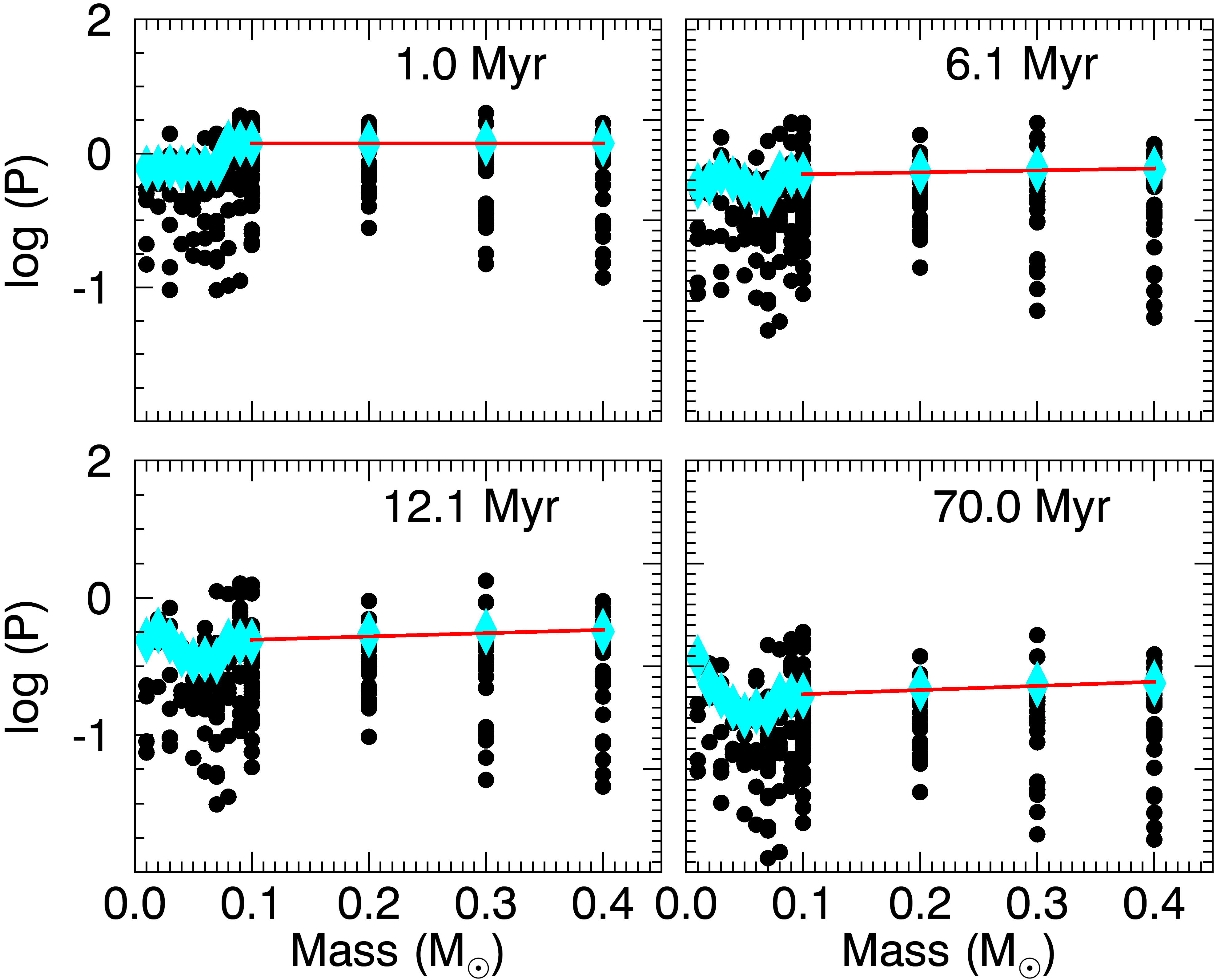}
\hspace{-0.1cm}
\includegraphics[width=0.42\textwidth]{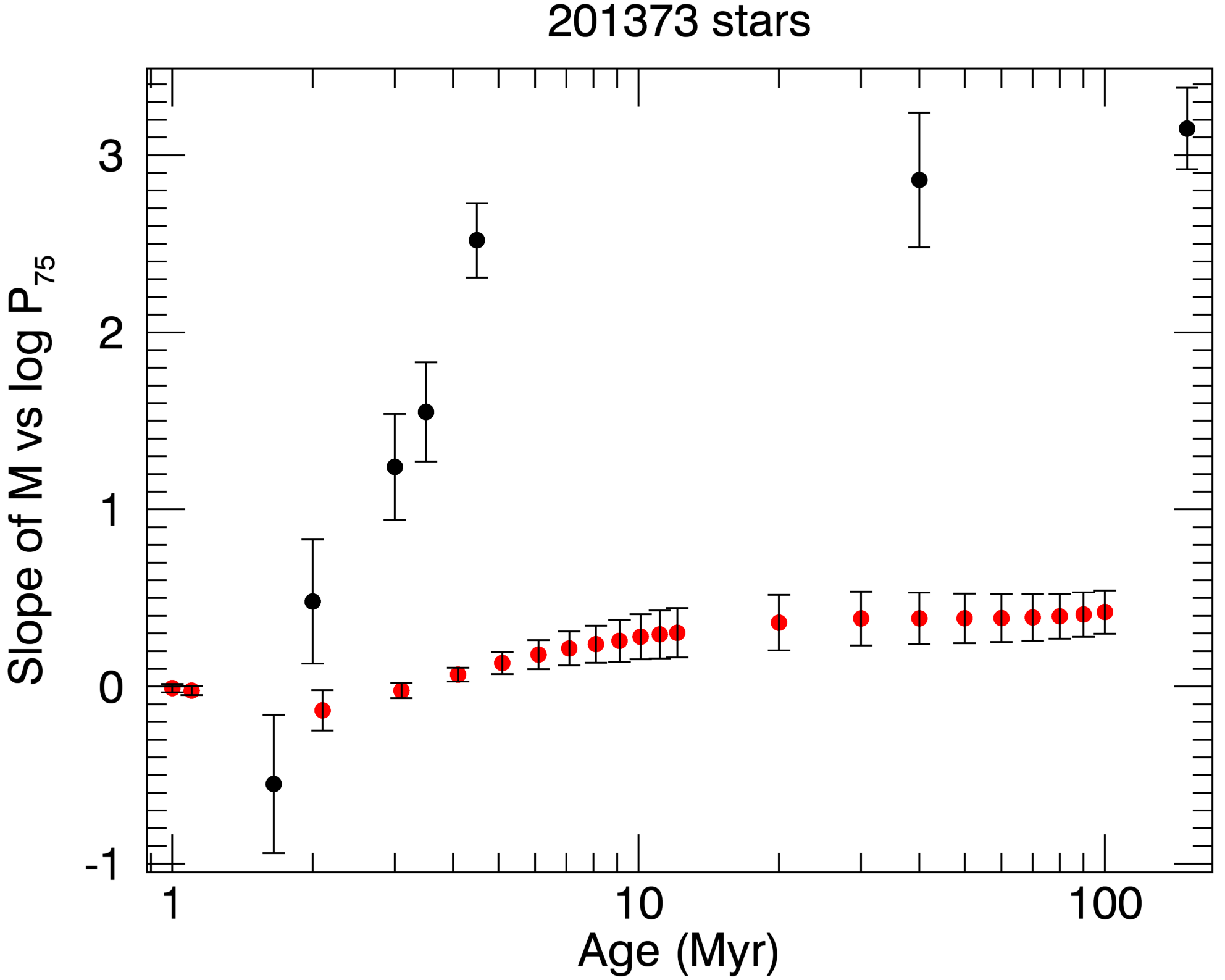}
\end{center}
\caption{{\it Left panels:} Mass {versus} the logarithm
of the rotational period of a sample of 200 stars extracted from
model M1 (black circles) at four different ages superimposed by the
75$^\mathrm{th}$ percentiles of the $\log \; (P)$ (cyan diamonds).
The best linear fits of the percentiles for 0.1 M$_\odot$ $\leq
M_\ast \leq 0.4$ M$_\odot$ are shown as red lines. {\it Right
panel:} Slope of the best linear fits plotted as a function of age
(red circles) superimposed with the values obtained by
\citetads{2012ApJ...747...51H} for seven clusters of different ages
(black circles). \label{massPrel}}
\end{figure*}

We also investigate the period-mass relation for model M1. In Figure
\ref{massPrel}, we plot mass x $\log P$ for 200 objects randomly
chosen from model M1.  We also show the 75$^\mathrm{th}$ percentiles
of $\log P$, the best linear fits for these percentiles, and the
slopes of the fits from 1 Myr to 100 Myr. These percentiles and the
fits were calculated for all objects with 0.1 M$_\odot \leq$ M $\leq
0.4$ M$_\odot$.  The error bars of the slopes are equal to the
standard errors of the estimate of the least-square fits. We see
that the slopes are approximately equal to 0 at all ages and we
conclude that there is no dependency of the period with mass in
model M1. For comparison, we plot the slopes obtained by
\citetads{2012ApJ...747...51H} for NGC 6530 (1.65 Myr), ONC (2.0
Myr), NGC 2264 (3.0 Myr), NGC 2362 (3.5 Myr), IC348 (4.5 Myr), NGC
2547 (40 Myr), and NGC 2516 (150 Myr) for the  low-mass stars  from
0.1 M$_\odot$ to 0.5 M$_\odot$, and we see a clear positive correlation
that increases with age.  \citetads{2010ApJS..191..389C} also found
a positive correlation between period and mass in their sample of
$\sigma$ Ori stars (see also Bouvier et al.
\citeyearads{2014prpl.conf..433B}).  In \PaperI\ we  analysed the
period-mass relation for stars from 0.3 M$_\odot$ to 0.5 M$_\odot$
and we  concluded that a correlation only exists if we relax the
disk locking hypothesis for the lowest mass stars in those simulations,
i.e. the 0.3 M$_\odot$ stars.  In model M4 (section \ref{M4}) we
relax the disk locking hypothesis for all stars except those of 0.4
M$_\odot$  and we  verify whether we can obtain a correlation similar
to that obtained in \PaperI.

\begin{figure}
\centerline{\includegraphics[width=0.4\textwidth]{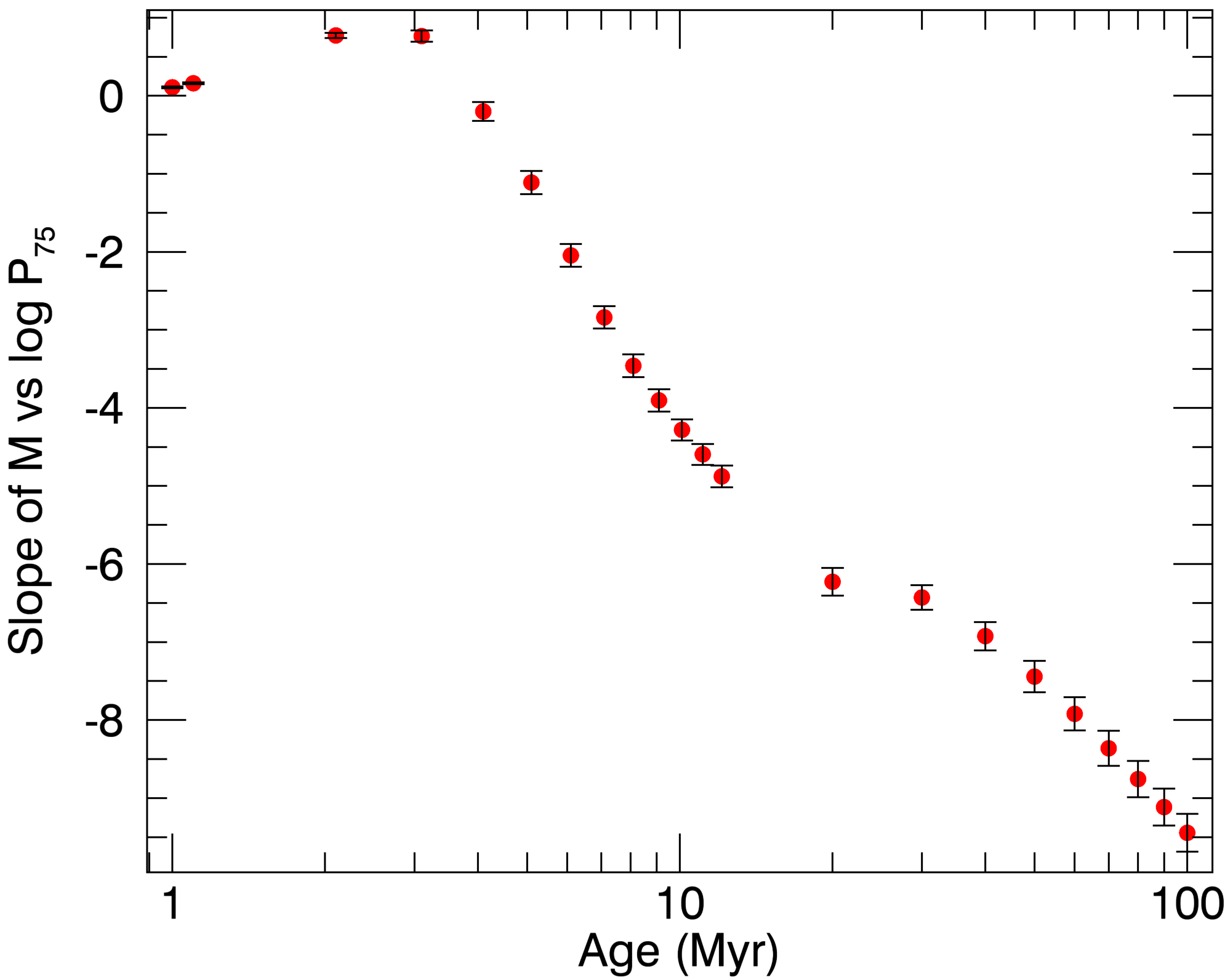}}
\caption{Slopes of the best linear fits to $M_\ast \times \log \;
(P)_{75^\mathrm{th}}$ plots as a function of age only for BDs from
model M1. \label{PMrelBD}}
\end{figure}

\begin{figure}
\includegraphics[width=0.39\textwidth]{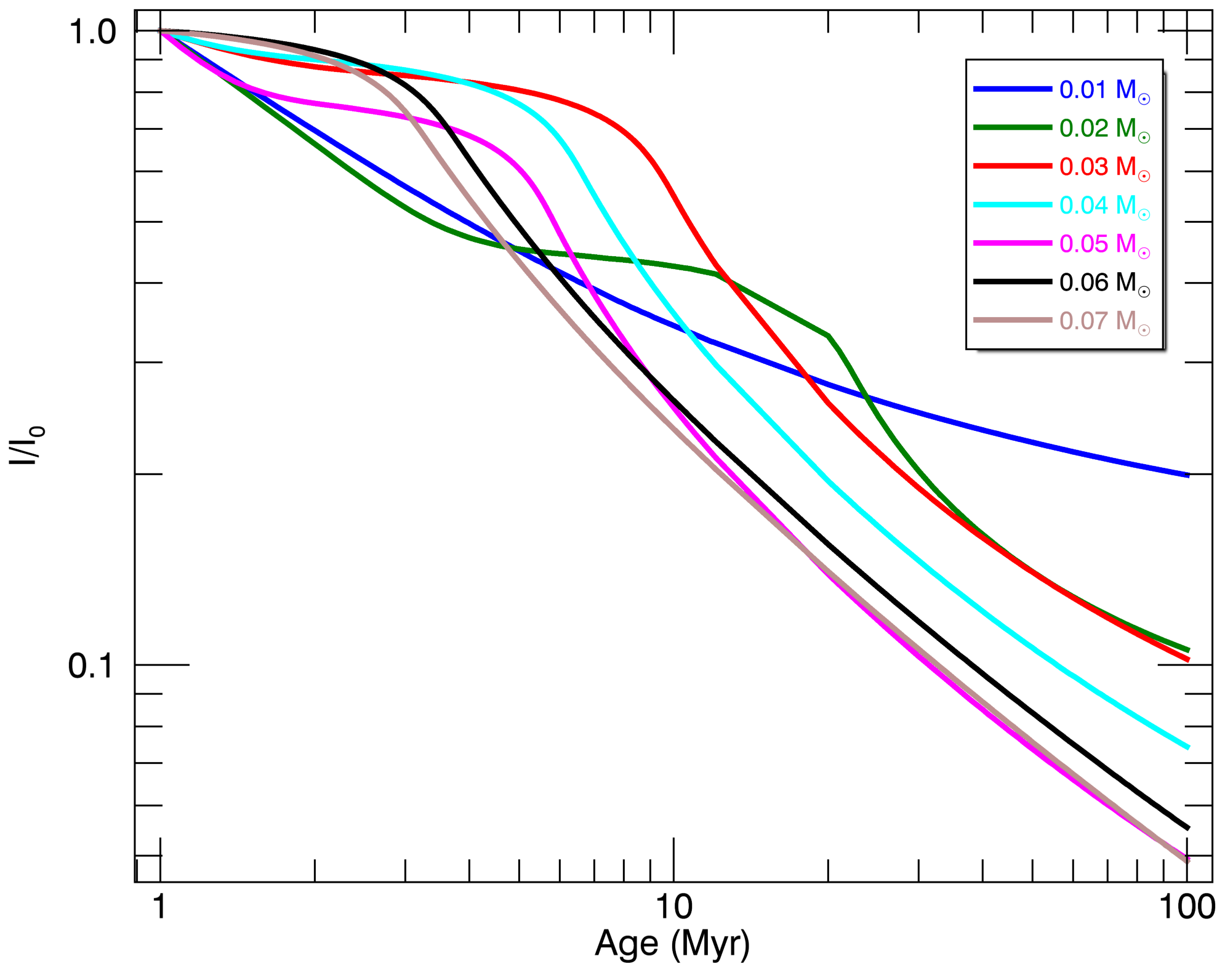}
\includegraphics[width=0.39\textwidth]{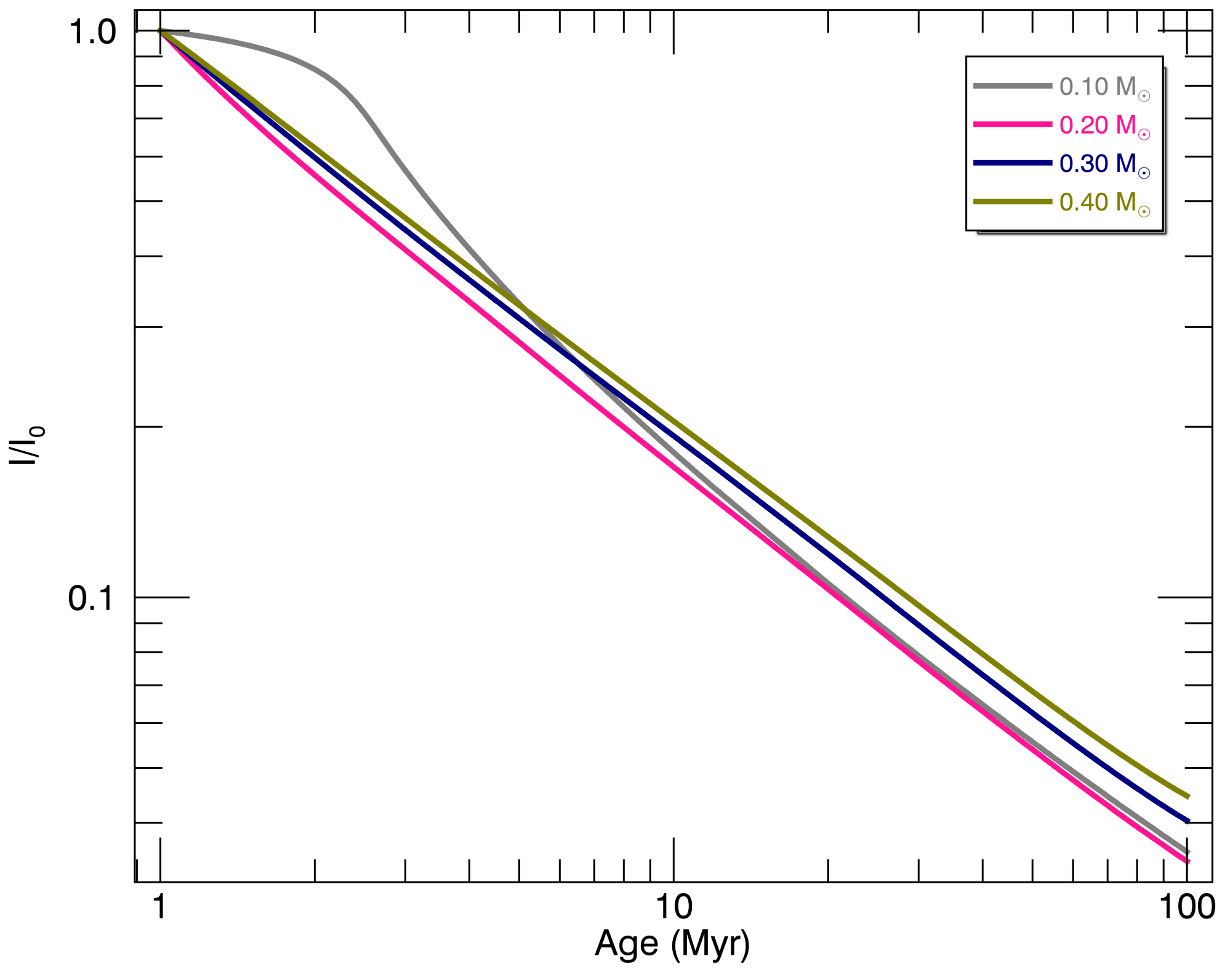}
\caption{Moments of inertia obtained from \citetads{1998A&A...337..403B}
normalized to the moment of inertia at 1.0 Myr for BDs (top panel) and
stars from 0.1 to 0.4 M$_\odot$ (bottom panel). Different colours show
different mass values. \label{momI}}
\end{figure}

Interestingly, when we construct a period-mass relation for BDs,
the slope we obtain is the opposite observed for stars, as can be
seen in Fig. \ref{PMrelBD}. The slope in this case is the fit of
the 75$^\mathrm{th}$ percentiles of the $\log \; (P)$ for M$_\ast
\leq 0.07$ M$_\odot$. This is caused by the more rapid decrease in
the moment of inertia $I$ of higher mass BDs than of lower mass
ones, as illustrated in Figure \ref{momI}.  On the other hand, this
does not happen at the 0.1 - 0.4 M$_\odot$ mass interval where $I$
decreases at the same rate for all masses.

\begin{figure*}
\centerline{\includegraphics[width=0.45\textwidth]{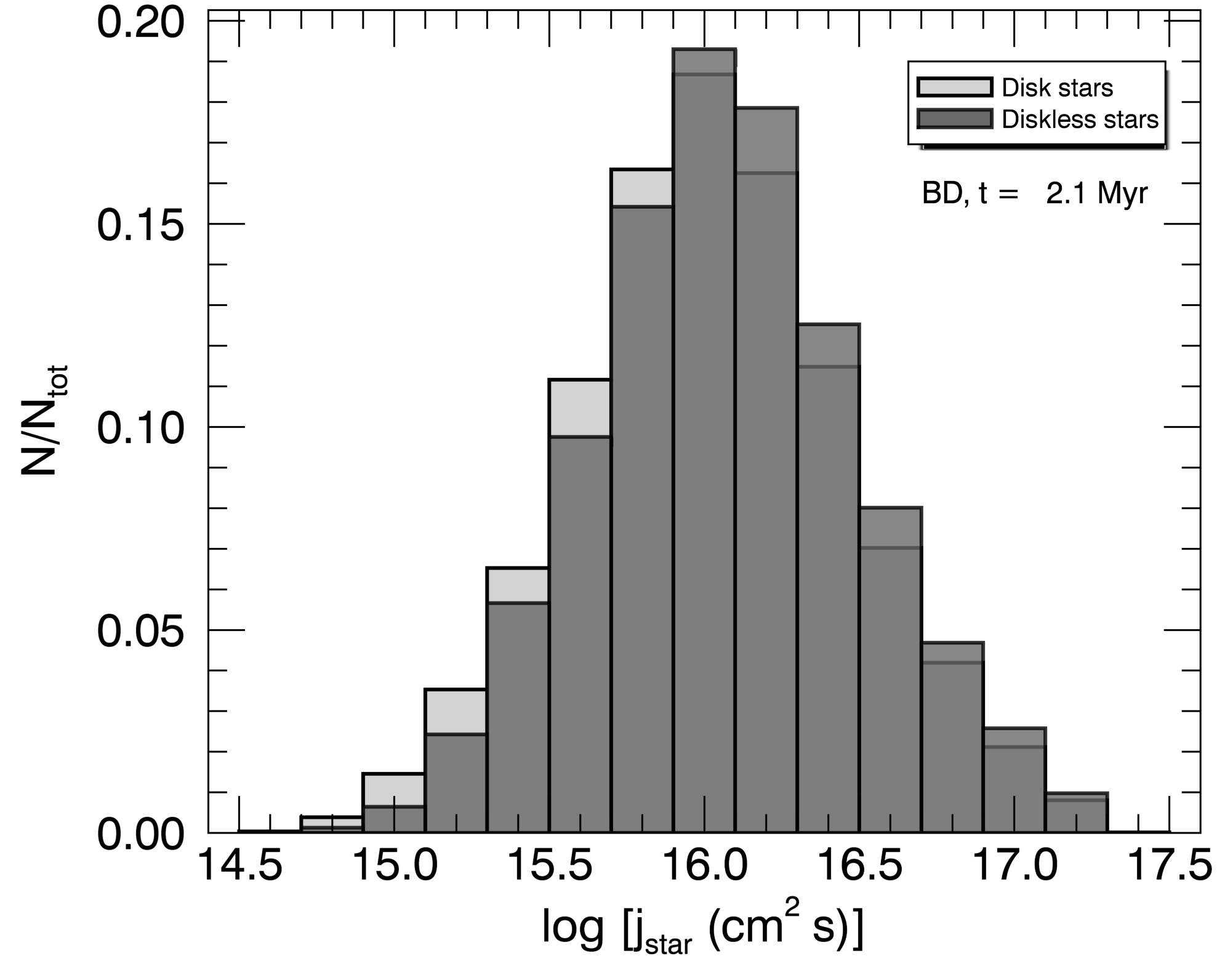}
            \includegraphics[width=0.45\textwidth]{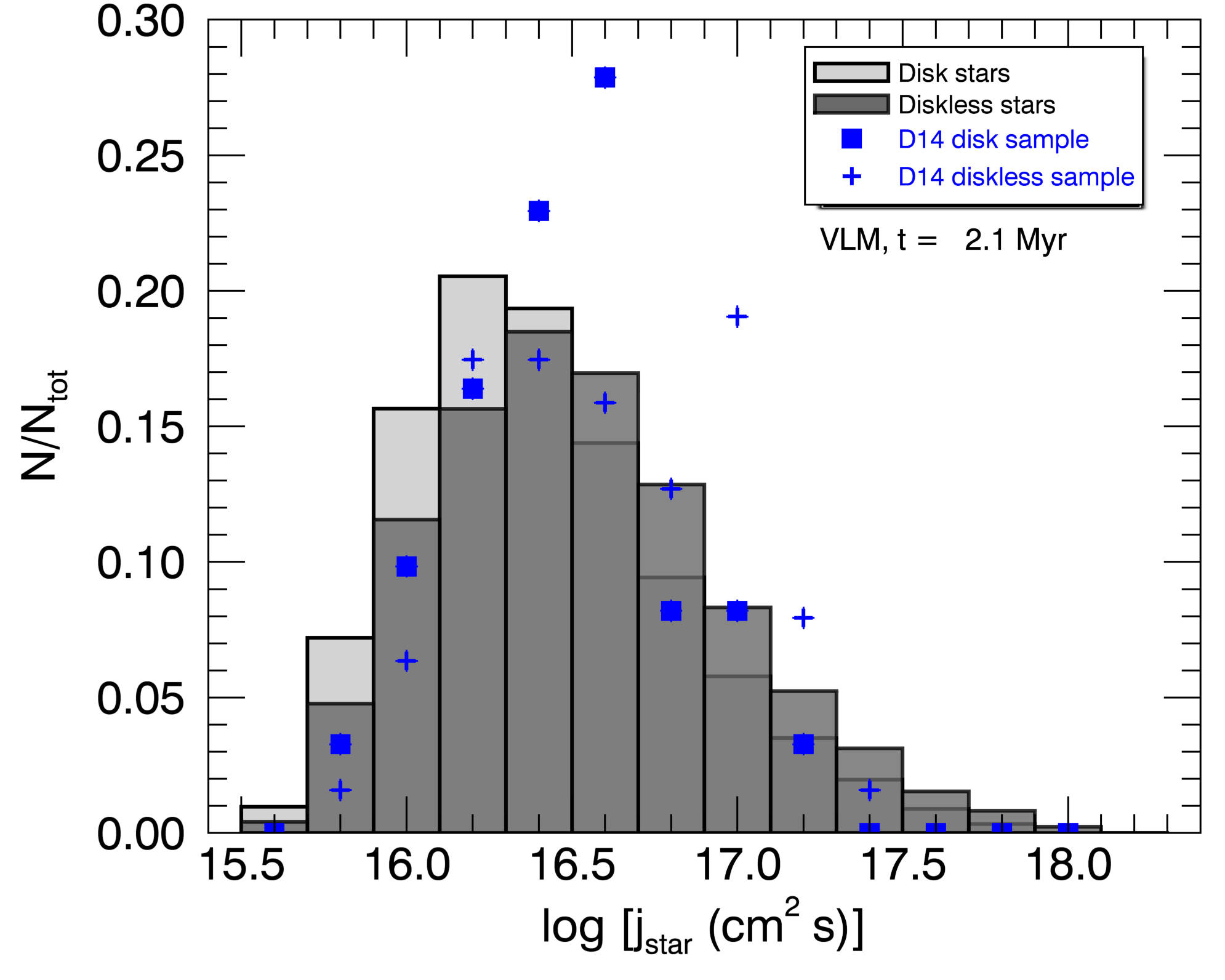}}
\caption{Distribution of the logarithm of the specific angular
momentum of BD (left panel) and VLM (right panel) disk (light grey
bars) and diskless (grey bars) objects at 2.1 Myr obtained from
model M1.  We superimpose data from the \DXIV\ ONC sample for disk
(squares) and diskless (crosses) stars of spectral type later than
M2. \label{distjamM1}}
\end{figure*}

In Fig. \ref{distjamM1} we show the specific angular momentum
distributions for BD and VLM stars at 2.1 Myr. We also show data
from \DXIV\ for stars in the ONC with spectral type later than M2.
These were calculated using period values from \DXIV, but gyration
and stellar radii were obtained from \citetads{1998A&A...337..403B}
models interpolated for the ages and masses of the \DXIV\ ONC sample.

The observational distributions show peaks at higher $\log j$ values
when compared to the simulations.  The medians of $\log j$ are also
displaced by $\sim 0.1$. Since the agreement between the model's
and \DXIV's period medians is very good (as seen in Fig.  \ref{distPM1})
the discrepancies stem from the broader range of masses and ages
found in the sample compared to those present in the model. In the
numerical distributions, which for BDs is single-peaked with a
maximum at $\log j \simeq$ 15.9 - 16.1.  For VLM stars, the peaks
of the distributions of disk and diskless objects are different:
$\log j$ = 16.1 - 16.3 for stars with disks and  $\log j = 16.3 -
16.5$ for diskless stars.  This is related to the different internal
structure of BD and VLM objects since the period distributions for
both are single-peaked (cf. Fig.  \ref{distPM1}). Indeed, Fig.
\ref{momI} shows that low-mass BDs do not contract much over the
first few Myr so that their moment of inertia remains constant,
while this is not the case for VLM stars which steadily contract
over time and therefore start to spin up earlier.

From K - S tests, we obtain that the specific angular momentum
distributions for the stars with disks from the ONC and from this
work have a  6\%  probability of sharing the same origin, while for
the diskless distributions the probability is  20\%.

\begin{figure*}[!ht]
\centerline{\includegraphics[width=0.4\textwidth]{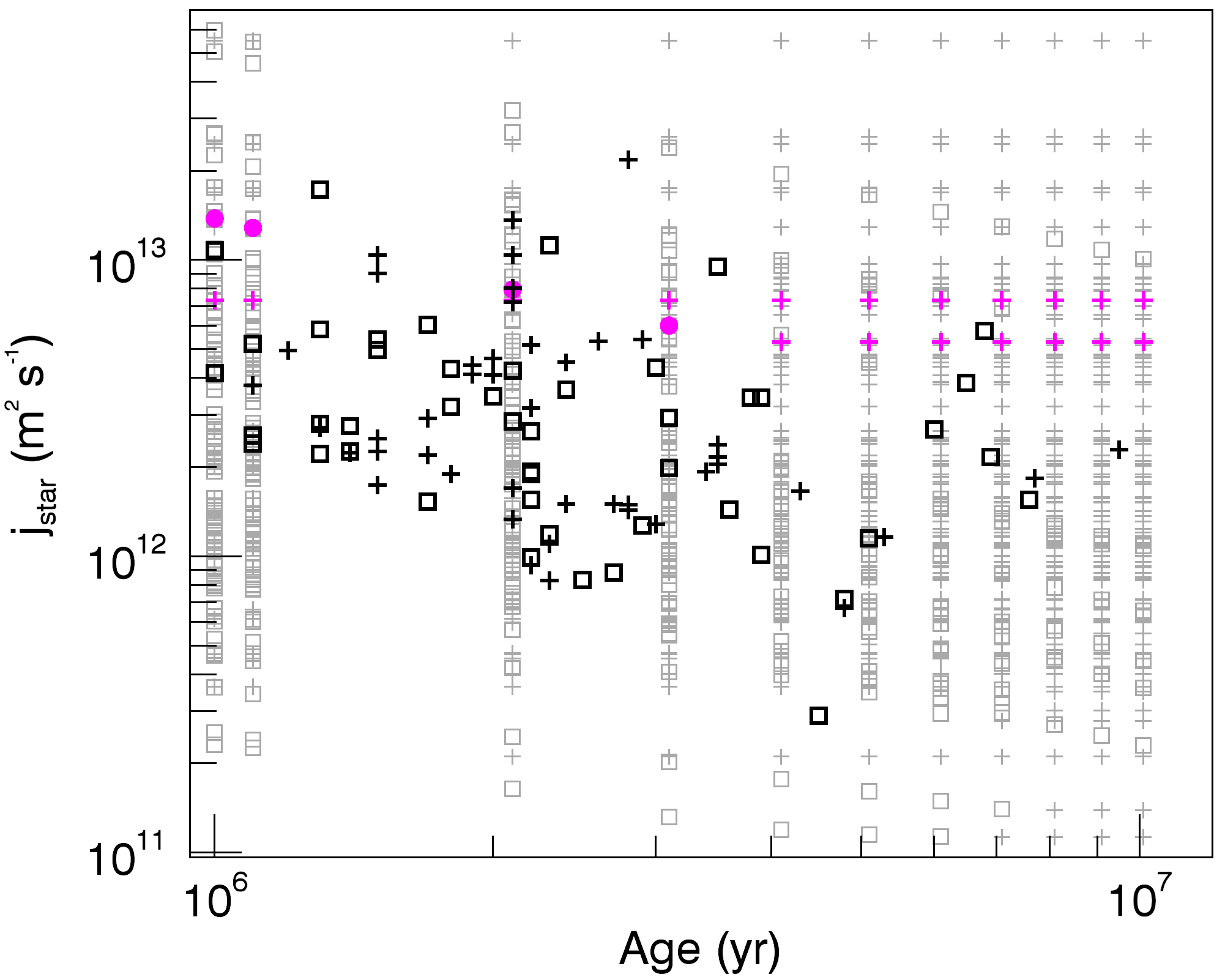}
\includegraphics[width=0.4\textwidth]{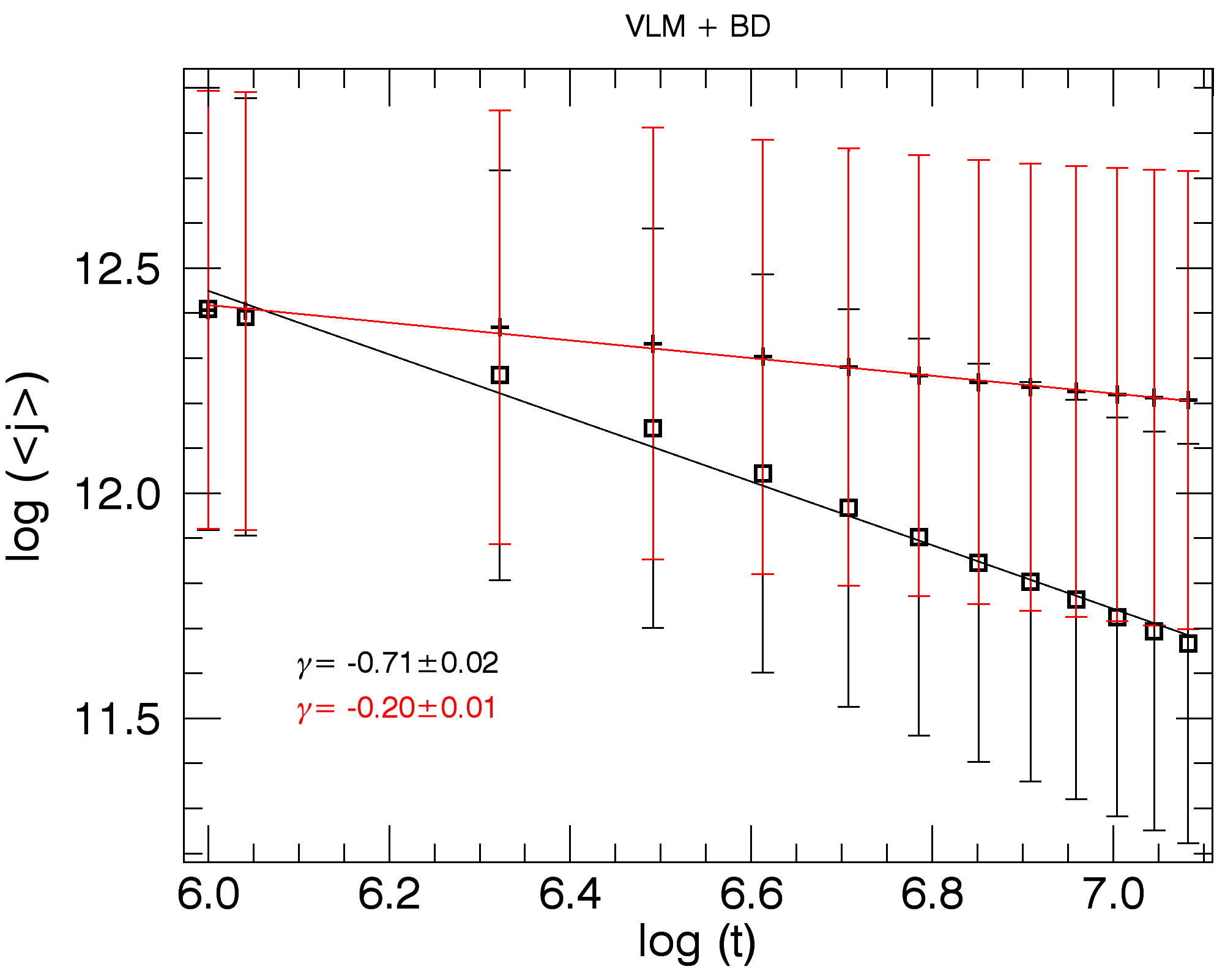}}
\caption{{\it Left:} Specific angular momentum evolution for 100
objects randomly chosen from model M1. Simulated data are shown as
light grey symbols, squares for disk objects and crosses for
diskless ones. Two stars are highlighted as magenta filled circles
(when in the disk phase) or crosses (when in the diskless phase)
to illustrate the evolution of a single star. The black symbols are
for the \DXIV\ ONC sample of stars of spectral type later than M2
whose $j$ values were recalculated using gyration and stellar radii
from the stellar evolutionary models of \citetads{1998A&A...337..403B}.
Squares represent Class II stars, crosses Class III stars. {\it
Right:} Logarithm of age {versus} the logarithm of median $j$
for disk objects (black squares) and diskless objects (black crosses). Error
bars (black for disk objects, red for diskless ones) show the rms
of $\log \; \langle j \rangle$ values around the median. The $\gamma$
values express the time dependency $\langle j \rangle \propto
t^\gamma$. \label{plotj}}
\end{figure*}

According to the conditions imposed in our simulations (equations
\ref{Pdisk} - \ref{Pdiskl}), the specific angular momentum is
constant for diskless objects, but it decreases for disk ones owing
to the stellar radius contraction at the pre-MS phase.  This can
be seen in Fig. \ref{plotj} where we plot $j$ as a function of age
for 100 objects extracted from model M1.  We overplot the specific
angular momenta of \DXIV\ ONC sample taking into account the different
ages of the stars.  Although initially the specific angular momentum
has the same value for disk and diskless stars, which is  expected
since they share the same initial rotational conditions, in the
course of the system's evolution the median values become lower for
disk objects, which decrease with time as $\langle j \rangle \propto
t^{\gamma}$ with $\gamma = - 0.71 \pm 0.02$, while for diskless
ones we obtain $\gamma = -0.20 \pm 0.01$, a much weaker time
dependency. This is not the same result we  obtained in \PaperI\
for solar-type stars where $\gamma_\mathrm{d} = -0.65$ and
$\gamma_\mathrm{dl} = -0.53$ for disk and diskless stars, respectively.
Also, the time exponent for stars with disks is smaller than is
expected from polytropic models ($j \propto t^{-0.66}$).  For
diskless stars it is the opposite; the value shows a much weaker
time dependency, closer to the constancy expected individually.  As
can be seen in Fig. \ref{momI}, for M$_\ast \geq 0.2$ M$_\odot$ the
moment of inertia falls off steadily, while for less massive stars
and BDs it remains constant for a time scale that increases with
decreasing stellar mass. Then, a VLM star with a disk will experience
a stronger decrease in its specific angular momentum than a BD disk
object. As said before, this is due to the different internal
structure evolution of BDs and VLM stars. We also examine separately
the slopes for BDs and stars with M$_\ast \geq 0.2$ M$_\odot$.  We
obtain a slope for stars with disks equal to $-0.690 \pm 0.005$,
not very far from the value expected from the polytropic case,   but
this exponent is  $-0.56 \pm 0.05$ for BDs. We thus observe  a
flatter $j$ decrease for accreting BDs since their momentum of
inertia hardly changes up to 3 Myr.

In summary,  model M1 is able to reproduce i) the period distributions
and medians of the ONC provided by \DXIV, of NGC 2362, and of the
Upper Sco association; ii) the period $\times$ disk fraction for
the ONC obtained by \RLX\ for BDs and by \DXIV\ for VLM stars; and
iii) the absence of correlation between the mass accretion rate and
period for stars with disks in agreement with \CX. However, it fails
to reproduce i) the period distributions and medians of the ONC
provided by \RLX\ and of the combined sample of $\sigma$ Ori; ii)
the period $\times$ disk fraction for the same sample; and iii) the
period-mass relation. In order to try to improve the agreement with
these three points, we will investigate other models that address
these discrepancies.

\subsubsection{Model M2: initial bimodal period distributions}
\label{M2}

\begin{figure*}
\centerline{\includegraphics[width=0.35\textwidth]{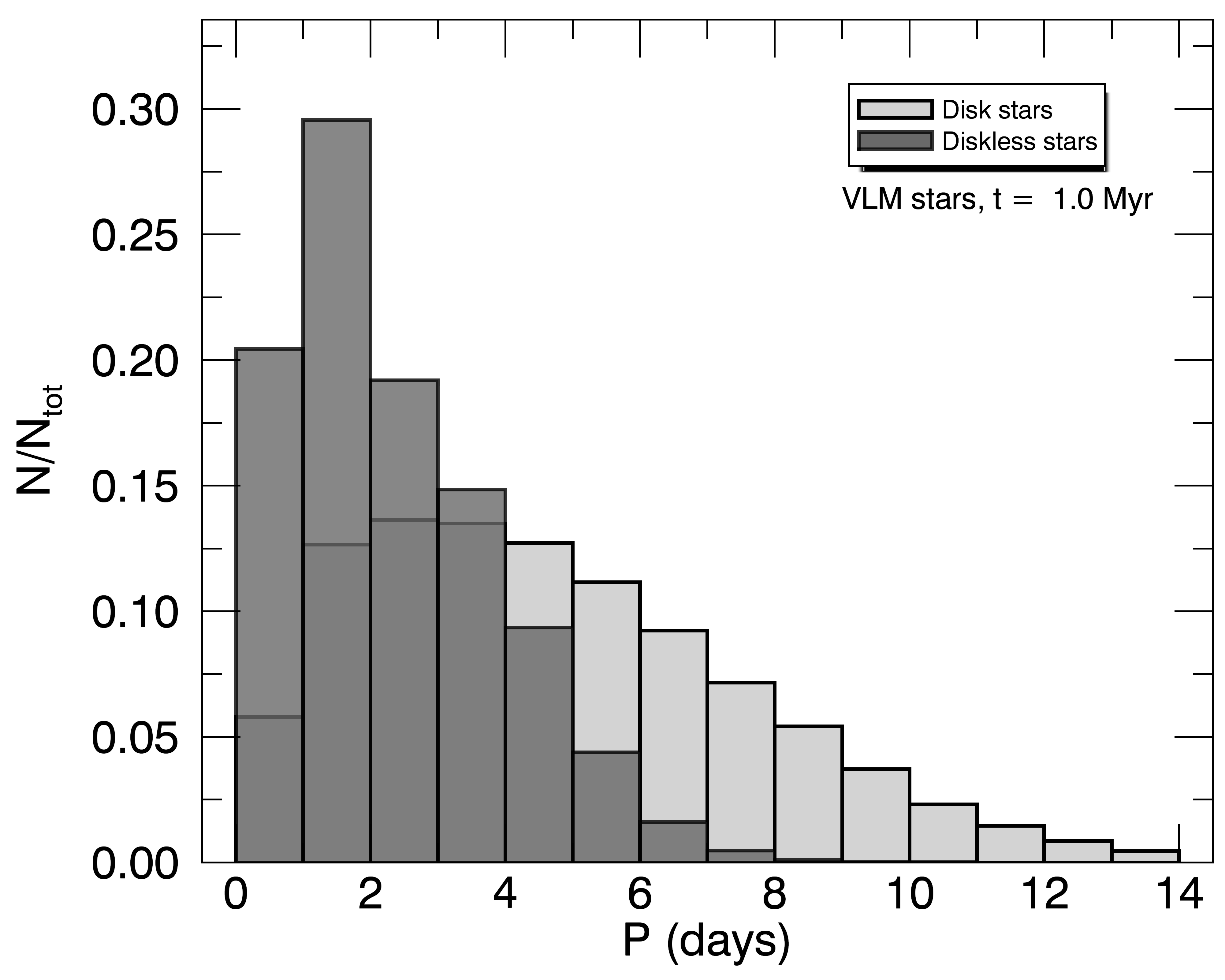}
\includegraphics[width=0.35\textwidth]{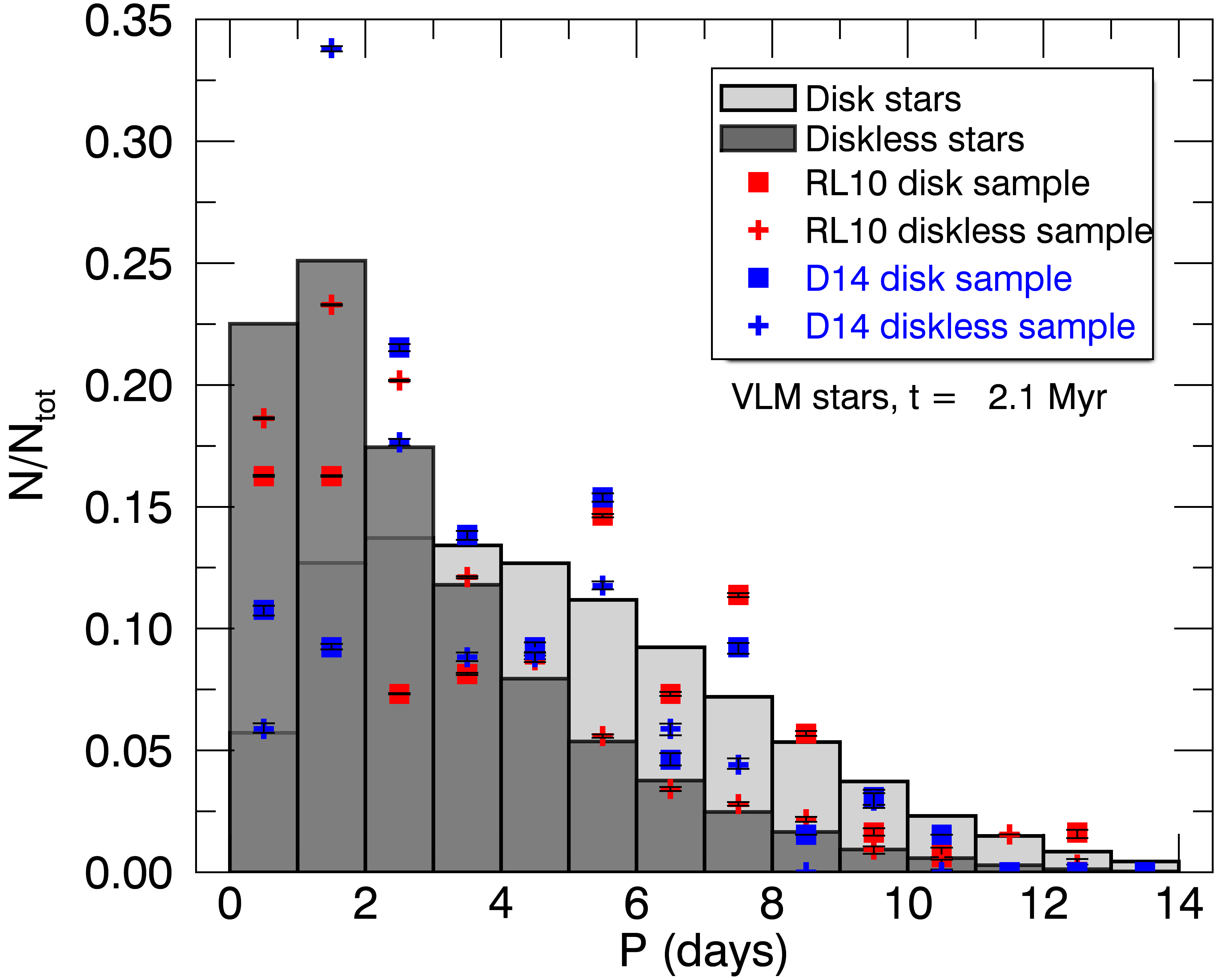}}
\centerline{\includegraphics[width=0.35\textwidth]{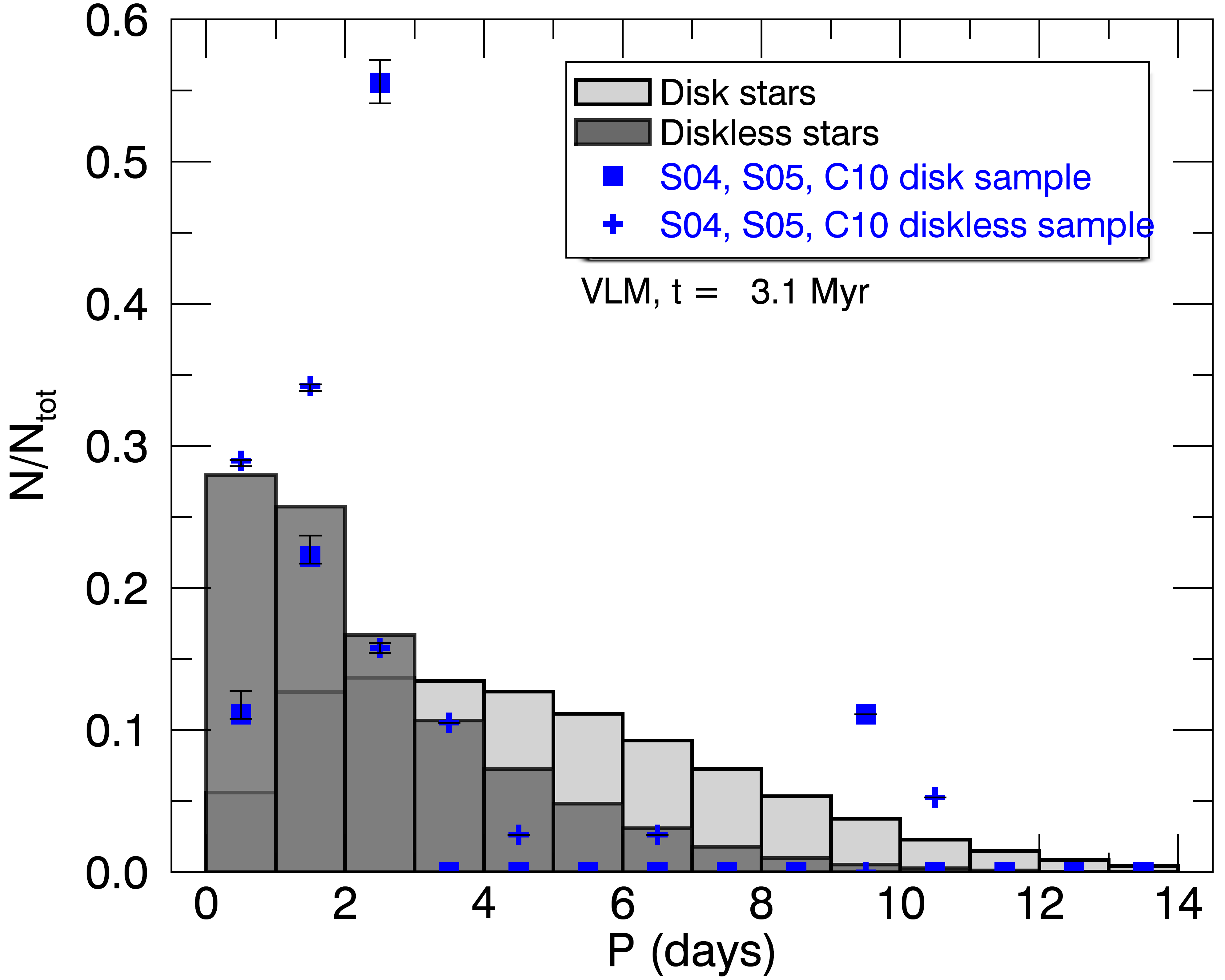}
\includegraphics[width=0.35\textwidth]{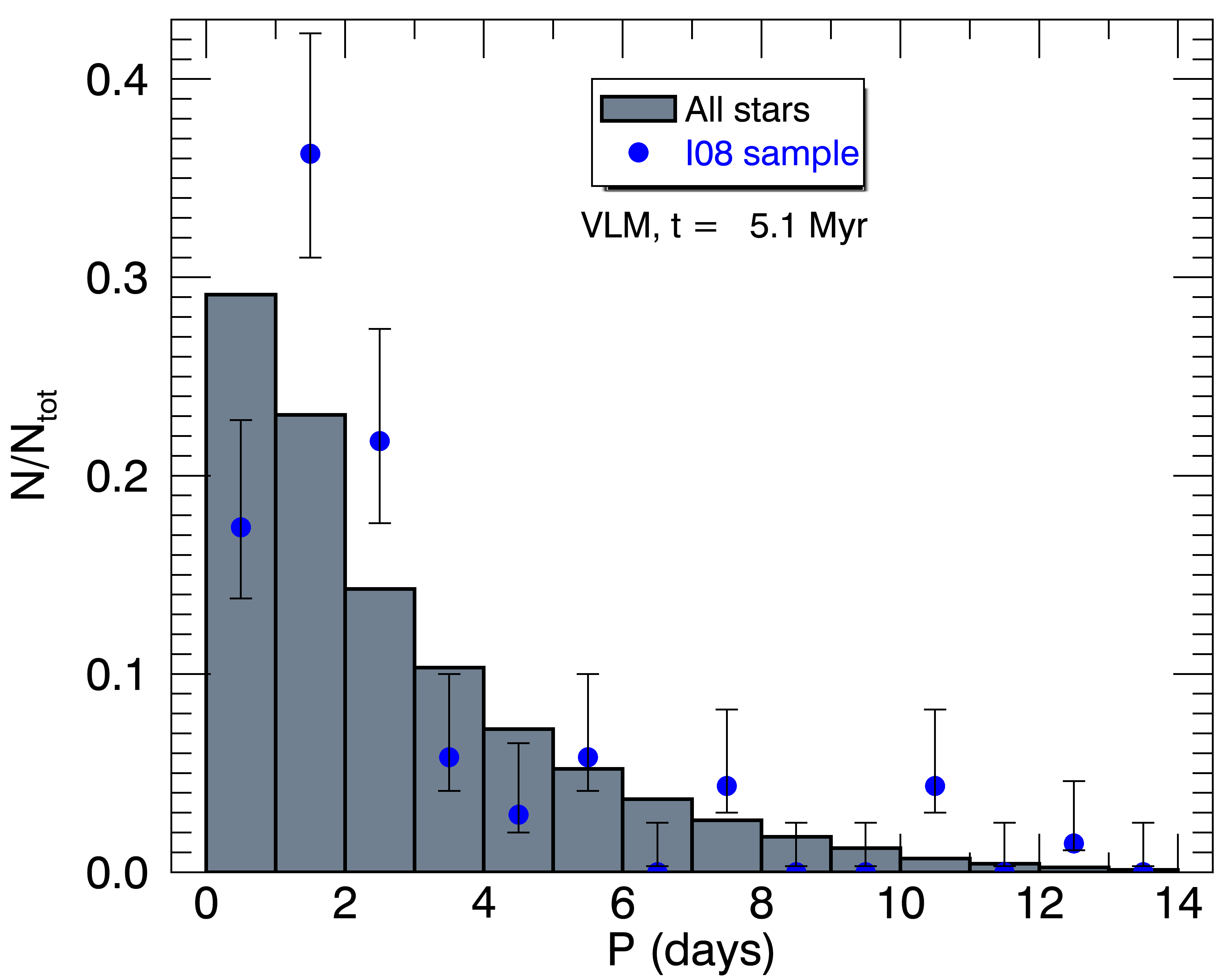}}
\caption{Period distributions for the VLM range for model M2 at 1.0
Myr, 2.1 Myr (two upper panels), 3.1 Myr, and  5.1 Myr (two lower
panels).  Bars, symbols, and error bars have the same colour and
form coding of Fig. \ref{distPM1}.  \label{distPM2}}
\end{figure*}

In order to try to reproduce the bimodal period distribution of VLM
stars in the ONC seen by \citetads{2010A&A...515A..13R} we ran model
M2, which differs from model M1 by the initial period distributions
(see Table \ref{paramod}).  We set a bimodal period distribution
at 1.0 Myr, but only for VLM stars. For BDs the initial period
distribution is the same used in model M1. For VLM stars the mean
period of the diskless distribution is  2.0 days and the minimum
period is  0.2 day.  For stars with disks, the mean period is 3.0
days and the minimum period is  0.5 day. In Fig.  \ref{distPM2} we
show the distributions of rotational periods for VLM stars at 1.0
Myr, 2.1 Myr, 3.1 Myr, and 5.1 Myr.  The bimodality imposed at the
beginning of the simulations is clearly visible at later times. In
terms of medians (see Table \ref{PmKol}), there is an improvement
between the model and the observational data from \RLX, but this
is not the case at 3.1 Myr and for the $\sigma$ Ori data. At 5.1
Myr there is no change in what was seen  with model M1.

The $\chi^2$ test between model M2 and the \RLX\ data shows that
the fit between the diskless VLM histogram data is now acceptable,
but this is not true for the disk VLM histogram data which again
do not match the \RLX\ observations. There was no change for the BD
distribution.  The K-S test for \DXIV\ data now show probabilities
of 40\% that the disk distribution and  0.4\% that the diskless
distribution and the numerical distributions are drawn from the
same population against 59\% and 18\%, respectively, obtained from
model M1 (see discussion at section \ref{M1}).

At 3.1 Myr, the K-S probability degraded for the VLM disk distribution
which now is only  0.3\%, while with model M1 it is 2\%. For the
diskless distribution, to the contrary, it is greatly improved,
reaching 49\%  agreement with the observational $\sigma$ Ori data
(against 8\% using model M1).  At 5.1 Myr, the probability that the
numerical VLM period distribution and the sample from \IVIII\ are
the same is  6\%, smaller than the 16\% value obtained with model
M1.  Since we have not changed the distributions for the BD objects,
the probabilities at 10.1 Myr have not changed.

When we analyse disk fractions as a function of the period for VLM
stars we note that -- compared to model M1 -- the agreement is
better with the \RLX\ sample and worse with the \DXIV\ one, although
the RL10 data still do not fit the numerical data.  The match with
the \DXIV\ data is still better than it is with the \RLX\ data that
justified model M2.

At 3.1 Myr, the $\chi^2$ value slightly decreases and the hypothesis
that the two values match cannot be discarded.

Thus, model M2 reproduces the bimodality at the VLM range at 2.1
Myr, but the agreement with the \RLX\ data is not as good as is
desirable. The bimodal period distribution of VLM stars is
observationally controversial since neither \DXIV\ nor \CX\ found
it for their samples.  Therefore, we consider that model M2 with
its initial bimodal distribution is not adequate.  Instead, in
models M3 and M4 we  keep the same initial period distributions for
disk and diskless stars. In these models we  investigate different
disk lifetimes for BD and VLM stars and address the disk locking
hypothesis.

\subsubsection{Model M3: different disk lifetimes for BD and VLM stars}
\label{M3}

Several observations point to a longer lifetime for disks around
BDs (e.g. Luhman et al. \citeyearads{2008ApJ...675.1375L}; Luhman
\& Mamajek \citeyearads{2012ApJ...758...31L}  and Downes et al.
\citeyearads{2015MNRAS.450.3490D};  see section \ref{secDF} and
Table \ref{tabdiskfracBD}). To examine its impact on our simulations,
we ran model M3 (see Table \ref{paramod}). The disk fraction of all
objects as a function of time is shown in Fig.  \ref{diskfrac} and
the disk lifetimes for BD and VLM stars are depicted in Fig.
\ref{diskfrac3}. Except for this property, the set-up of model M3
is the same as model M1.

\begin{figure*}
\centerline{
\includegraphics[width=0.35\textwidth]{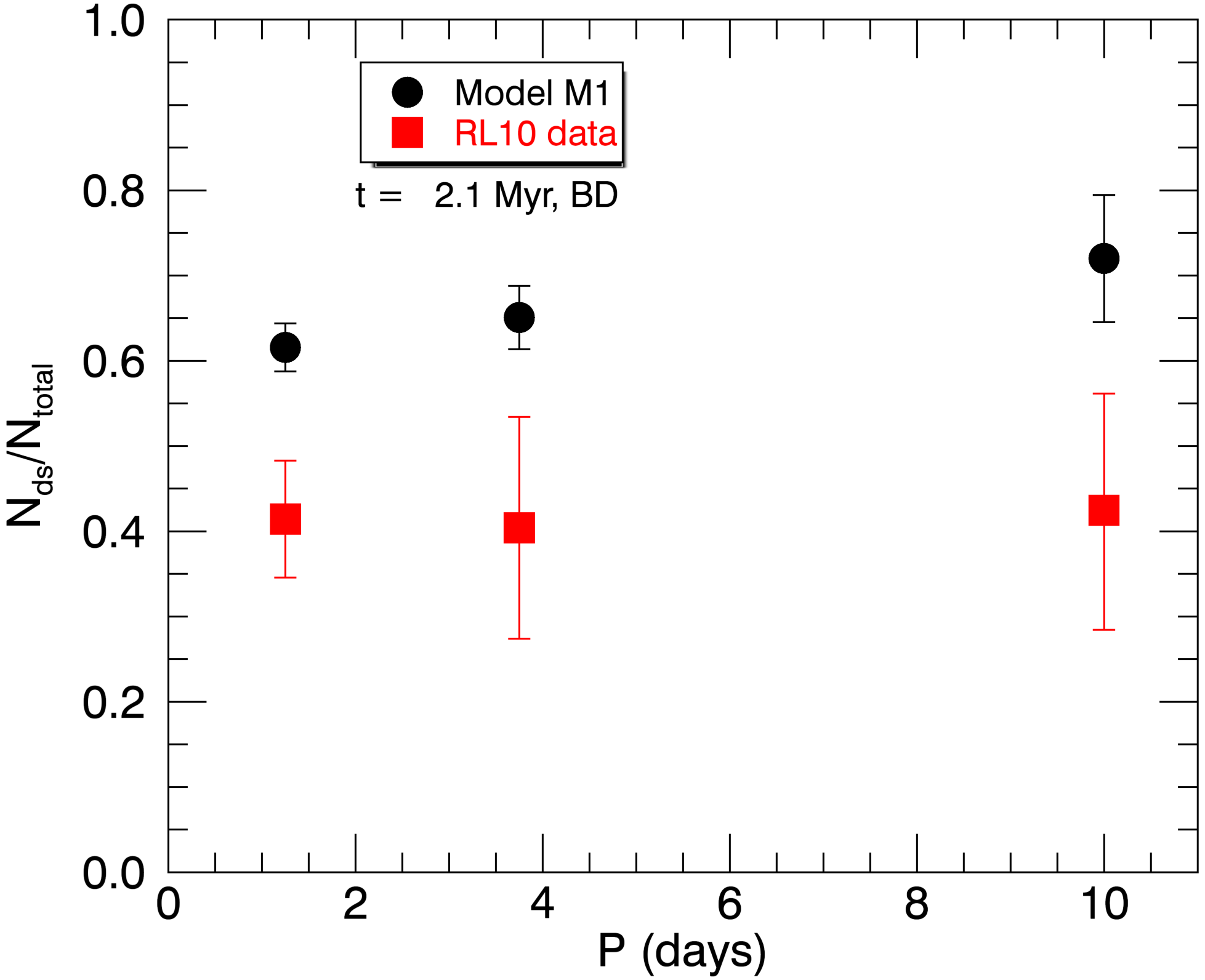}
\includegraphics[width=0.35\textwidth]{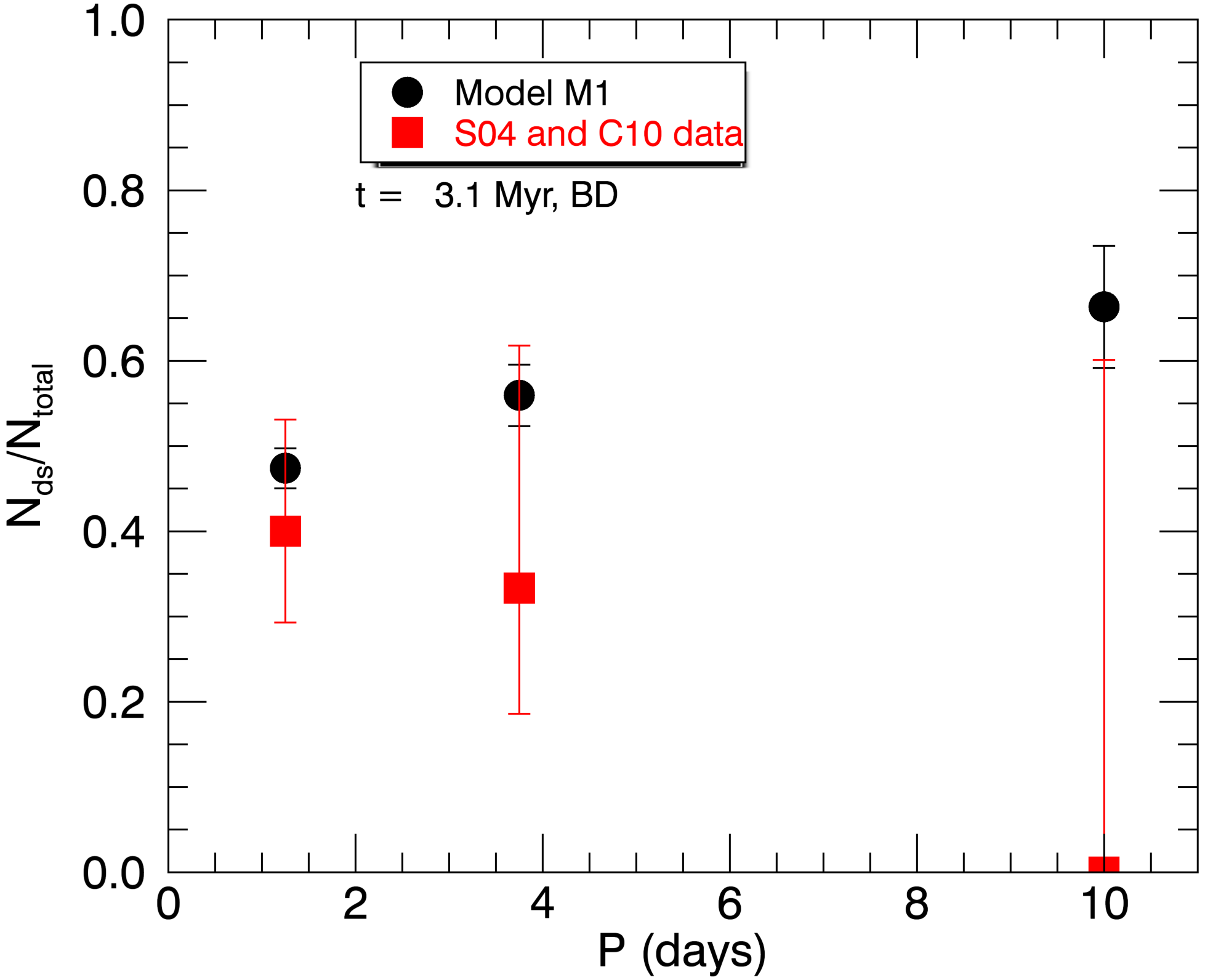}}
\caption{Disk fractions as a function of the period for BDs from
model M3. \label{diskfPM3}}
\end{figure*}

In comparison to model M1 in Figure \ref{distPM1}, the period
distributions are practically unaltered except that there are more
disk than diskless BDs. When we analyse the disk fractions as a
function of the period, there is no change for VLM stars compared
to model M1 (cf. Fig. \ref{diskfPM1}), but the disk fractions change
at the BD regime (Figure \ref{diskfPM3}). Now they are higher than
in model M1. At 2.1 Myr, they reach 70\% which is well above the
\RLX\ fractions, which are around 40\%. At 3.1 Myr, the numerical
fractions show a smooth rising towards longer periods and present
values between 45\% and 65\%,  far from the $\sigma$ Ori fractions.
This is a direct consequence of the longer disk lifetime at the BD
mass regime.  Again, we do not obtain any correlation between the
mass accretion rate and the period.

We conclude that the significant increase in the number of BDs with
disks with short periods predicted by model M3 contradicts the
observations.  Model M3 does not adequately represent the rotational
evolution of BDs.

\subsubsection{Model M4: no disk locking for stars less massive than 0.3
M$_\odot$ \label{M4}}

Model M4 imposes no disk locking for all objects except the 0.4
M$_\odot$ stars.  As shown in Table \ref{paramod}, this model has
the same initial period distributions as model M1 and also the same
mass accretion rates and disk lifetimes.

\begin{figure*}
\centerline{\includegraphics[width=0.35\textwidth]{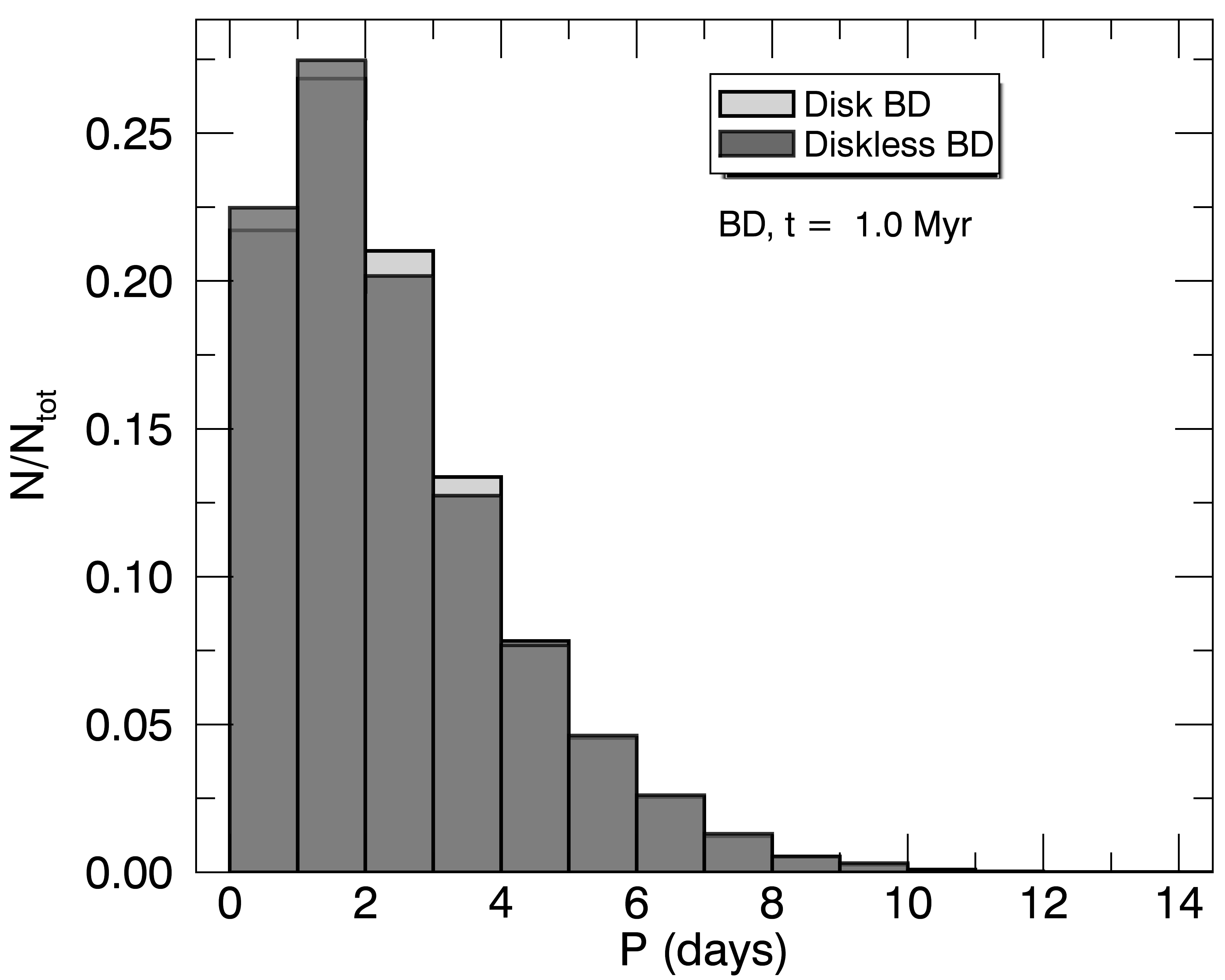}
\includegraphics[width=0.35\textwidth]{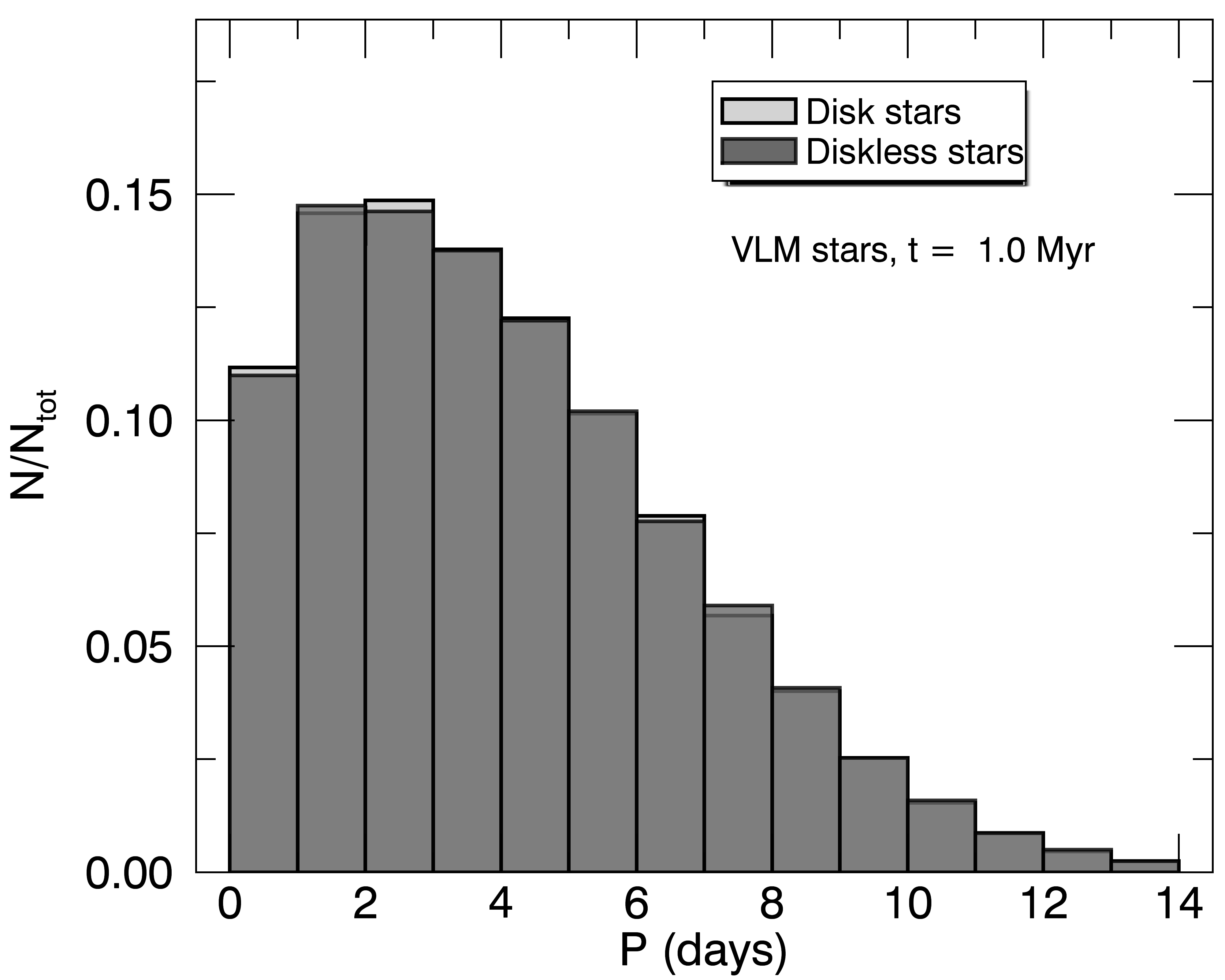}}
\centerline{\includegraphics[width=0.35\textwidth]{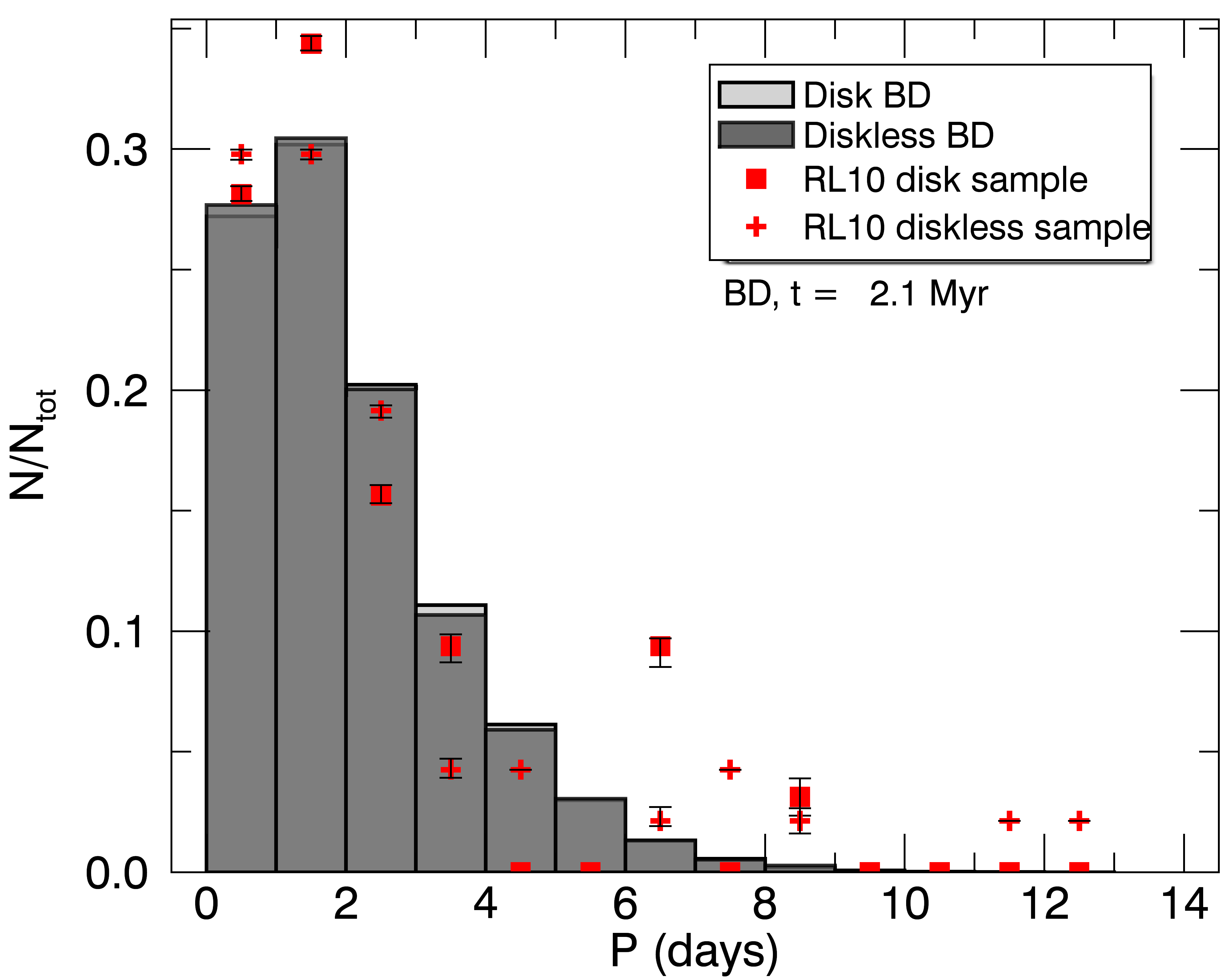}
\includegraphics[width=0.35\textwidth]{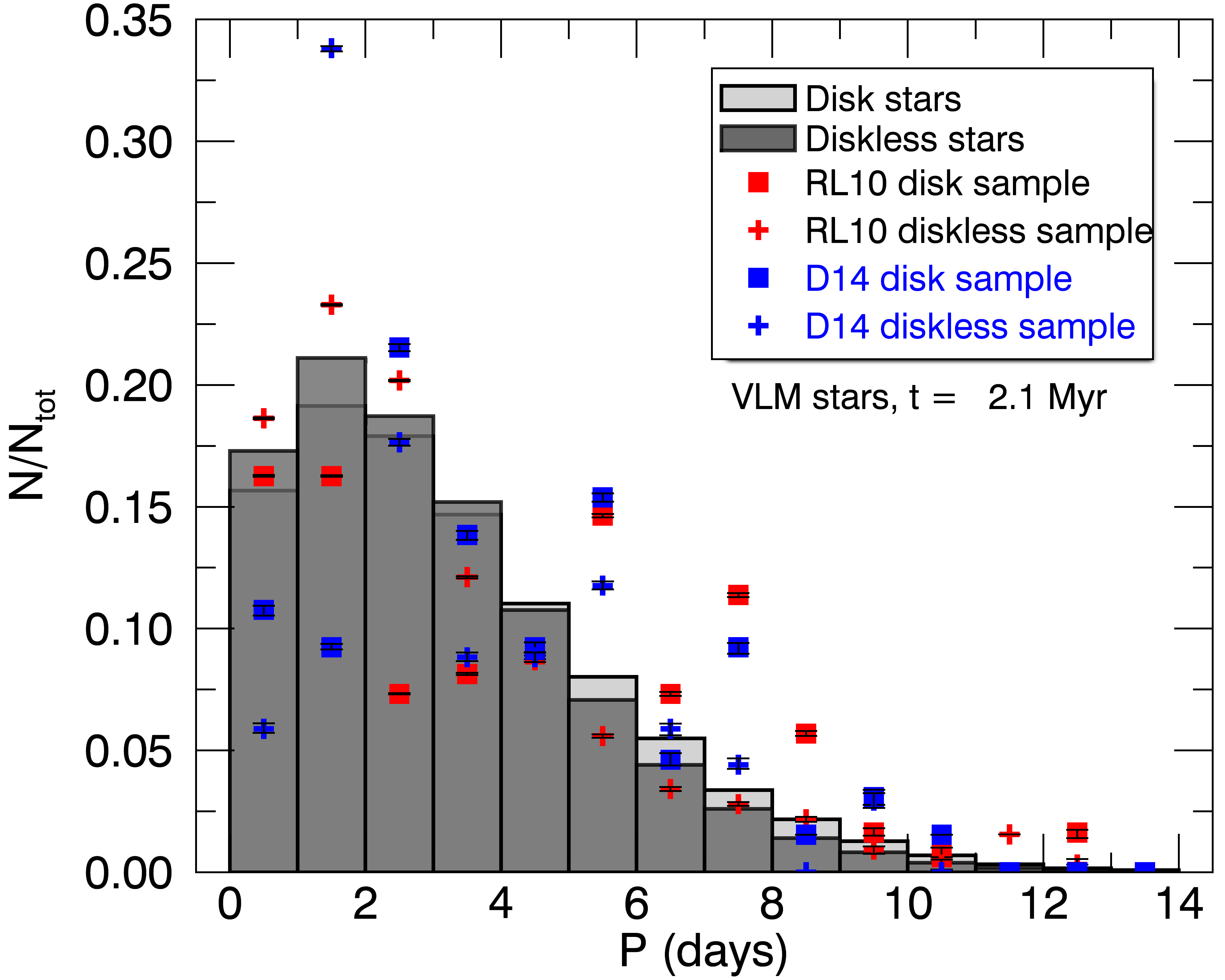}}
\centerline{\includegraphics[width=0.35\textwidth]{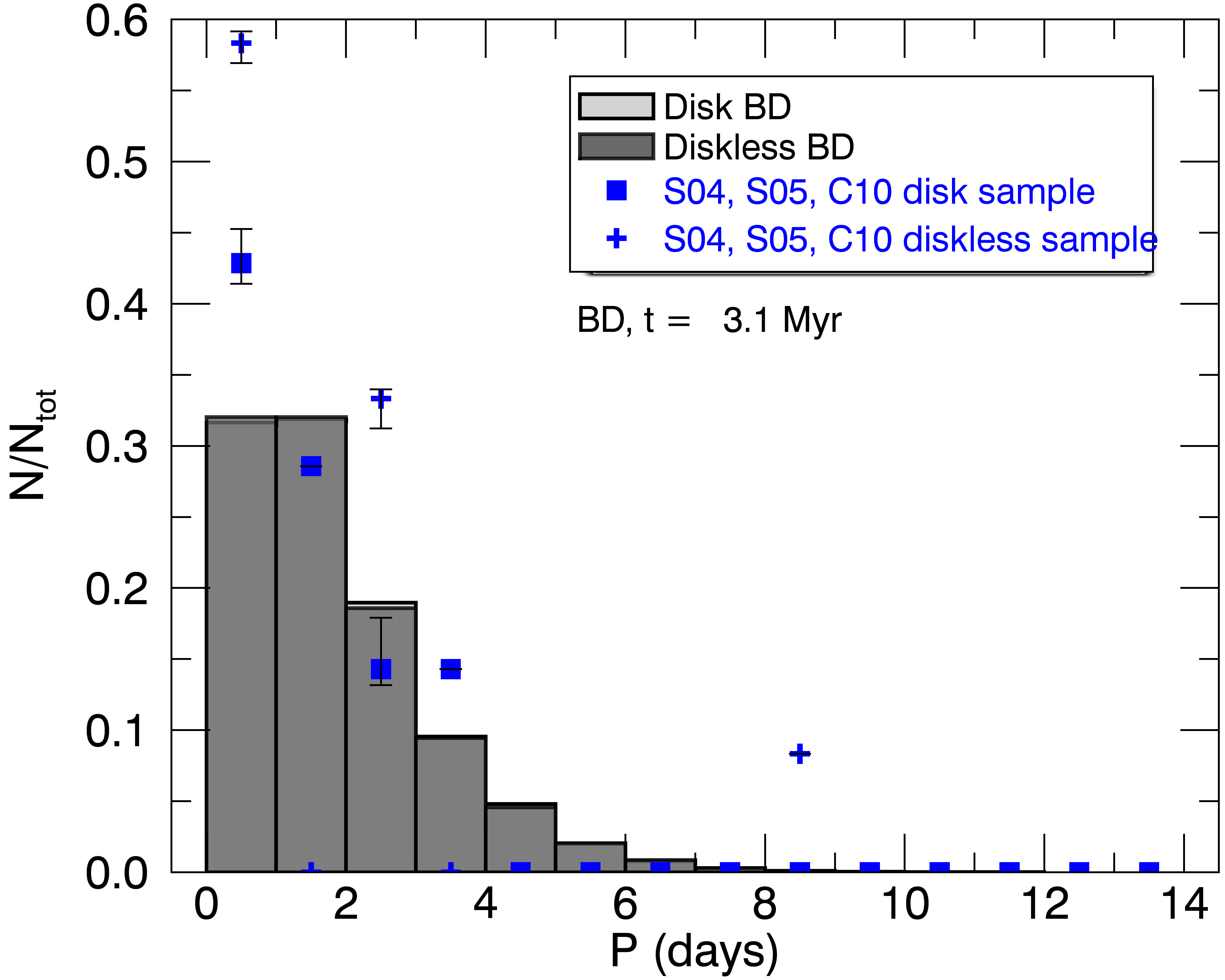}
\includegraphics[width=0.35\textwidth]{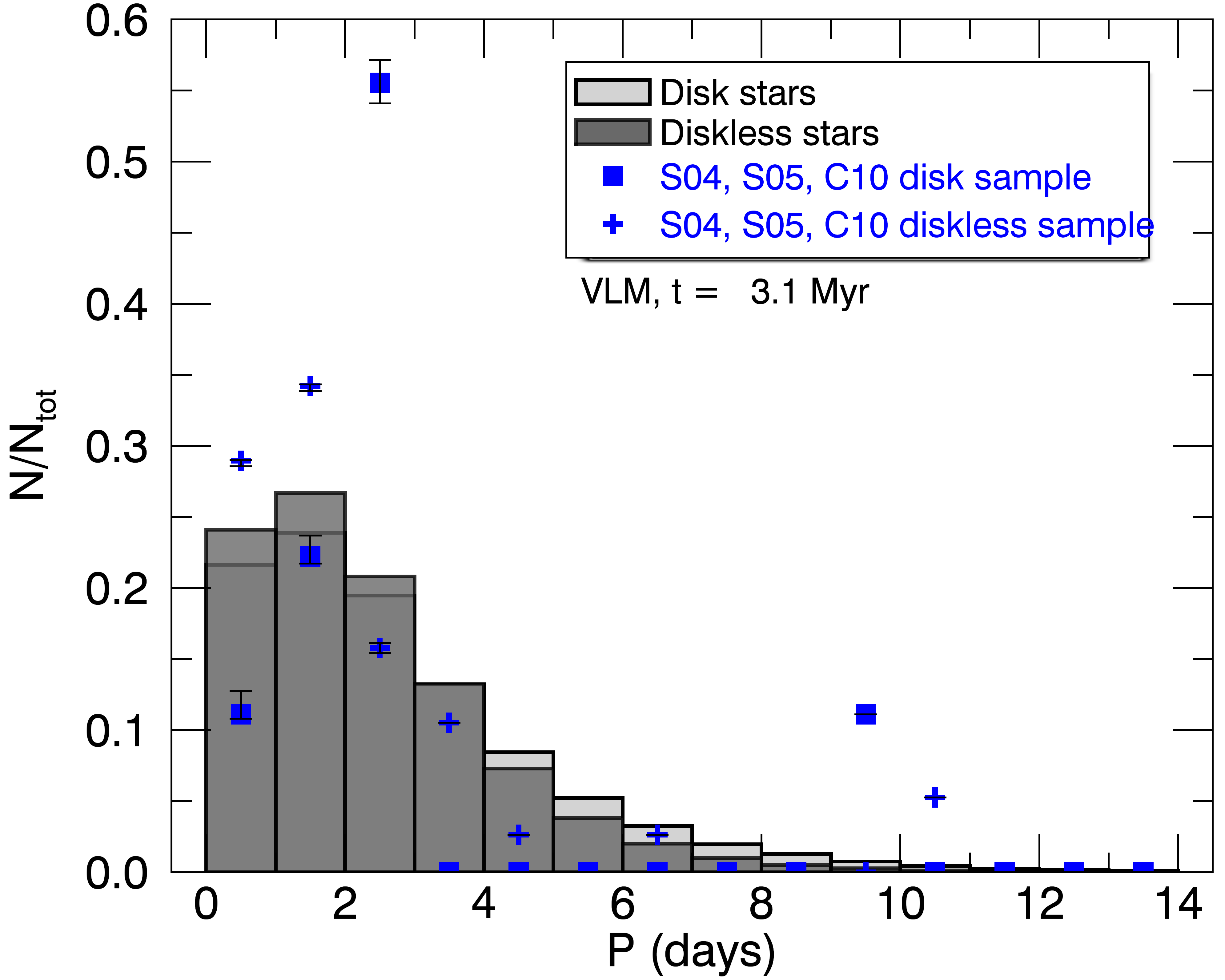}}
\centerline{\includegraphics[width=0.35\textwidth]{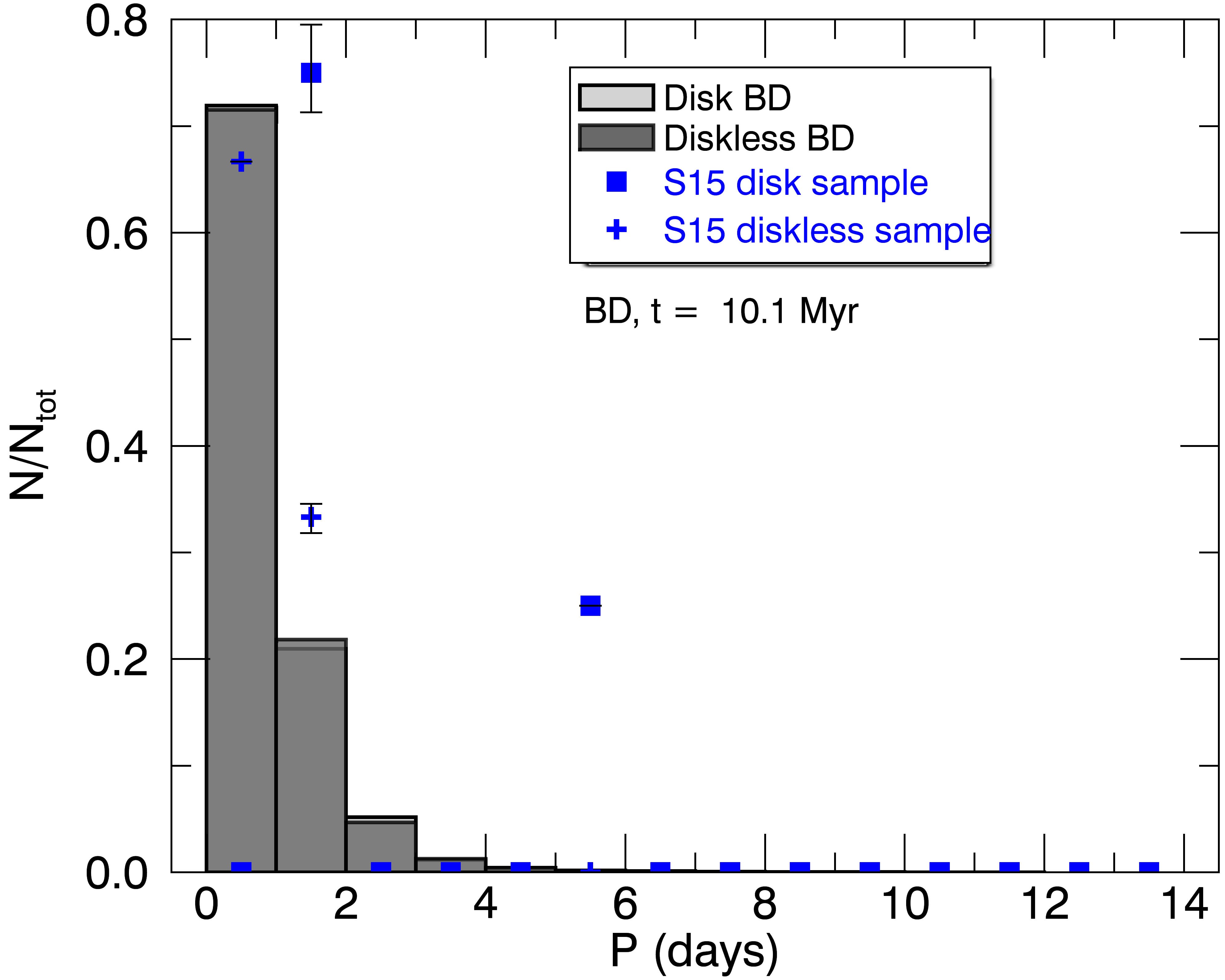}
\includegraphics[width=0.35\textwidth]{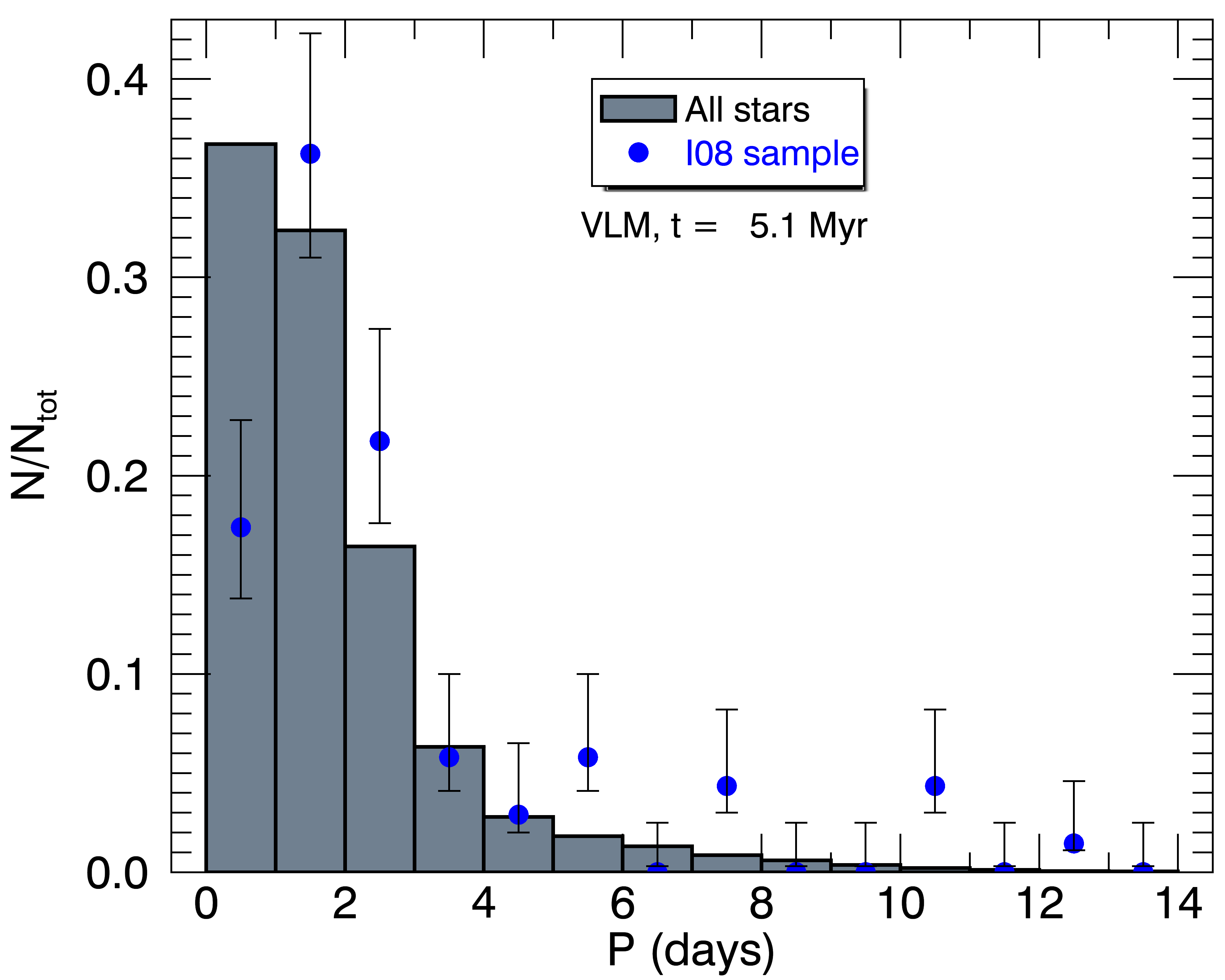}}
\caption{Same as Fig. \ref{distPM1} but for model M4. \label{distPM4}}
\end{figure*}

Owing to the results of models M1 to M3, for model M4  we should
expect  the total disk fraction to follow the curve shown in Fig.
\ref{diskfrac}, which is indeed the case. The period distributions,
however, are different from those of previous models (Fig.
\ref{distPM4}).  The free spin-up of all BDs erases the differences
between the medians of disk and diskless BD objects and the two
curves show the same values at all ages (see Table \ref{PmKol}).
As can be seen, in comparison with the observations, the agreement
is improved at 2.1 Myr with the \RLX\ BD medians, but at 3.1 Myr
and at 10.1 Myr it becomes worse. At 3.1 Myr, the diskless BD median
is greater than the observational value, while at 10.1 Myr the
medians have moved to shorter periods and thus farther away from
the observational ones.

Concerning the period distributions, the agreement with \RLX\ data
seen through the $\chi^2$ tests is below the confidence level. The
K-S tests show a probability of only 3\% of compatibility with the
\DXIV\ period distribution for stars with disks and of 11\% with
the same distribution but for diskless stars.

At 3.1 Myr the probabilities of sharing the same population are
90\% and 16\% among observational and numerical BD disk and diskless
period distributions, and  47\% and 30\% for disk and diskless VLM
stars, respectively.

At 5.1 Myr the disagreement between the distributions is great: the
K-S test provide a probability of 0.02\% that the observational and
theoretical samples are the same.

Finally, at 10.1 Myr we obtain a probability of 0.2\% that the disk
BD period distributions agree and of 46\% that the diskless BD
distributions are the same.

So, except at 3.1 Myr, model M4 is not as good as model M1 in
reproducing the observational period distributions.

\begin{figure*}
\centerline{
\includegraphics[width=0.35\textwidth]{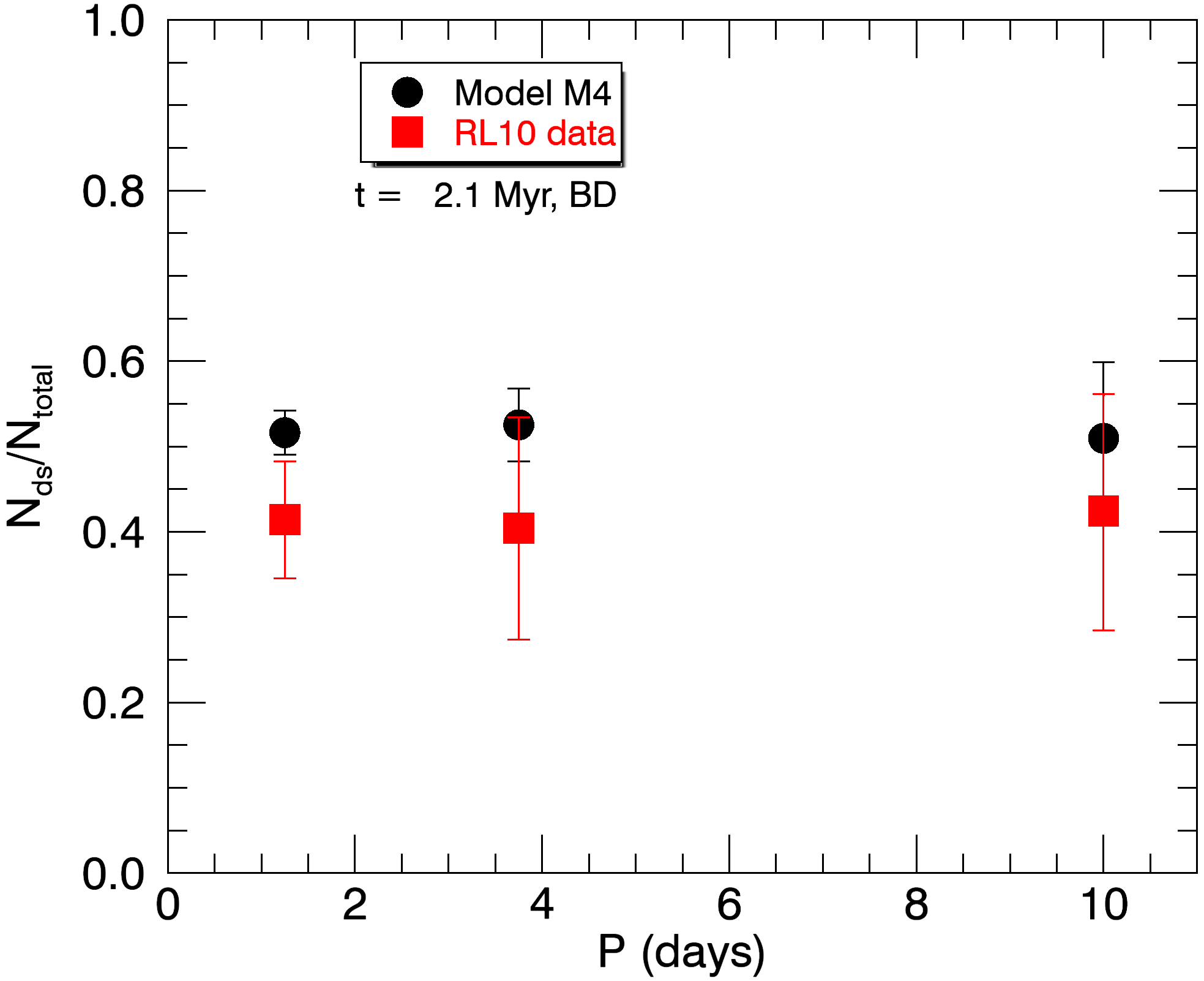}
\includegraphics[width=0.35\textwidth]{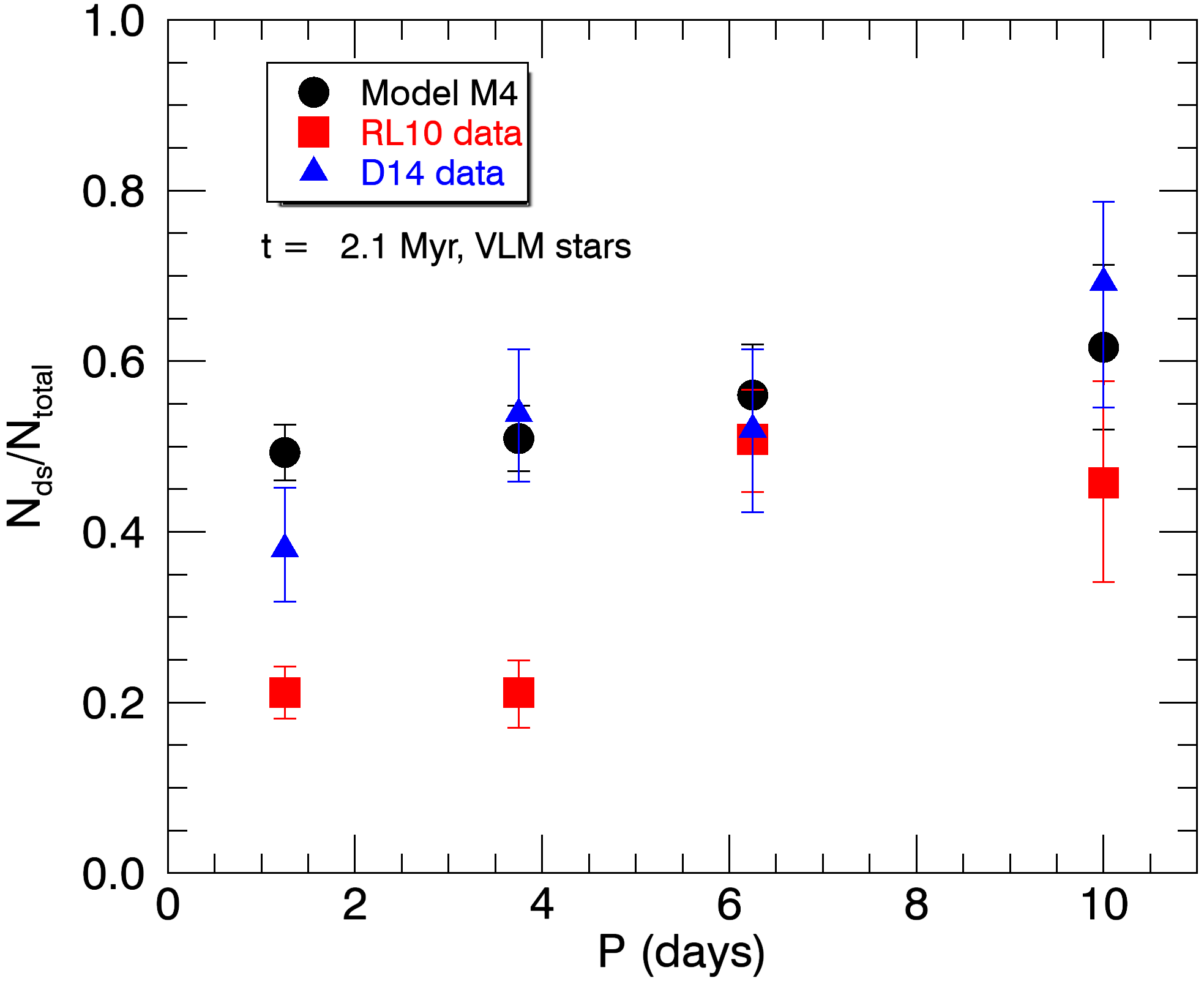}}
\centerline{
\includegraphics[width=0.35\textwidth]{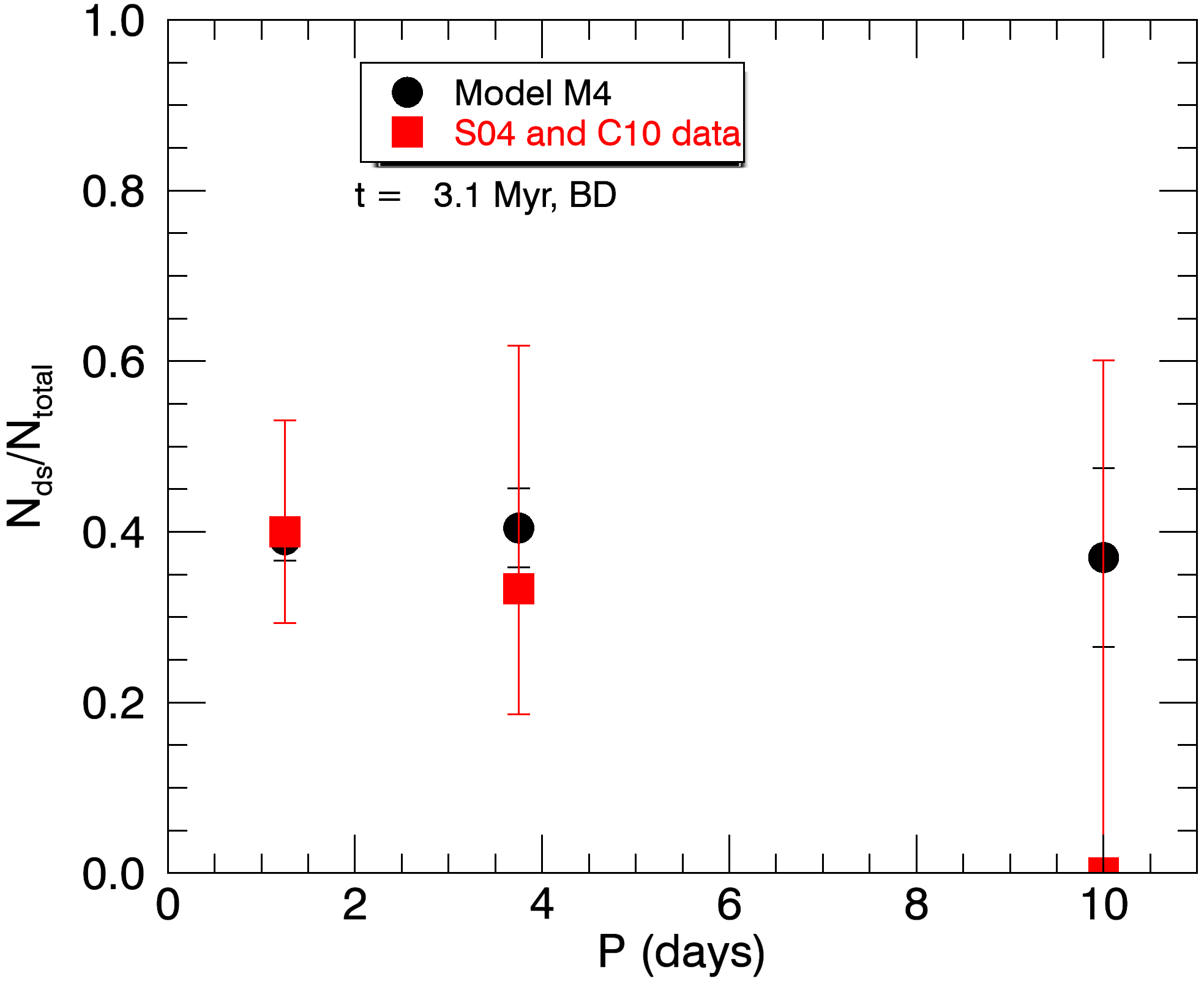}
\includegraphics[width=0.35\textwidth]{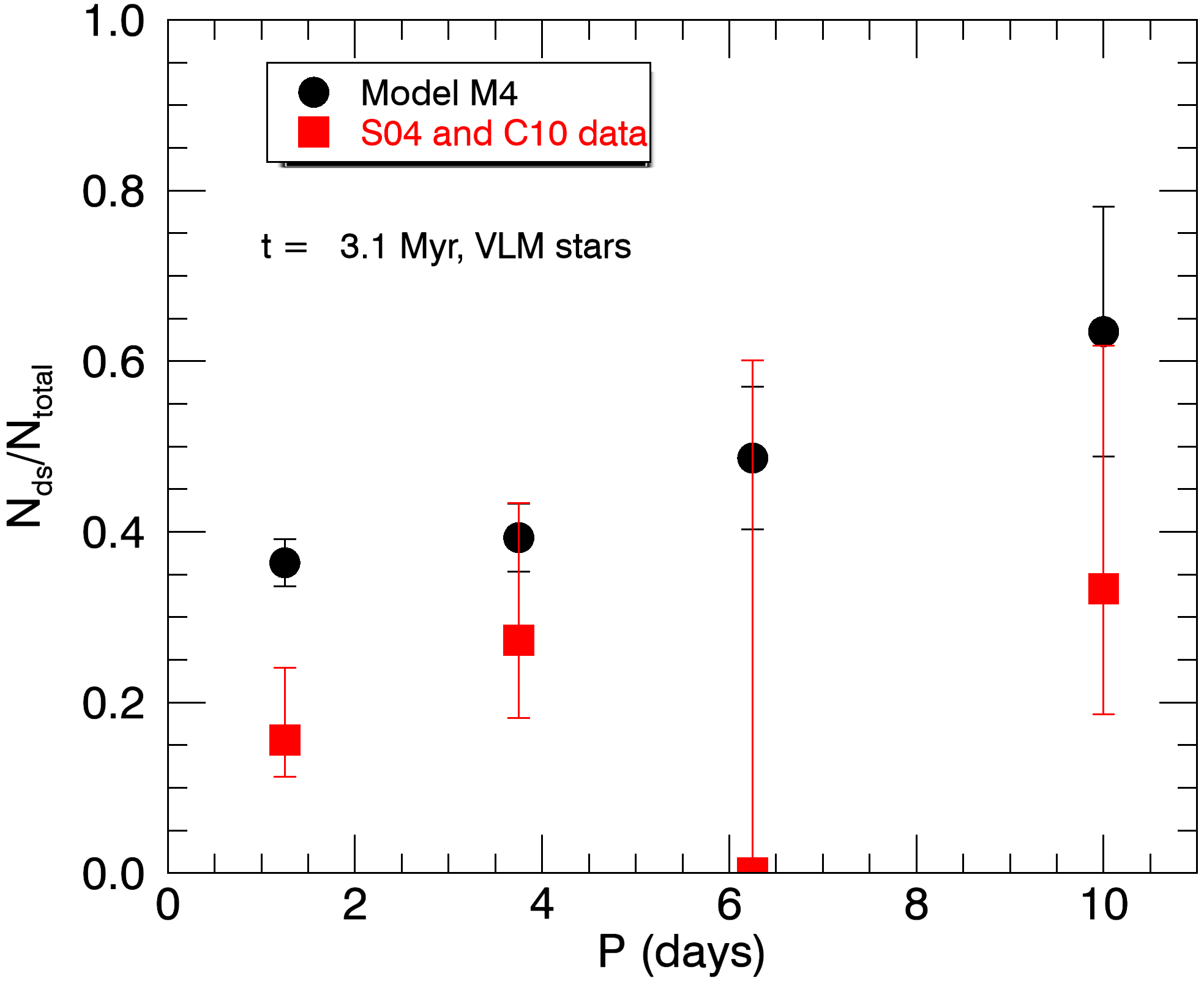}}
\caption{Same as Fig. \ref{diskfPM1} but for model M4. \label{diskfPM4}}
\end{figure*}

In Figure \ref{diskfPM4} we show plots of disk fraction versus
period. The results are different from what was seen in model M1
and in model M3. The disk fractions for BDs remains approximately
constant around 50\% at 2.1 Myr in agreement with the observations
and around 40\% at 3.1 Myr. For VLM stars, we do not observe the
high disk fractions at longer periods that are seen in model M1,
but this is not very far from what is observed. The $\chi^2$ tests
show that model M4 fits the BD and VLM disk fractions from the
$\sigma$ Ori sample at 3.1 Myr.  At 2.1 Myr, the agreement with
\RLX\ BD data is inside the confidence level and the same occurs
with the fit between our data and the \DXIV\ sample. However, this
is not the case with VLM data from \RLX.

The specific angular momentum evolution on the other hand is
drastically different.  The time dependency exponents are now very
low, and for the BD sample they are practically zero in accordance
with the simulation's assumption of constant specific angular
momentum for diskless stars. In contrast with model M1 for which
the sequential release of a group of stars from the disk causes a
decrease in the mean specific angular momentum, in model M4 -- since
all stars except those of 0.4 M$_\odot$   are already free to spin
up -- the group as a whole evolves at almost constant angular
momentum.  In Table \ref{tabgamma2} we summarize the $\gamma$
exponents obtained so far with our models.  The probability that
the numerical and \DXIV\ specific angular momentum distributions
are similar rises to 31\% for disk VLM stars and 41\% for diskless
VLM stars, against 6\% and 20\%, respectively, obtained with model
M1.

\begin{table}
\caption{Specific angular momentum time dependency exponent ($\gamma$)
\label{tabgamma2}}
\centering
{\small
\begin{tabular}{ccc}
\hline\hline
Model & $\gamma_\mathrm{d}$ & $\gamma_\mathrm{dl}$ \\
\hline
M1 & $-0.71 \pm 0.02$ & $-0.20 \pm 0.01$ \\
M2 & $-0.71 \pm 0.02$ & $-0.37 \pm 0.02$ \\
M3 & $-0.74 \pm 0.02$ & $-0.26 \pm 0.01$ \\
M4 & $-0.156 \pm 0.004$ & $-0.084 \pm 0.002$ \\
\hline
\end{tabular}
\tablefoot{$\gamma_\mathrm{d}$ gives the specific angular momentum time
dependency for stars with disks while $\gamma_\mathrm{dl}$ shows the
same quantity for diskless ones.}}
\end{table}

\begin{figure}
\begin{center}
\includegraphics[width=0.39\textwidth]{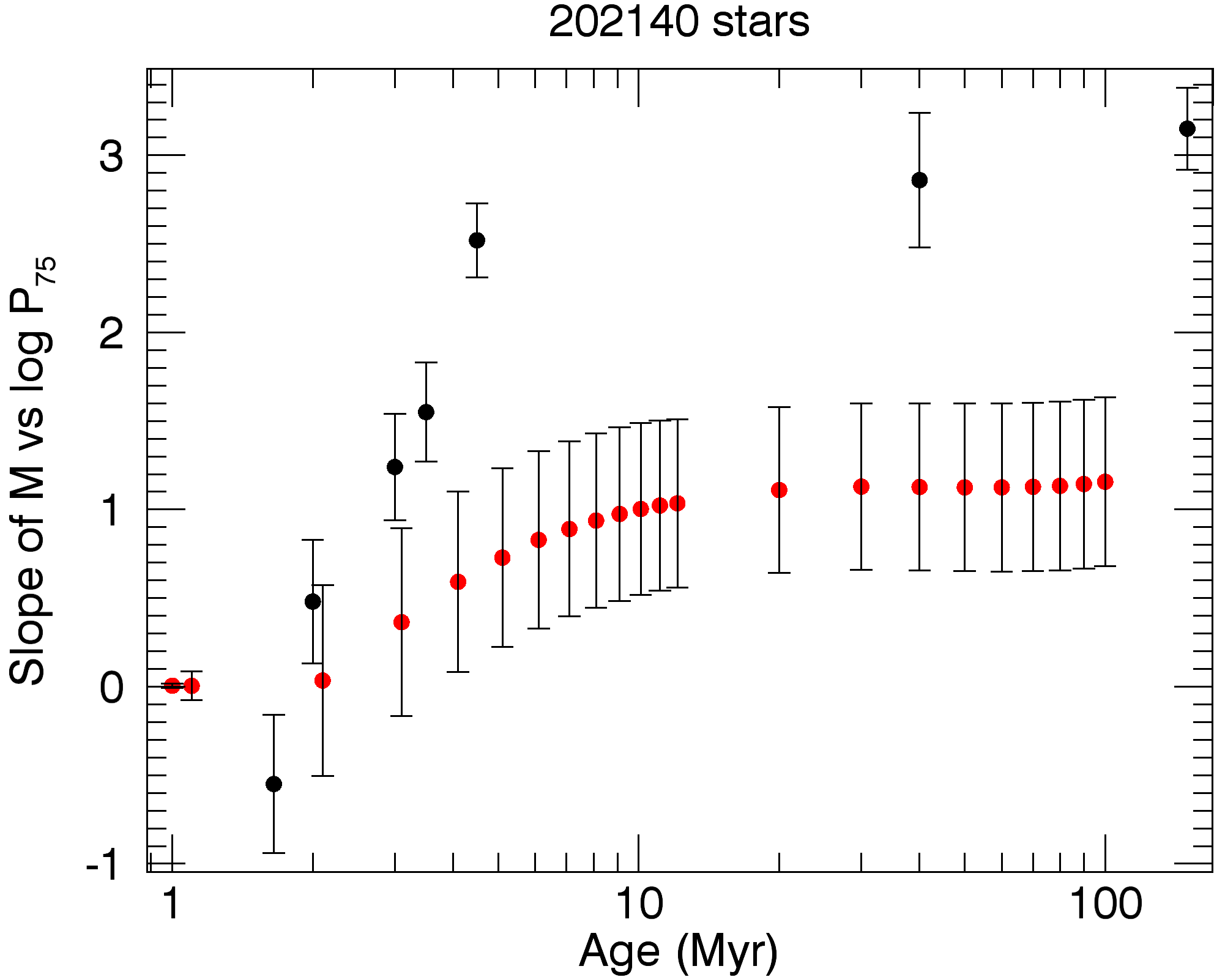}
\end{center}
\caption{Slopes of mass {versus} the logarithm of the
75$^\mathrm{th}$ percentile of the rotational period plots at ages
ranging from 1.0 Myr to 100 Myr obtained from model M4. \label{massPrelM4}}
\end{figure}

We analysed the period-mass relation in model M4, with no disk
locking for stars less massive than 0.4 M$_\odot$. In Fig.
\ref{massPrelM4} we show  the slopes of plots $M \, \times \, \log
P_{75}$ (cf.  Fig. \ref{massPrel}). In comparison to what was
obtained with model M1, we  observe a correlation between the
rotational period and the mass for ages older than 3.0 Myr in
agreement with the \citetads{2012ApJ...747...51H} trend with,
however, a shallower slope. We thus confirm the results of \PaperI\
that suggest that VLM stars and BDs are not as efficiently disk
locked as solar-type stars when accreting from their disks.

In summary, model M4 is worse in reproducing the period distributions,
but is equivalent in accounting for the disk fractions as a function
of the period in comparison to model M1. Model M4 predicts practically
no angular moment evolution, which is in disagreement with the
observations of \DXIV. However, the stronger correlation between the
mass and the period favours this model.

\section{Conclusions} \label{conclusions}

We have analysed four Monte Carlo models in order to investigate
the rotational behaviour of very low-mass stars and brown dwarfs
with masses below 0.4 M$_\odot$ during the early pre-main sequence.

Stars in this mass range rotate differently from their  solar mass
counterparts. In order to fit the period distributions for BDs, we
had to consider that they rotate faster than VLM stars, whose period
distribution is also centred at a shorter period compared to that
of the solar mass stars considered in \PaperI. There is no need to
consider bimodal distributions for BDs and VLM stars at the beginning
of the simulations, contrary to what was obtained for solar mass
stars.  All this suggests a weaker influence of the disk on the
angular momentum evolution of BD  and VLM stars.  It is possible
to  speculate that the different magnetic field topology found among
the lowest mass stars is the cause of their faster rotation rate.

Of all four models outlined here, model M1  (with an initial
single-peak period distribution for disk and diskless stars and BDs
as faster rotators) best reproduces the observations.  Considering
an initial bimodal period distribution as  was done in model M2
does not improve the agreement with the observational data. Model
M1 nevertheless has shortcomings; most notably, it does not reproduce
the period-mass relationship observed for VLM stars. Changing the
average disk lifetime of BDs (model M3) does not improve the results.
Relaxing the disk locking hypothesis for the lowest mass stars, as
was required in \PaperI, yields a period-mass trend in the same
direction as observed, but is poorer in reproducing the evolving
rotational distributions. Hence, none of the models developed here
can reproduce all available observational constraints simultaneously,
though they offer useful guidelines regarding initial period
distributions, disk lifetimes, and the disk locking efficiency at
very low masses. We also note that the observational results do not
always seem consistent, and much remains to be done to properly
characterize the rotational properties of young VLM stars and brown
dwarfs.

\begin{acknowledgements}

MJV would like to thank the financial support provided by CAPES
(fel\-low\-ship n. 2565-13-7) un\-der the pro\-gram ``Sci\-ence
with\-out borders'' and by the project PRO\-CAD – CNPq/CAPES number
552236/2011-0.  JB ac\-knowl\-edges the sup\-port of ANR grant 2011
Blanc SIMI5-6 020 01 {\it Toupies: Towards understanding the spin
evolution of stars} (http://ipag.osug.fr/Anr\_Toupies/).  The authors
thank the referee for very useful suggestions that helped to improve
the quality of the paper.

\end{acknowledgements}

\bibliographystyle{ms} 
\bibliography{ref} 

\end{document}